\documentclass[12pt]{article}
\pdfoutput=1

\usepackage{draft} 
\usepackage{hyperref}
\usepackage{graphicx,color,subfig}
\usepackage{cite}
\usepackage{mciteplus}
\usepackage{skak}
\DeclareFontFamily{OT1}{pzc}{}
\DeclareFontShape{OT1}{pzc}{m}{it}{<-> s * [1.10] pzcmi7t}{}
\DeclareMathAlphabet{\mathpzc}{OT1}{pzc}{m}{it}

\def\be#1\ee{\begin{align}#1\end{align}}

\begin{document}

\unitlength = .8mm

\begin{titlepage}
	\rightline{ CALT-TH 2016-030 }
	 \rightline{ MIT-CTP/4824 }
	 \rightline{ PUPT-2511 }
\begin{center}

\hfill \\
\hfill \\
\vskip 1cm

\title{$(2,2)$ Superconformal Bootstrap in Two Dimensions}

\author{Ying-Hsuan Lin,$^{\textsymrook\textsymbishop}$ Shu-Heng Shao,$^{\textsymrook\textsympawn}$ Yifan Wang,$^{\textsymknight\textsymking}$ Xi Yin$^\textsymrook$}

\address{
$^\textsymrook$Jefferson Physical Laboratory, Harvard University, \\
Cambridge, MA 02138 USA
\\
$^\textsymbishop$Walter Burke Institute for Theoretical Physics, California Institute of Technology, \\ Pasadena, CA 91125, USA
\\
$^\textsympawn$School of Natural Sciences, Institute for Advanced Study, \\Princeton, NJ 08540, USA
\\
$^\textsymknight$Center for Theoretical Physics, Massachusetts Institute of Technology, \\
Cambridge, MA 02139 USA
\\
$^\textsymking$Joseph Henry Laboratories, Princeton University, \\Princeton, NJ 08544, USA}

\email{yhlin@caltech.edu, shuhengshao@gmail.com, yifanw@princeton.edu,
xiyin@fas.harvard.edu}

\end{center}

\abstract{ We find a simple relation between two-dimensional BPS ${\cal N}=2$ superconformal blocks and bosonic Virasoro conformal blocks, which allows us to analyze the crossing equations for BPS 4-point functions in unitary $(2,2)$ superconformal theories numerically with semidefinite programming. We constrain gaps in the non-BPS spectrum through the operator product expansion of BPS operators, in ways that depend on the moduli of exactly marginal deformations through chiral ring coefficients. In some cases, our bounds on the spectral gaps are observed to be saturated by free theories, by ${\cal N}=2$ Liouville theory, and by certain Landau-Ginzburg models. }

\vfill

\end{titlepage}

\eject

\tableofcontents

\section{Introduction} 

The conformal bootstrap is based on the idea that a conformal field theory may be determined entirely by conformal symmetry, associativity of operator product expansion, unitarity, and certain basic assumptions on the spectrum of operators and on the structure of OPE. The method has been surprisingly successful in solving a variety of CFTs in various spacetime dimensions \cite{Belavin:1984vu,Rattazzi:2008pe, Rychkov:2009ij, Poland:2010wg, Poland:2011ey, ElShowk:2012ht,Kos:2013tga, Beem:2013qxa, Beem:2014zpa, El-Showk:2014dwa,Chester:2014fya,Chester:2014mea,Chester:2014gqa,Bae:2014hia,Chester:2015qca,Iliesiu:2015qra,Kos:2015mba,Beem:2015aoa,Lemos:2015awa,Lin:2015wcg}. In this paper, we explore ${\cal N}=(2,2)$ superconformal theories (SCFT) in two dimensions using the bootstrap method, extending the results of \cite{Lin:2015wcg}.

$(2,2)$ SCFTs play a central role in the study of two-dimensional CFTs and string compactifications \cite{Candelas:1985en, Gepner:1987qi, Gepner:1987vz}. Typical constructions of such theories are based on supersymmetric nonlinear sigma models on Calabi-Yau manifolds \cite{Candelas:1985en,Eguchi:1988vra,Witten:1993yc}, Landau-Ginzburg models \cite{Vafa:1988uu,Lerche:1989uy,Cecotti:1989jc,Cecotti:1989gv,Martinec:1989in,Vafa:1990mu,Cecotti:1990kz}, and orbifolds \cite{Dixon:1986qv,Vafa:1989xc}. They often admit exactly marginal deformations \cite{Dixon:1987bg, Seiberg:1988pf,Kutasov:1988xb}, and the generic points on their moduli spaces are expected to give irrational theories \cite{Gukov:2002nw}. While the BPS operator spectra and their OPEs in $(2,2)$ SCFTs have been extensively studied \cite{Eguchi:1988af, Eguchi:1988vra, Lerche:1989uy,Cecotti:1989jc,Cecotti:1989gv,Cecotti:1990kz, Cecotti:1991me}, much less is known about the non-BPS spectrum, known to exhibit highly nontrivial moduli dependence \cite{Ooguri:1995wj,Seiberg:1999xz} and control the massive spectrum in models of string compactifications. With the available analytic methods, the non-BPS spectrum is accessible only at special solvable points in the moduli space \cite{Dixon:1986qv,Gepner:1987qi, Gepner:1987vz, Lunin:2000yv}, and through conformal perturbation theory \cite{Kutasov:1988xb} at the vicinity of these points or in the large volume (weak coupling) limit \cite{Douglas:aa}. 

The goal of this paper is to constrain the non-BPS spectrum across the entire moduli space of $(2,2)$ SCFTs. There are two known (computable) ways to encode the moduli dependence in the CFT data: through the chiral ring relations \cite{Cecotti:1990kz, Cecotti:1991me}, and through the spectrum of boundary states (D-branes) \cite{Ooguri:1996ck}. Here we consider the former, since the chiral ring relations can be straightforwardly incorporated into the conformal bootstrap based on sphere 4-point functions. Imposing the crossing equation, while assuming unitarity (reality of OPE coefficients), we will be able to constrain the scaling dimensions of non-BPS operators that appear in the OPE of BPS operators through the chiral ring data.

To begin with, let us recall that the BPS representations of the ${\cal N}=2$ superconformal algebra (SCA) are known as chiral or anti-chiral primaries
that saturate the BPS bound $h=|q|/2$, where $h$ is the conformal weight and $q$ the $U(1)_R$ charge. The ${1\over 2}$-BPS operators of the SCFT involve BPS representations of both the left and the right ${\cal N}=2$ SCAs, and depending on whether these representations are chiral or anti-chiral, are referred to as $(c,c)$ and $(c,a)$, as well as their Hermitian conjugate, $(a,a)$ and $(a,c)$, operators. The BPS operators of the same type have non-singular OPEs, and form a ring with respect to products at coincident points, known as the $(c,c)$ ring or the $(c,a)$ ring \cite{Lerche:1989uy}. The set of $(c,c)$ and $(c,a)$ operators are exchanged under mirror symmetry, which amounts to flipping the right $U(1)_R$ charge \cite{Candelas:1990rm, Witten:1991zz}. Of particular interest are $c=9$ $(2,2)$ SCFTs with spectral flow symmetry, that are described by supersymmetric nonlinear sigma models on Calabi-Yau threefolds, where the $(c,c)$ ring and $(c,a)$ ring capture the geometry of the quantum K\"ahler and complex structure moduli spaces, respectively \cite{Lerche:1989uy}.


In this paper, we focus on BPS operators of the $(c,c)$ type and their Hermitian conjugate $(a,a)$ operators, and investigate the non-BPS spectra in their OPEs. Of course the exactly same analysis may be applied to $(c,a)$ and $(a,c)$ operators, but we do not consider OPE of $(c,c)$ with $(c,a)$ operators here. The reason is that it is more difficult to incorporate the chiral ring data in analyzing 4-point functions of a mixture of $(c,c)$ and $(c,a)$ operators. Thus, without further specification, we will refer to $(c,c)$ operators as ``chiral primaries" and $(a,a)$ operators as ``anti-chiral primaries". We will also restrict our attention to BPS operators of equal left and right $U(1)$ R-charge, although the generalization to cases with unequal left and right R-charges would be straightforward.

Let $\phi$ be a  $(c,c)$ primary with R-charge $q=\bar q>0$. Its Hermitian conjugate $\overline\phi$ is an $(a,a)$ primary. The OPE $\phi \overline\phi$ contains the identity representation as well as R-charge neutral non-BPS representations of the ${\cal N}=2$ superconformal algebra.
We will refer to the $\phi\overline\phi$ OPE as the \textit{chiral-antichiral} ({\it CA}) {\it channel}, and denote by $\Delta_{gap}^{CA}$ the scaling dimension of the lowest non-BPS superconformal primaries appearing in this OPE.

On the other hand, in the $\phi\phi$ OPE, the lightest operator is a $(c,c)$ primary $\phi_{2q}$ of twice the R-charge of $\phi$. We denote by $\lambda$ the coefficient of  $\phi_{2q}$ in the OPE $\phi\phi$, where $\phi$ and $\phi_{2q}$ are respectively normalized with unit two-point functions. $\lambda$ will be referred to as the \textit{chiral ring coefficient}. We will refer to the  $\phi\phi$ OPE as the \textit{chiral-chiral} ({\it CC}) {\it channel}, and define $\Delta^{CC}_{gap}$ to be the gap in the scaling dimensions between $\phi_{2q}$ and the lightest operator in the $\phi\phi$ OPE that does not belong to a $(c,c)$ multiplet. The operators appearing in the CC channel may be ${1\over 2}$-BPS, ${1\over 4}$-BPS (that is, BPS on the left, non-BPS on the right, or vice versa), or non-BPS (that is, non-BPS on both left and right). Furthermore, non-BPS representations that carry nonzero R-charges in a suitable range may be degenerate \cite{Boucher:1986bh,Eguchi:1988af,Hosomichi:2004ph}.\footnote{The role of such short (degenerate) but non-BPS representations will be clarified in the next section.} Note that in the CC channel, the lightest state in a non-BPS representation that appears on the left or right of either a ${1\over 4}$-BPS operator or a non-BPS operator, is always a superconformal {\it descendant}, rather than a primary (see Subsection \ref{sec:selection} for the selection rules in the OPE of BPS operators).   

The BPS four-point function $\left\langle  \overline\phi(z,\bar z)\overline\phi(0)\phi(1) \phi(\infty) \right\rangle$ can be decomposed in terms of ${\cal N}=2$ superconformal blocks, in three different ways related by crossing symmetry.  Two of the three channels are
\ie\label{introcrossing}
\left\langle  \overline\phi(z,\bar z)\overline\phi(0)\phi(1) \phi(\infty) \right\rangle 
&= |\lambda|^2 {\cal F}_{(c,c)}^{\text{CC}}(z,\bar z) + \sum_{\Delta,s} (C^{\text{CC}}_{\Delta,s})^2 {\cal F}^{\text{CC}}_{\Delta,s}(z,\bar z)
\\
&= {\cal F}_\text{vac}^{\text{CA}}(1-z,1-\bar z) + \sum_{{\Delta\geq \Delta_{gap}^{\text{CA}}, s}}( C^{\text{CA}}_{\Delta,s})^2 {\cal F}^{\text{CA}}_{\Delta,s}(1-z,1-\bar z)\,,
\fe
while the third one comes from the OPE  channel $\bar\phi(z,\bar z)\phi(\infty)$.  The functions ${\cal F}^{\text{CC}}$ and ${\cal F}^{\text{CA}}$ are the appropriate ${\cal N}=2$ superconformal blocks, to be described in detail in Section~\ref{Sec:Block}. The subscripts vac, $(c,c)$, and $(\Delta,s)$ indicate respectively the vacuum, $(c,c)$, and a generic representation (${1\over 4}$-BPS or non-BPS) labelling a superconformal primary of dimension $\Delta$ and spin $s$. $\lambda$ is the chiral ring coefficient as already mentioned, while $C^{\rm CC}_{\Delta,s}$ and $C^{\rm CA}_{\Delta,s}$ are the OPE coefficients for the other representations in the CC and CA channels. In a unitary theory, the latter OPE coefficients can be taken to be real (by a choice of phase of the operators in question), hence so are their squares appearing in (\ref{introcrossing}). By exploiting the non-negativity of the coefficients $(C^{\text{CC}}_{\Delta,s})^2$ and $(C^{\text{CA}}_{\Delta,s})^2$, we can constrain the allowed set of values for $(\Delta,s)$ in the CC and CA channels, in a way that depends on the value of $\lambda$, which in turn varies over the moduli space of exactly marginal deformations of the SCFT. The simplest example of such a constraint is an upper bound on the gap in the spectrum, e.g. an upper bound on $\Delta_{gap}^{CA}$ as a function of $\lambda$ and  $\Delta^{CC}_{gap}$.

Constraints on the spectrum of this sort can be found numerically through semidefinite programming \cite{Simmons-Duffin:2015qma}, provided that we can compute the ${\cal N}=2$ superconformal blocks to high precision. While the bosonic Virasoro conformal blocks can be efficiently computed using Zamolodchikov's recurrence relation \cite{zamolodchikov1987conformal}, the analogous formula for the general ${\cal N}=2$ blocks are not yet available.\footnote{On the other hand, the analogous recurrence relation for $\cN=1$ superconformal blocks has been worked out in \cite{Belavin:2006zr,Hadasz:2006qb,Hadasz:2008dt}.} Fortunately, there exists a simple relation between BPS ${\cal N}=2$ blocks (BPS external operators and non-BPS internal operators)\footnote{The $\cN=2$ blocks with vacuum or BPS internal operators can be obtained as limits of the blocks for the non-BPS channels. This is in contrast with modular bootstrap, where the analogous statement does not hold for Virasoro characters.} of central charge $c={3(k+2)\over k}$, and bosonic Virasoro blocks of central charge $c=13+6k+{6\over k}$ with appropriately shifted weights on the external as well as internal primaries. We will derive this relation by consideration of BPS 4-point functions in the ${\cal N}=2$ cigar SCFT \cite{Ooguri:1995wj, Giveon:1999px, Aharony:2003vk, Aharony:2004xn,Chang:2014jta}, and confirm the result at low levels with computer algebra. 


Our numerical investigation of the OPE spectrum will focus on two cases. The first case involves a marginal BPS operator $\phi$ (which is necessarily exactly marginal \cite{Dixon:1987bg,Seiberg:1988pf}), namely one with conformal weight $h={1\over 2}$ and R-charge $q=1$ on both left and right. Without making any assumption on the chiral ring coefficients or the CC channel operator content, apart from unitarity constraints on the representations of ${\cal N}=2$ SCA,\footnote{Such constraints are particularly nontrivial when non-BPS degenerate representations are present.} we can already bound  the gap among the R-charge  neutral non-BPS operators in the CA channel. We will determine numerically an upper bound on $\Delta_{gap}^{CA}$ as a function of central charge $c$, for $3\leq c\leq 9$. Interestingly, for several values of $c$ that lie between 3 and ${18\over 5}$, the bound is saturated by OPEs in products of certain ${\cal N}=2$ minimal models (that happen to admit a marginal deformation, and are conveniently described by Landau-Ginzburg models), and we conjecture that the bound on $\Delta_{gap}^{CA}$ is linear in $c$ in this range.

The second case of our investigation concerns the OPE of BPS operators with R-charge\footnote{Note that for this value of external R-charge, the internal chiral primary in the CC channel may be related by (diagonal) spectral flow to an anti-chiral primary with the opposite R-charge as the external primary. 
In the analysis of the crossing equation, however, we do not make use of nor assume spectral flow symmetry.} $q=c/9$, for central charges $c=3,6,9$. 
We will bound $\Delta_{gap}^{CA}$ as a function of the chiral ring coefficient $\lambda$ and $\Delta_{gap}^{CC}$. In the $c=3$ case, rather strikingly, our bound is saturated by the OPE of twist fields in the $T^2/\mathbb{Z}_3$ orbifold SCFT along certain loci on its conformal manifold, for {\it all possible values} of $\lambda$ and $\Delta^{CC}_{gap}$.

Perhaps of most interest is the case $c=9$ and $q=1$, which may be applied to the OPE of marginal BPS operators in a Calabi-Yau threefold sigma model, yielding nontrivial moduli dependent constraints on the mass spectrum of string compactification in the quantum regime that have been uncomputable with known analytic methods. We compare our bounds on the gaps with the OPE of K\"ahler moduli (which belong to the $(c,c)$ ring) operators in the quintic threefold model, and the OPE of twist fields in the Z-manifold $T^6/\mathbb{Z}_3$. It is observed that, in a rather nontrivial manner, the large volume limits converge to the kinks on the boundary of the allowed domain in the space of OPE gaps $\Delta_{gap}^{CA}$, $\Delta_{gap}^{CC}$, and the chiral ring coefficient $\lambda$. The gap below the continuum of states that arise in the conifold limit, which admits a description in terms of the $\mathcal{N}=(2,2)$ Liouville theory (or its T-dual cigar SCFT) \cite{Ooguri:1995wj, Hori:2001ax}, appears to saturate our bound in the asymptotic region of large $\lambda$. Various Gepner models and free orbifolds are seen to satisfy the bounds but do not lead to saturation. Much of the allowed domain of our superconformal bootstrap analysis remains unexplored, and we will comment on the future perspectives at the end of the paper.

\section{The ${\cal N}=2$ Superconformal Algebra and Its Representations}

The two-dimensional ${\cal N}=2$ superconformal algebra (SCA) is generated by the stress-energy tensor $T(z)$, the superconformal currents $G^\pm(z)$, and the $U(1)_R$ current $J(z)$. Their Fourier modes in radial quantization obey the commutation relations
\ie
 {}[L_m, L_n] =& (m-n) L_{m+n} + {c \over 12} (m^3 - m) \delta_{m, -n}\,,
\\
 {} [L_m, G^\pm_r] =& \left({m \over 2} - r\right) G^\pm_{m+r}\,,
\\
{}  [L_m, J_n] =& -n J_{m+n}\,,
\\
{} \{ G^+_r, G^-_s \} =& 2 L_{r+s} + (r-s) J_{r+s} + {c \over 3} \left(r^2 - {1 \over 4}\right) \delta_{r, -s}\,,
\\
{} \{ G^+_r, G^+_s \} =& \{ G^-_r, G^-_s \} = 0\, ,
\\
{}[J_n, G^\pm_r] =& \pm G^\pm_{r+n}\, ,
\\
{} [J_m, J_n] = &{c \over 3} m \delta_{m, -n}\, ,
\fe
where $r,s$ are  integers in the R sector and  half-integers in the NS sector.

\subsection{Unitary Representations}

From now on we will focus on the NS sector. An irreducible highest weight representation of the $\mathcal{N}=2$ superconformal algebra is labeled by the weight $h$ and  the R-charge $q$ of its primary operator.  A representation is unitary provided that one of the following two conditions is satisfied \cite{Boucher:1986bh,Eguchi:1988af,Hosomichi:2004ph}:
\begin{align}
g_r(h,q)\geq 0 \,,~~~~~~\forall\, r\in \mathbb{Z}+\frac12\,,
\label{urep1}
\end{align}
or
\begin{align}
g_r(h,q)=0\,,~~~~~g_{r+\text{sgn}(r)} (h,q)< 0\,~~~~\text{and}~f_{1,1}(h,q)\ge0\,,~~~~~\text{for some $r\in\mathbb{Z}+\frac12$}\,.
\label{urep2}
\end{align}
Here the functions $g_r(h,q)$ and $f_{m,n}(h,q)$ are defined as
\ie
&g_r (h,q)\equiv 2h  -2 r q +\left(\frac c3 -1\right)\left(r^2 -\frac14\right)\,,~~~~~~~~~~~~~~~~~~~~~~~~~~~~~~~~~~~~~~~~~~r\in \mathbb{Z}+\frac12\,,\\
&f_{m,n}(h,q)  \equiv 2\left( \frac c3 -1\right) h -  q^2 -\frac14 \left(\frac c3 -1\right)^2 +\frac14 \left[ \left(\frac c3-1\right)m+2n\right]^2\,,~~~~~~~m,n\in\mathbb{Z}_{\ge0}\,.
\fe

A unitary representation is called  \textit{non-degenerate} if
\ie\label{ndrep}
g_r(h,q)> 0 \,,~~~~~~~\forall\, r\in \mathbb{Z}+\frac12\,,
\fe
and  \textit{degenerate} otherwise.  

In particular, a degenerate primary is called  \textit{chiral} if $g_{1/2}(h,q)=0$, i.e. if $h=q/2$.  Similarly, a degenerate primary is called  \textit{antichiral}  if $g_{-1/2}(h,q)=0$, i.e. if $h= -q/2$.  The chiral and antichiral primaries are superconformal primaries that are annihilated by $G^+_{-1/2}$ and $G^-_{-1/2}$ respectively.  Either a chiral or an antichiral primary generates a {\it BPS representation}. A {\it non-BPS representation}, on the other hand, refers to one that is generated either by a non-degenerate primary, or by a non-BPS degenerate primary that satisfies $g_{r}=0$ for some $r\neq \pm1/2$.  We think of the latter as non-BPS because they are not annihilated by the global supercharges.

Note that \eqref{ndrep} is generally a stronger condition than $h>|q|/2$.  In other words, there is generally a gap between the chiral primary  and non-BPS primaries of the same R-charge.  We will come back to this when we discuss the gap in the chiral-chiral channel in Section \ref{sec:CCgap}.

Based on our definition of BPS and non-BPS representations, independently in the left and right sector, there are four different types of superconformal primaries.  A \textit{$\frac12$-BPS primary} involves BPS representations on both left and right. A \textit{$\frac14$-BPS primary} involves a BPS  representation on the left, and a  non-BPS representation on the right, or vice versa. A \textit{non-BPS primary} involves non-BPS representations on both left and right.

\subsection{$\cN=2$ Selection Rules}\label{sec:selection}

We now describe the selection rules for the OPE of a pair of BPS primaries $\phi_{q_1}$ and $\phi_{q_2}$ of R-charges $q_1$ and $q_2$, which can be derived from superconformal Ward identities on three point functions along the lines of \cite{Baggio:2012rr,Lin:2015wcg}. These selection rules will apply independently to the left and right moving sectors.
Here we shall denote by $\phi_q$ a BPS primary of R-charge $q$, and by $O_q$ a non-BPS one. Without loss of generality, it suffices to consider three distinct cases:

\begin{enumerate}[(a)]

\item\label{SelectionRule(a)}  $q_1>0$, $q_2>0$, and $q_1+q_2>1$. In this case, the only multiplets that can appear in the OPE are those that contain either a chiral primary $\phi_{q_1+q_2}$ (of R-charge $q_1+q_2$) or a non-BPS primary $O_{q_1+q_2-1}$ (of R-charge $q_1+q_2-1$).  The operators that actually appear in the OPE would be the chiral primary $\phi_{q_1+q_2}$ itself or the level-${1\over2}$ descendant of the non-BPS primary, $G^+_{-1/2} O_{q_1+q_2-1}$, along with higher level superconformal descendants of the same R-charge.

\item  $q_1>0$, $q_2>0$, with $q_1+q_2<1$. In this case, in addition to the multiplets that appear in (\ref{SelectionRule(a)}), another BPS multiplet that contains an anti-chiral primary $\phi_{q_1+q_2-1}$ may also appear in the OPE. The actual operators in the OPE that belong to this multiplet are the level-${1\over2}$ descendant $G^+_{-1/2} \phi_{q_1+q_2-1}$ and higher level superconformal descendants with the same R-charge.

\item  $q_1>0$, $q_2<0$. In this case, the only multiplets that can appear in the OPE are those of an (anti)chiral primary $\phi_{q_1+q_2}$ and of a non-BPS primary $O_{q_1+q_2}$. 

\end{enumerate}
The rules in cases where $q_1<0$, $q_2<0$ are similar to those of (a) and (b). These selection rules are summarized in the following table. 

\begin{table}[h]\label{Tab:Selection}
\centering
\begin{tabular}{|c|c|c|c|c|}
\hline
$q_1$ & $q_2$ & $q_1+q_2$ & Multiplets & Lowest weight operators in OPE
\\\hline\hline
$>0$ & $>0$ & $>1$ & $\phi_{q_1+q_2}$, ~$O_{q_1+q_2-1}$ & $\phi_{q_1+q_2}$, ~$G^+_{-1/2}O_{q_1+q_2-1}$
\\
$>0$ & $>0$ & $<1$ & ~$\phi_{q_1+q_2}$,~ $\phi_{q_1+q_2-1}$,~ $O_{q_1+q_2-1}$ ~& ~$\phi_{q_1+q_2}$, ~$G^+_{-1/2}\phi_{q_1+q_2-1}$,~ $G^+_{-1/2}O_{q_1+q_2-1}$~
\\
$>0$ & $<0$ & & $\phi_{q_1+q_2}$, ~$O_{q_1+q_2}$ & $\phi_{q_1+q_2}$, ~$O_{q_1+q_2}$
\\\hline
\end{tabular}
\end{table}


\subsection{Spectral Flow}

The spectral flow \cite{Lerche:1989uy} transforms the generators of the ${\cal N}=2$ SCA according to 
\ie
J_n \to J_n + \eta {c\over 3} \delta_{n,0},~~~ L_n \to L_n + \eta J_n + \eta^2 {c\over 6}\delta_{n,0},~~~G_r^\pm \to G^\pm_{r\mp \eta},
\fe
where $r\in \mathbb{Z}/2$. In particular, the spectral flow with half integer $r$ relates NS and R sector states. A chiral primary $\phi_q$ with $U(1)_R$ charge $q\geq 0$ is annihilated by $G_{r\geq {1\over 2}}^\pm$ and $G^+_{-{1\over 2}}$. The $\eta=-1$ spectral flow takes $\phi_q$ to an anti-chiral primary of R-charge $q-{c\over 3}$, which must be non-positive. This is guaranteed by the aforementioned unitarity bound $f_{1,1}\geq 0$ in \eqref{urep2} which implies $|q| \leq {c \over 3}$.

While the spectral flow by an integer $\eta$ is an automorphism of the ${\cal N}=2$ SCA, it need not be a symmetry of the SCFT. Calabi-Yau models admit independently left and right spectral flow symmetries by integer $\eta$; in particular, the $\eta=1$ spectral flow maps the identity operator to a chiral primary of R-charge $q=c/3$ (associated with the holomorphic top form on the Calabi-Yau target space). Such spectral flow symmetries enlarge the ${\cal N}=2$ superconformal algebra, and put strong additional restrictions on the unitary representations \cite{Odake:1988bh, Odake:1989ev} (in particular, on the possible R-charges of the superconformal primaries); they played an important role in the modular bootstrap analysis of \cite{Keller:2012mr, Friedan:2013cba}.

In our analysis of the OPE through the crossing equation, however, the spectral flow symmetry does not play a significant role, due to the already existing selection rule on the R-charge of the internal primaries. Unless otherwise stated for specific models, we will not assume the spectral flow symmetry in this paper.

\subsection{The Minimal Gap in the Chiral-Chiral Channel}\label{sec:CCgap}

In the  OPE of a pair of identical chiral primaries $\phi_q$, there is generally a nonzero gap  $\Delta^{CC}_{gap}$ between the scaling dimensions of the chiral primary $\phi_{2q}$ and of the lightest operator (necessarily a level ${1\over 2}$ descendant, rather than a primary) that belongs to a different representation.\footnote{
This definition allows for a smooth limit of the CC channel superconformal block when the gap is taken to zero.
} In this subsection, we will describe a lower bound on $\Delta^{CC}_{gap}$ that follows from unitary representations of the $\cN=2$ SCA, which depends on the central charge $c$ and the external R-charge $q$.  Later when analyzing the crossing equation, this lower bound on $\Delta^{CC}_{gap}$ will be assumed.


A nontrivial lower bound on $\Delta^{CC}_{gap}$ exists when the unitarity bound \eqref{ndrep} for the non-degenerate multiplets,
\ie\label{nonBPS2}
g_r (  h,2q-1)  =  2h  -2r(2 q-1)  + \left( \frac c3 -1\right) \left( r^2 - \frac 14\right) >0\,,~~~~~\forall\, r\in \mathbb{Z}+\frac12\,,
\fe
is stronger than $h>q -\frac12$ (assuming $q>0$). 
For central charges $c>3$, this occurs when
\ie\label{nonzeroCCgap}
0<q<\frac12\,,~~~~~~\text{or}~~~~~~q>\frac c6. 
\fe
Note that for $c=3$, the non-degeneration condition \eqref{nonBPS2} can never be satisfied unless the R-charge of the external operator is $q=1/2$. That is, there is no non-degenerate primary with nonzero R-charge in $c=3$ theories.

We will be interested in SCFTs that admit marginal BPS primaries ($q=1$) with central charge $c\geq 3$.
Firstly, note that when $c<6$, the internal chiral primary of charge $2q=2$ would be forbidden by the unitarity bound. We nonetheless define $\Delta^{CC}_{gap}$ to be the gap above this internal non-unitary R-charge 2 chiral primary (which is absent from the OPE).

\begin{figure}[h!]
\centering
\includegraphics[width=1\textwidth ]{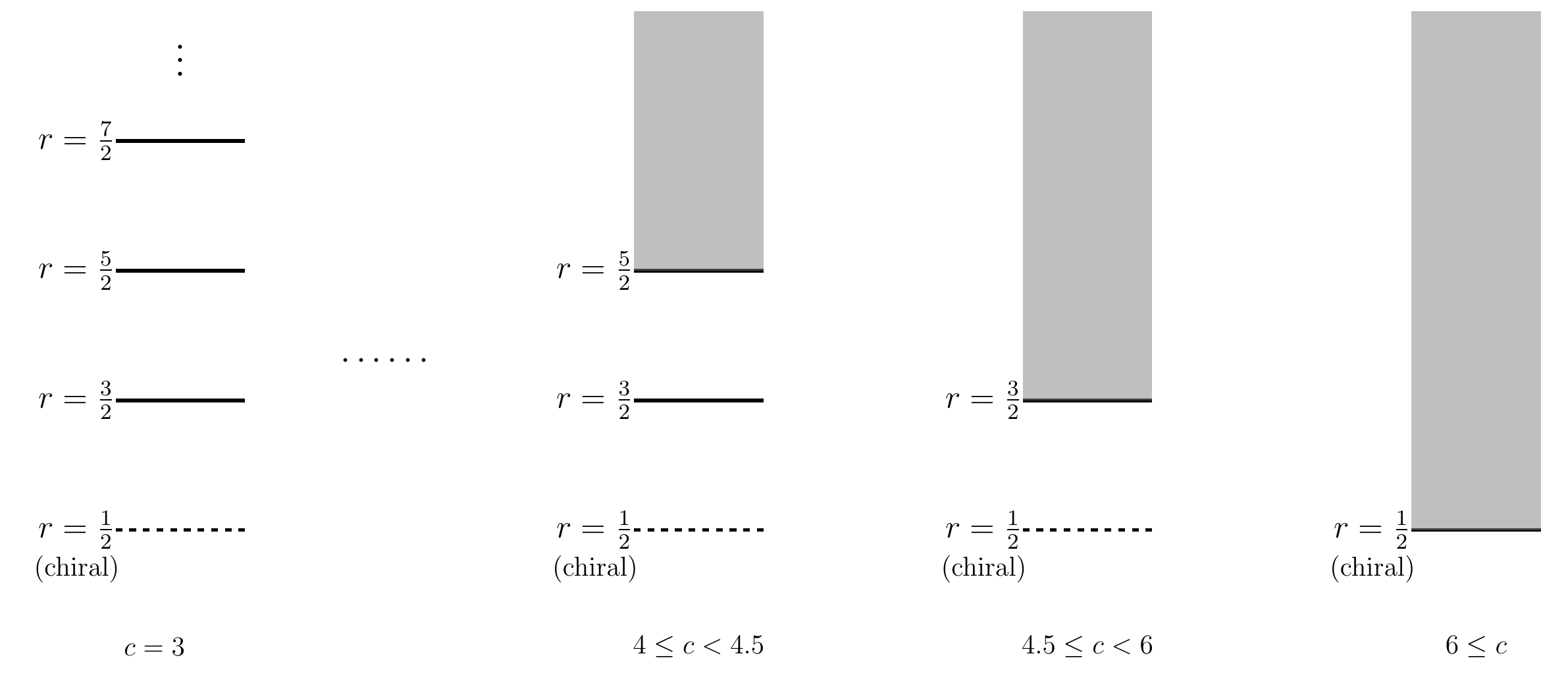}
\caption{Possible unitary multiplets allowed by the $\mathcal{N}=2$ superconformal symmetry in the $\phi_1\phi_1$ OPE between two chiral primaries of R-charge $q=1$.   The solid/dashed lines are the degenerate multiplets satisfying $g_r =0$.  In particular, the $r=\frac12$ degenerate primary is the chiral primary of R-charge $2q=2$, while all the non-BPS primaries carry R-charge $2q-1=1$. The internal chiral primary is shown in dashed lines for $3\le c<6$ because it violates the unitarity bound $|2q|\le c/3$ and is not present in unitary CFTs.  The gray shaded region corresponds to the continuum of non-degenerate multiplets. Note there is necessarily a gap in the weight above the chiral primary if $3\le c < 6$.}
\label{fig:q=1rep}
\end{figure}

When $3\le c <6$, there are discrete non-BPS degenerate primaries satisfying $g_r=0$ with $r=\frac 32 , \frac52, \cdots  ,r_0$ in
  the gap between the allowed range of non-degenerate non-BPS operators and the internal chiral primary.   Here $r_0 = \lceil {1\over  c/3-1}\rceil-\frac12$.  In particular, when $c=3$, there are no non-degenerate primaries with $q=1$; only degenerate primaries are present in the CC channel.  The lowest weight operator in the CC channel is the level-${1\over 2}$ descendant of the $g_{r=3/2}=0$ non-BPS degenerate primary. When $c\ge6$, there is no lower bound on $\Delta^{CC}_{gap}$ from $\cN=2$ representation theory.  See Figure \ref{fig:q=1rep}.


The presence of a level $\frac12$ descendant of the $g_{r=\frac32}=0$ non-BPS degenerate primary $O_{r=\frac32}$ in the OPE requires the three-point function
\ie\label{higherselection}
\langle \phi_1 (z_1) \phi_1(z_2) G^-_{-\frac12}\cdot O_{r=-\frac 32} (z_3)\rangle
\fe
to be consistent with the existence of null states in the relevant non-BPS degenerate representation. The first null operator $\chi(z)$ in the $g_{r=-\frac32}=0$ non-BPS degenerate multiplet\footnote{We switch to the Hermitian conjugate of the $g_{r=\frac32}=0$ non-BPS degenerate primary $O_{r=\frac32}$ for the consideration of the three-point function.} occurs at level $\frac32$,
\ie
\chi  \equiv \left(  
{c-6 \over 3 } \,G^-_{-3/2} +J_{-1} G^-_{-1/2} +L_{-1} G^-_{-1/2} \right) \cdot O_{r=-\frac32} \,.
\fe
Demanding 
\ie\label{3pt0}
\langle \phi_1 (z_1) \phi_1(z_2) \chi (z_3)\rangle=0\,,
\fe
we arrive at a differential equation on  $\langle \phi_1 (z_1) \phi_1(z_2) G^-_{-\frac12}\cdot O_{r=-\frac 32} (z_3)\rangle$ which itself is a three-point function of Virasoro primaries.\footnote{Here we have used a contour deformation trick to replace $G^-_{-3/2}$ by ${z_1+z_2-2z_3\over (z_1-z_3)(z_2-z_3)} G^-_{-1/2}$ in the three-point function \eqref{3pt0} with two chiral primaries. More specifically, we used
\ie
 G^-_{+1/2} -(z_1+z_2-2z_3)G^-_{-1/2} +(z_1-z_3)(z_2 -z_3) G^-_{-3/2} =\oint_{z=z_3} {dz\over 2\pi i} G^-(z) {(z-z_1) (z-z_2)\over z-z_3} \to 0\,,
\fe
in the three-point function \eqref{3pt0}.  See for instance \cite{Lin:2015wcg} for more details.
}
It turns out that this equation is trivially satisfied for all $c$, hence the appearance of the $g_{r=3/2}$ non-BPS degenerate primary in the OPE is consistent with the selection rule.

Therefore, for $3\le c< 6$ and $q=1$, the gap in the CC channel is at least that of the $g_{r=3/2}$ non-BPS degenerate primary, whose weight is $h=\frac 52-\frac c3$ and R-charge 1.  The actual operator that appears in the OPE is the level $\frac12$ descendant with R-charge 2. It follows that the gap in the holomorphic weight is $h^{CC}_{gap}=2-\frac c3$ if $3\le c< 6$, and $h^{CC}_{gap}=0$ if $6\le c$.

Finally, we need to combine the holomorphic and antiholomorphic weights to determine the gap in the scaling dimension.  Let us examine the possibility of a primary that is $g_{r=\frac32}=0$ non-BPS degenerate primary on the left, and chiral primary on the right in the range of $3\le c\le6$. The actual operator that appears in the OPE is a level $(\frac12,0)$ descendant of this primary, with weight $h=3-\frac c3$, $\bar h = 1$ and R-charge $q=\bar q=2$.  In the OPE between two identical scalars $\phi_1(z,\bar z)$, only even spin Virasoro primaries are allowed.  Hence the above level $(\frac12,0)$ descendant can appear only when $2-\frac c3 \in 2\mathbb{Z}$, which does not occur for $3\leq c<6$.  
Hence we may take the lower bound on the CC dimension gap to be simply $\Delta^{CC}_{gap}=2h^{CC}_{gap}$.

Another special case that will be of interest is $c=3$ and $q=1/3$. The lowest dimensional BPS primary is an $(a,a)$ primary with $g_{r=-\frac12}=0$ on the left and on the right, giving a gap $\Delta^{CC}_{gap} = 2/3$ in this case.  Furthermore, as we saw in Section \ref{sec:selection}, this internal $(a,a)$ primary is not ruled out by the $\mathcal{N}=2$ selection rule, so the gap $\Delta^{CC}_{gap}=2/3$ may be saturated.  An example of this, based on twist fields in the supersymmetric orbifold $T^2/\mathbb{Z}_3$, is discussed in detail in Appendix \ref{Sec:T2Z3}.


We conclude this subsection by recording the minimal values of $\Delta^{CC}_{gap}$ allowed by the $\mathcal{N}=2$ representation theory for various values of $c$ and $q$ that will be analyzed in the superconformal bootstrap analysis later on:
\ie\label{minimalCCgap}
&3\le c<6 \,,&q=1\,,~~~& \Delta^{CC}_{gap} =  4-{2c\over3}\,,\\
&c\geq 6 \,,&q=1\,,~~~&\Delta^{CC}_{gap}=0\,,\\
&c=3\,,&q=\frac13\,,~~~&\Delta^{CC}_{gap} = \frac23\,,\\
&c=6,9\,,&q=\frac{c}{9}\,,~~~&\Delta^{CC}_{gap}=0\,.
\fe

\section{${\cal N}=2$ Superconformal Blocks and Virasoro Blocks}\label{Sec:Block}

In this section we will discuss the sphere four-point ${\cal N}=2$ superconformal block with four external BPS primaries of R-charge $\pm q$, with either  BPS or non-BPS internal states.\footnote{For a technical simplification, the external BPS primaries will be taken to have R-charges of the same absolute value.}  In particular, we will  present an interesting  relation between the $\cN=2$ superconformal block and the bosonic Virasoro block of a different central charge, generalizing the results of \cite{Lin:2015wcg}.

We will start with the superconformal blocks with either a  non-BPS internal representation. There are two distinct cases as discussed in Section \ref{sec:selection}. The first one is the chiral-chiral (CC) block, where two chiral primaries of R-charge $q$ fuse into descendants of a  non-BPS primary of R-charge $2q-1$. The second one is the chiral-antichiral (CA) block, where a chiral and an anti-chiral primary of R-charge $q$ and $-q$ fuse into a R-charge neutral non-BPS primary and its descendants. The CC block will be denoted by
\ie
{\cal F}^{{\rm CC}, c}_{-q,-q,q,q|h}(z),
\fe
where $c$ is the central charge of the ${\cal N}=2$ SCA, $h$ is the weight of the internal non-BPS primary of R-charge $2q-1$, and $z$ is the cross ratio of the four external vertex operators.
 We emphasize here again that only the descendants of charge $2q$ actually appear in the OPE. The CA block will be denoted by
\ie
{\cal F}^{{\rm CA}, c}_{-q,q,q,-q|h}(z),
\fe
where $h$ is the weight of the R-charge-neutral internal non-BPS primary.  The vacuum block can be obtained as a limit of the non-BPS block,
\ie
\label{VacuumLimit}
{\cal F}^{{\rm CA}, c}_{-q,q,q,-q|\text{vac}}(z) = {\cal F}^{{\rm CA}, c}_{-q,q,q,-q|h = 0}(z).
\fe

\begin{figure}[h!]
\begin{tabular}{cc}
\includegraphics[width=.5\textwidth]{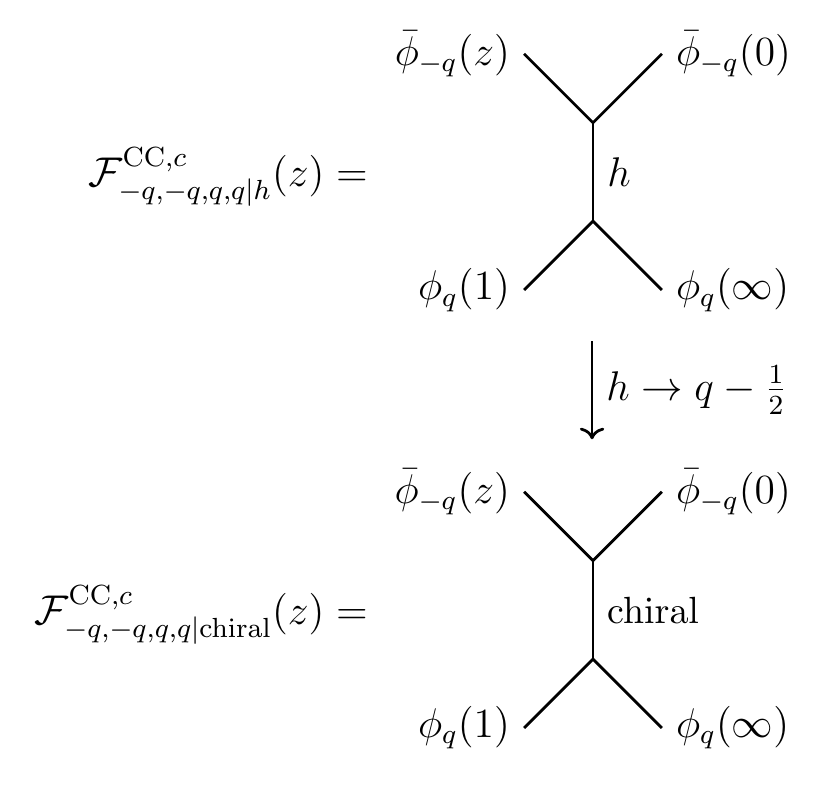}~
\includegraphics[width=.5\textwidth]{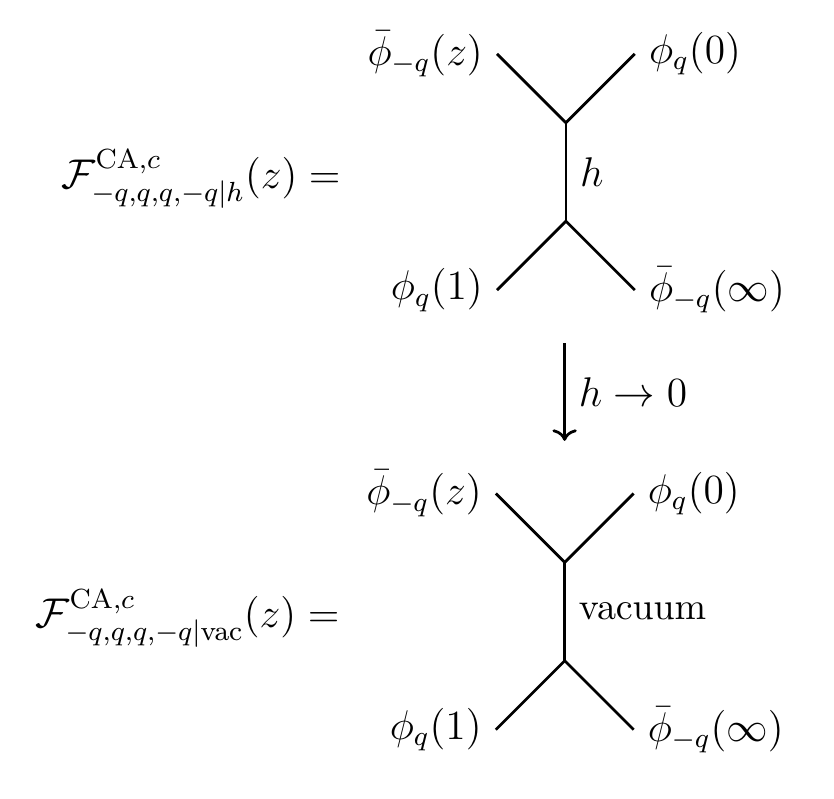}
\end{tabular}
\caption{The limits that relate ${\cal N}=2$ super-Virasoro blocks with BPS external primaries and non-BPS as well as BPS internal representations.}
\end{figure}

The CC block with an internal chiral primary with charge $2q$ can be obtained from a limit of the non-BPS block,
\ie
\label{ChiralLimit}
{\cal F}^{{\rm CC}, c}_{-q,-q,q,q|\text{chiral}}(z) = {\cal F}^{{\rm CC}, c}_{-q,-q,q,q|h=q-{1\over 2}}(z).
\fe
 In the case $0<q<{1\over 2}$, there is another possible internal antichiral primary  of weight $h={1\over 2}-q$ and R-charge $2q-1$ in the CC channel (see Section~\ref{sec:selection} for the selection rule).  Its CC block can also be obtained as a limit of the non-BPS block,
 \ie
\label{AntichiralLimit}
{\cal F}^{{\rm CC}, c}_{-q,-q,q,q|\text{antichiral}}(z) = {\cal F}^{{\rm CC}, c}_{-q,-q,q,q|h={1\over 2}-q}(z).
\fe
We checked \eqref{VacuumLimit}, \eqref{ChiralLimit} and \eqref{AntichiralLimit} by brute-force computation of the $\mathcal{N}=2$ superconformal blocks to the $z^4$ order using computer algebra.\footnote{That is, we work with the oscillator representation of the descendant operators, and computing their OPE coefficients with the external primaries and the relevant Gram matrices, order by order in the conformal cross-ratio $z$.}

The $\cN=2$ superconformal blocks with external BPS primaries of charge $\pm q$ can be related to the bosonic Virasoro conformal blocks of different central charges.  To understand this relation, let us consider the ${\cal N}=2$ $SL(2)_k/U(1)$ cigar coset model, of central charge $c=3(k+2)/k$. In this theory, there is a family of superconformal primaries of the form $\Phi_{j,m,\bar m}$, that descend from bosonic $SL(2)_{k+2}$ primaries, of left and right weights and R-charges
\ie
& h = {-j(j+1)+m^2\over k},~~~~ \bar h =  {-j(j+1)+\bar m^2\over k},
\\
& q = {2m\over k},\hspace{1.1in} \bar q = {2\bar m\over k}.
\fe
The quantum numbers $m,\bar m$ are subject to the constraints $m- \bar m \in \mathbb{Z}$, $m + \bar m \in k\mathbb{Z}$. There is a set of normalizable states that correspond to certain discrete real values of $j$, among which the (anti)chiral primaries are of the form $\Phi_{j,m,\bar m}$ with $m=\bar m$, $j=|m|-1$. If we assume that $k$ is a positive integer, the condition $m+\bar m\in k\mathbb{Z}$ may be relaxed to $m+\bar m\in\mathbb{Z}$ if we consider twisted sector states of the orbifold $(SL(2)_k/U(1))/\mathbb{Z}_k$, where $\mathbb{Z}_k$ acts by rotation along the circle direction of the cigar.

The correlation functions of operators of the form $\Phi_{j,m,\bar m}$ that conserve the total $m$ and $\bar m$ quantum numbers can be computed directly from the bosonic $SL(2)_{k+2}$ WZW model, by factoring out the $U(1)$ part of the vertex operators.  The correlators of $SL(2)$ primaries can further be related to those of a bosonic Liouville theory of central charge $c=1+6(\sqrt{k} + {1\over \sqrt{k}})^2$, via \cite{Ribault:2005wp}. In \cite{Chang:2014jta} the sphere four-point function of the (anti)chiral primaries of $(SL(2)_k/U(1))/\mathbb{Z}_k$ are rewritten in terms of four-point functions in Liouville theory. It was further observed in \cite{Lin:2015wcg} that the ${\cal N}=2$ superconformal block decomposition of the former coincides with the bosonic Virasoro conformal block decomposition of the latter. This leads to the following relations between the non-BPS ${\cal N}=2$ superconformal blocks and Virasoro conformal blocks. For the CC block, we have
\ie
\label{CCRelation}
{\cal F}_{-q,-q,q,q|h}^{{\rm CC}, c = {3(k+2)\over k}}(z) = z^{{k\over 2} q^2} (1-z)^{{k\over 2} q(1-q)} F^{\rm Vir}_{c=13+6k+{6\over k}} (h_-, h_-, h_+, h_+; h+{1\over 2} + kq(1-q) ; z),
\fe
where $F^{\rm Vir}_c$ is the Virasoro block with central charge $c$, and
\ie
h_- = {q (2k-kq+2)\over 4},~~~ h_+ = {(q+1)(k-kq+2)\over 4}.
\fe
\begin{figure}[h!]
\begin{tabular}{c}
\includegraphics[width=.9\textwidth]{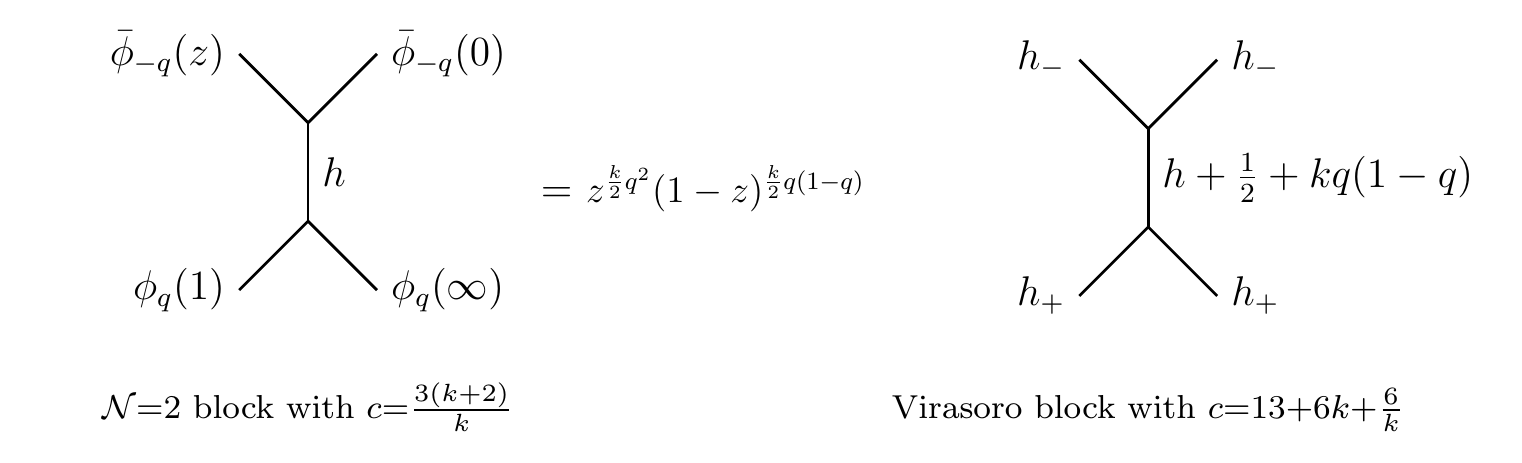}~
\\
\includegraphics[width=.9\textwidth]{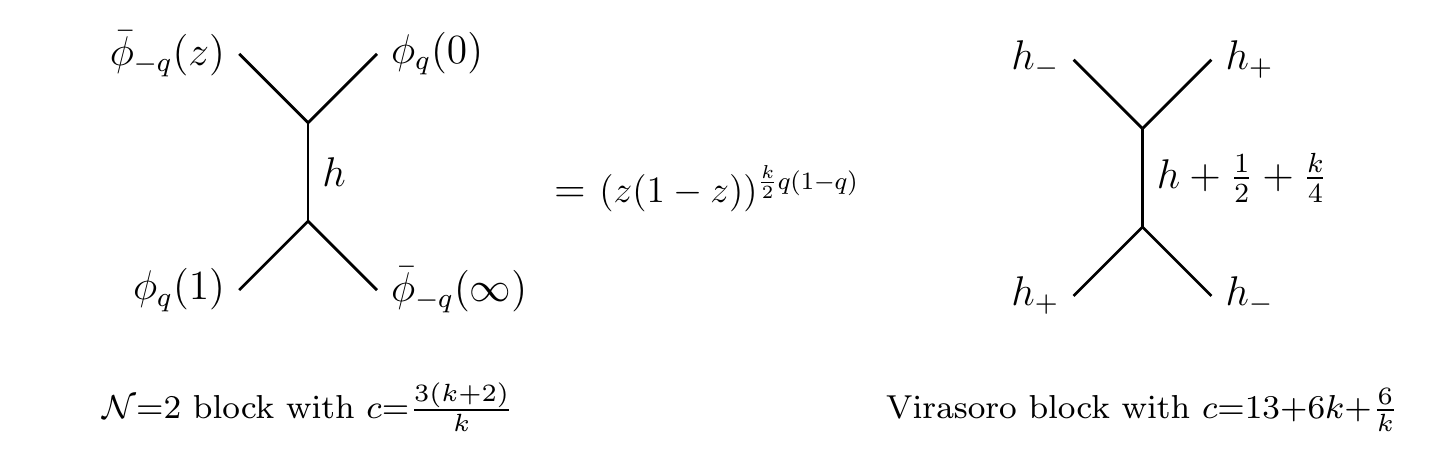}
\end{tabular}
\caption{Relation between $\mathcal{N}=2$ super-Virasoro blocks with external BPS primaries and bosonic Virasoro blocks.}
\end{figure}
For the CA block,
\ie
\label{CARelation}
{\cal F}_{-q,q,q,-q|h}^{{\rm CA}, c = {3(k+2)\over k}}(z) = (z(1-z))^{{k\over 2} q(1-q)} F^{\rm Vir}_{c=13+6k+{6\over k}} (h_-, h_+, h_+, h_-; h+{1\over 2} + {k\over 4}; z).
\fe
The vacuum and the BPS blocks are also related to Virasoro conformal blocks via \eqref{VacuumLimit}, \eqref{ChiralLimit} and \eqref{AntichiralLimit}.  The relations \eqref{CCRelation} and \eqref{CARelation}  have been checked by brute-force computations of (super)conformal blocks to the $z^4$ order.


Having equipped with the relation between the $\mathcal{N}=2$ blocks and the bosonic Virasoro blocks, we can now  compute the former to high precision efficiently.  This is achieved through Zamolodchikov's recurrence relation \cite{Zamolodchikov:1985ie,Zamolodchikov:1995aa}, which computes the (bosonic) Virasoro block as a series expansion in the ``nome'' $q(z)$, defined as
\ie
q(z) \equiv \exp(i\pi\tau(z)),
\quad
\tau(z) \equiv {i F(1-z) \over F(z)}, \quad F(z) = {}_2F_1({1/2}, {1/2}, 1 | z).
\fe
The holomorphic Virasoro block for a four-point function $\langle {\cal O}_1(z) {\cal O}_2(0) {\cal O}_3(1) {\cal O}_4(\infty) \rangle$ with central charge $c$, external weights $h_i$, and internal weight $h$ has the following representation
\ie
F_c^{Vir}(h_i; h; z) &= {[16 q(z)]^{h - {c-1 \over 24}} z^{{c-1 \over 24} - h_1 - h_2}} (1-z)^{{c-1 \over 24} - h_1 - h_3}
\\
& \hspace{1in} \times [\theta_3(q(z))]^{{c-1 \over 8} - 4(h_1 + h_2 + h_3 + h_4)} H( \lambda_i^2, h | q(z)).
\fe
If we define
\ie
c = 1 + 6Q^2, \quad Q = b + {1 \over b}, \quad h_{m,n} = {Q^2\over 4} - \lambda_{m,n}^2, \quad \lambda_{m,n} = {1\over 2} ({m\over b} + nb),
\fe
then $H( \lambda_i^2, h | q(z))$ satisfies Zamolodchikov's recurrence relation
\ie\label{recH}
H(\lambda_i^2, h | q(z)) = 1 + \sum_{m,n\geq 1} {[q(z)]^{mn} R_{m,n}(\{\lambda_i\}) \over h - h_{m,n} }
H(\lambda_i^2, h_{m,n} + mn| q(z) ),
\fe
where $h_{m,n}$ are the conformal weights of degenerate representations of the Virasoro algebra, and $R_{m,n}(\{\lambda_i \})$ are given by
\ie
R_{m,n}(\{\lambda_i \}) = 2 {\prod_{r,s} (\lambda_1+\lambda_2 - \lambda_{r,s}) (\lambda_1-\lambda_2 - \lambda_{r,s}) (\lambda_3+\lambda_4 - \lambda_{r,s}) (\lambda_3-\lambda_4 - \lambda_{r,s}) \over \prod_{k,\ell}' \lambda_{k,\ell} }.
\fe
The product of $(r,s)$ is taken over
\ie\label{rsrange}
& r = -m+1, -m+3, \cdots, m-1,
\\
& s = -n+1, -n+3, \cdots, n-1,
\fe
and the product of $(k,\ell)$ is taken over
\ie
& k = -m+1, -m+2, \cdots, m,
\\
& \ell = -n+1, -n+2, \cdots, n,
\fe
{\it excluding} $(k,\ell)=(0,0)$ and $(k,\ell)=(m,n)$.

\section{Bounding the Gaps in the OPE of BPS Operators}\label{sec:bound}

Our objective is to constrain the spectrum of non-BPS operators in the OPE of a pair of BPS primaries, either of the form $\phi_q(z,\bar z)\overline\phi_{-q}(0)$ (CA channel), or $\phi_q(z,\bar z)\phi_q(0)$ (CC channel), by analyzing the ${\cal N}=2$ superconformal block decomposition of the BPS 4-point function $\left\langle  \overline\phi_{-q}(z,\bar z)\overline\phi_{-q}(0)\phi_q(1) \phi_q(\infty) \right\rangle$. The latter can be decomposed in either the chiral-chiral channel or the two chiral-antichiral channels. The equivalence of these decompositions gives the following set of crossing equations,
\ie
\label{crossing}
| {\cal F}^{CA}_{-q,q,q,-q|vac}(z) |^2 + 
\sum _{\substack{h, \bar h\\ h+\bar h \ge \Delta^{CA}_{gap}}} (C^{CA}_{h, \bar h})^2 | {\cal F}^{CA}_{-q,q,q,-q|h}(z) |^2 &
\\
& \hspace{-2.5in} =| {\cal F}^{CA}_{-q,q,q,-q|vac}(1-z) |^2 + \sum_{\substack{h, \bar h\\ h+\bar h \ge \Delta^{CA}_{gap}}}(C^{CA}_{h, \bar h})^2 | {\cal F}^{CA}_{-q,q,q,-q|h}(1-z) |^2,
\\
| {\cal F}^{CA}_{-q,q,q,-q|vac}(z) |^2 + \sum_{\substack{h, \bar h\\ h+\bar h \ge \Delta^{CA}_{gap}}} (C^{CA}_{h, \bar h})^2 (-)^{h-\bar h} | {\cal F}^{CA}_{-q,q,-q,q|h}(z) |^2 &
\\
& \hspace{-2.5in}= |\lambda|^2 | {\cal F}^{CC}_{-q,-q,q,q|chiral}(1-z) |^2 + \sum_{h, \bar h} (C^{CC}_{h, \bar h})^2 | {\cal F}^{CC}_{-q,-q,q,q|h}(1-z) |^2.
\fe
As discussed in Section \ref{sec:selection}, the sum in the chiral-antichiral channels includes only the non-BPS multiplets.  $\Delta^{CA}_{gap}$ is defined as the scaling dimension of the lowest non-BPS primary  in the chiral-antichiral channel.  

On the other hand, the spectrum in the chiral-chiral channel is more involved.  When $q>1/2$, the sum includes only the $\frac14$-BPS and non-BPS representations, while in the case of $0<q<1/2$,  $(a,a)$ primaries of weight $h=\bar h= \frac12 -q$ and R-charge $1-2q<0$ can also contribute.\footnote{Since the $\mathcal{N}=2$ block for this internal antichiral primary is a limit of the non-BPS block \eqref{AntichiralLimit}, we do not have to single out its contribution from the crossing equation \eqref{crossing} as we did for the internal $(c,c)$ primary.}    We define $\Delta^{CC}_{gap}$ to be the gap between the scaling dimension of the lightest operator that does not belong to a $(c,c)$ multiplet,\footnote{That is, the lightest operators in the second term of the second line in  \eqref{crossing}.} and that of a charge $2q$ $(c,c)$ primary.

Using the positivity of the coefficients $(C^\text{CC})^2$ and $(C^\text{CA})^2$, we will obtain numerical upper bounds on  $\Delta^{CA}_{gap}$.  The bound will depend on the chiral ring coefficient $\lambda$ and the gap $\Delta^{CC}_{gap}$ in the $\phi_q\phi_q$ OPE, the chiral-chiral channel.


\subsection{Semidefinite Programming}\label{sec:sp}

We now describe the method of using semidefinite programming to generate numerical upper bounds on the gap. Our first task is to write the crossing equations in a form that is convenient for the implementation of semidefinite programming. By defining (the operators are placed in the order $z, 0, 1, \infty$)
\ie
& G^{CA, -}_{h, \bar h}(z) \equiv |{\cal F}^{CA}_{-q,q,q,-q|h}(z)|^2 - |{\cal F}^{CA}_{-q,q,q,-q|h}(1-z)|^2,
\\
& \widehat G^{CA, \pm}_{h, \bar h}(z) \equiv |{\cal F}^{CA}_{-q,q,-q,q|h}(z)|^2 \pm |{\cal F}^{CA}_{-q,q,-q,q|h}(1-z)|^2,
\\
& G^{CC, \pm}_{h, \bar h}(z) \equiv |{\cal F}^{CC}_{-q,-q,q,q|h}(z)|^2 \pm |{\cal F}^{CC}_{-q,-q,q,q|h}(1-z)|^2,
\fe
the crossing equations can be packaged as \cite{Kos:2014aa}
\ie
0 = \sum_{h, \bar h} (C^{CA}_{h, \bar h})^2
\begin{pmatrix}
G^{CA, -}_{h, \bar h}(z)
\\
(-)^{h-\bar h} \widehat G^{CA, -}_{h, \bar h}(z)
\\
(-)^{h-\bar h} \widehat G^{CA, +}_{h, \bar h}(z)
\end{pmatrix}
+
\sum_{h, \bar h} (C^{CC}_{h, \bar h})^2
\begin{pmatrix}
0
\\
G^{CC, -}_{h, \bar h}(z)
\\
- G^{CC, +}_{h, \bar h}(z)
\end{pmatrix},
\fe
where the sum includes the vacuum multiplet in the CA channel and the charge 2$q$ chiral multiplet in the CC channel.  Next we act by a vector linear functional $\vec\A$ with three components, which we write as a sum $\vec\A_o + \vec\A_e$ where
\ie
\A^i_o \equiv \sum_{m + n \ {\rm odd}} \A^i_{m, n} \partial_z^m \partial_{\bar z}^n|_{z = \bar z = 1/2},
\quad
\A^i_e \equiv \sum_{m + n \ {\rm even}} \A^i_{m, n} \partial_z^m \partial_{\bar z}^n|_{z = \bar z = 1/2},
\fe
to put the crossing equations into the form
\ie
0 &= \sum_{h, \bar h} (C^{CA}_{h, \bar h})^2
\left( 
\A^1_o [ G^{CA, -}_{h, \bar h} ]
+
(-)^{h-\bar h} \A^2_o [ \widehat G^{CA, -}_{h, \bar h} ]
+
(-)^{h-\bar h} \A^3_e [ \widehat G^{CA, +}_{h, \bar h} ]
\right)
\\
& \hspace{2.5in} +
\sum_{h, \bar h} (C^{CC}_{h, \bar h})^2
\left(
\A^2_o [ G^{CC, -}_{h, \bar h} ]
-
\A^3_e [ G^{CC, +}_{h, \bar h} ]
\right).
\fe

A hypothetical spectrum in the CA and CC channels can be ruled out by unitarity if we can find an 
 $\vec\A$ satisfying
\ie
\label{vecApos}
& \A^1_o [ G^{CA, -}_{h, \bar h} ]
+
(-)^{h-\bar h} \A^2_o [ \widehat G^{CA, -}_{h, \bar h} ]
+
(-)^{h-\bar h} \A^3_e [ \widehat G^{CA, +}_{h, \bar h} ] \geq 0, \quad \forall ~(\Delta, s) \in {\cal I}^{CA},
\\
& \A^2_o [ G^{CC, -}_{h, \bar h} ]
-
\A^3_e [ G^{CC, +}_{h, \bar h} ] \geq 0, \quad \forall ~(\Delta, s) \in {\cal I}^{CC}\,,
\fe
where $\mathcal{I}^{CA}$ and $\mathcal{I}^{CC}$ are the sets of scaling dimensions and spins for the superconformal multiplets in the CA and CC channels, respectively.  In particular, we aim to  rule out hypotheses of the form
\ie
\label{hypo}
{\cal I}^{CA} = \{ \Delta = 0 \text{ or } \Delta \geq \widehat \Delta^{CA}_{gap} \}, \quad {\cal I}^{CC} = \{ \Delta = 2q \text{ or } \Delta \geq \widehat \Delta^{CC}_{gap} \},
\fe
and find the lowest $\widehat \Delta^{CA}_{gap}$ and $\widehat \Delta^{CC}_{gap}$ that can be ruled out to obtain the most stringent bound on the gaps.  Such a problem can be solved using the method of semidefinite programming.



\subsection{Some Comments on the Details of the Numerics}

We implement semidefinite programming using the  SDPB package~\cite{Simmons-Duffin:2015qma}.  In practice, to obtain an upper bound on the gaps, we need to truncate our basis of linear functionals at finite total derivative order $N_\A$ in $\partial_z$, $\partial_{\bar z}$.  The most stringent upper bound on the gaps is then bound by extrapolating to $N_\A \to \infty$.  We must also truncate the set of spins on which to impose positivity \eqref{vecApos}, and approximate the superconformal block in Zamolodchikov's representation by truncating \eqref{recH} to a finite series in the nome $q(z)$. The largest spin considered and the order of the $q(z)$-series are denoted by $s_{max}$ and $d_q$, respectively. We would like to emphasize here that whereas the truncations in the spins and $q(z)$-orders are (controlled) approximations, the truncation in derivative orders $N_\A$ always yields rigorous bounds (for sufficiently high $s_{max}$ and $d_q$).

The conformal blocks are computed numerically via Zamolodchikov's recurrence relation that was reviewed in Section~\ref{Sec:Block}. The blocks are computed separately for each value of the central charge, so that all inputs to the recurrence relation except for the internal weight $h$ are numerical numbers.
Since the conformal block for arbitrary internal weight $h$ is a combination of $H(\lambda^2_i, h_{m,n}+mn | q(z))$ for $m, n \geq 1$ via the recurrence relation \eqref{recH}, an efficient way to compute the general conformal block is to first compute $H$ at these special values of the internal weight. Moreover, in order to compute the general conformal block to ${\cal O}([q(z)]^{d_q})$, we only need $H(\lambda^2_i, h_{m,n}+mn | q(z))$ for $mn \leq d_q$. Denoting by $\vec H$ the column vector that contains this finite set of $H$ as entries, the recurrence relation \eqref{recH} implies a matrix equation of the form
\ie
\label{MatrixEqn}
({\bf I} - {\bf M}) \vec H = \vec 1 + {\cal O}([q(z)]^{d_q+1}),
\fe
where ${\bf I}$ is the identity matrix, $\vec 1$ is a column vector with every entry equal to 1, and ${\bf M}$ is a matrix with elements
\ie
~[{\bf M}]_{(p,q), (m,n)} = {[q(z)]^{mn} R_{m,n}(\{\lambda_i\}) \over h_{p,q} - h_{m,n} }.
\fe
It is then straightforward to invert ${\bf I} - {\bf M}$ to obtain $\vec H$.\footnote{This inversion is performed by writing $\vec H$ and ${\bf I} - {\bf M}$ both as series in $q(z)$, and matching the coefficients order by order. A direct matrix inversion would be extremely inefficient and unnecessary since \eqref{MatrixEqn} is only accurate to a finite order.
}


%
%

For a given derivative order $N_\A$, the dependence of the bound on $s_{max}$ and $d_q$
has the following behavior:  when the truncation order is small, an $\vec\A$ satisfying \eqref{vecApos} always exists even when the hypothetical gaps $(\widehat\Delta^{CA}_{gap}, \widehat\Delta^{CC}_{gap})$ are set to zero, thereby ruling out any hypothesis of the form \eqref{hypo}; as the truncation order exceeds some minimum, a bound on $(\widehat\Delta^{CA}_{gap}, \widehat\Delta^{CC}_{gap})$ starts to exist and stabilize as we go to higher truncation orders.  We adjust the truncation order to make sure that the bound has stabilized to within the desired numerical precision.  Empirically we find that while setting $s_{max} = d_q = N_\A + 4$ usually suffices, sometimes higher truncation orders are needed, for example when the chiral ring coefficient $\lambda$ is sent to infinity, or when the central charge is close to 3.

The bottleneck for the speed of the numerical computation is the truncation order of the $q(z)$-series.  This is because in the Zamolodchikov representation of the conformal block, the coefficients in the $q(z)$-expansion have denominators that are higher and higher-degree polynomials in $h$, and the degree of the polynomial is a key factor affecting the computation speed.  This imposes a limit on the highest derivative order $N_\A$ we can go to, since as mentioned in the previous paragraph, the derivative order must be somewhat lower than the $q(z)$-expansion order $d_q$.  We have chosen to only consider $d_q$ up to 28, and hence $N_\A$ up to 24 or less.

\section{$(2,2)$ Theories with Exactly Marginal Deformations}

In this section, we study constraints on the R-charge neutral non-BPS spectrum of $(2,2)$ SCFTs with exactly marginal deformations, by considering the OPE of a pair of BPS primaries of R-charge $\pm 1$ (on both left and right), in theories whose central charges lie in the range $3\le c \le9$. The $G^\mp_{-1/2}\tilde G^\mp_{-1/2}$ descendants of these primaries generate $\mathcal{N}=(2, 2)$-preserving exactly marginal deformations.  When there are more than one modulus for the $\mathcal{N}=(2,2)$ conformal manifold, we will consider the BPS four-point function associated to only one of them.

Let us comment on the chiral ring coefficient $\lambda$, which controls the contribution from the ${\cal N}=2$ superconformal block with an internal BPS representation (of R-charge 2 in this case) in the chiral-chiral channel. For $c<6$, an R-charge 2 chiral primary would be forbidden by the unitarity bound, and thus $\lambda=0$. For $c\geq 6$, $\lambda$ can be nonzero, and we will study the $\lambda$-dependence in the $c=9$ case in detail in the next section. If we introduce a nonzero $\lambda$ into the crossing equation, the bootstrap bounds will be strictly stronger than that of $\lambda=0$. This is because the contribution of a superconformal block with a BPS internal representation may be viewed as a limiting case of superconformal blocks with a non-BPS internal representation, as we have seen in Section \ref{Sec:Block}. Thus, we will simply set $\lambda=0$ for now, which amounts to not keeping track of the chiral ring coefficient.

We begin with the $c = 3$ SCFTs, which include the supersymmetric sigma model on $T^2$ and its orbifolds.  In this case the crossing equation can be trivially solved as follows.  The  CA block with external $q=\pm1$ BPS primaries and a R-charge neutral internal non-BPS primary of weight $h$ has the following closed form expression,\footnote{This expression is checked by computer algebra up to $z^6$ order. }
\ie
{\cal F}_{-1,1,1,-1|h}^{{\rm CA}, c = 3}(z) = { z^{h-1} \over (1-z)^{h+1} }\,.
\fe
It turns out that crossing symmetry constrains the four-point function with BPS primaries $\phi_{\pm1}$ of R-charge $q=\pm1$  in any unitary $c=3$ $(2,2)$ SCFT to be the square of the vacuum block,\footnote{For example, the four-point function of the fermion bilinears $\phi_1 (z,\bar z)\equiv \psi^+(z) \widetilde\psi^+(\bar z)$, $\overline{\phi}_{-1} (z,\bar z)\equiv \psi^-(z) \widetilde\psi^-(\bar z)$  in the $T^2$ CFT or its orbifolds can be readily computed to be \eqref{c=3q=1}. }
\ie\label{c=3q=1}
\langle \overline{\phi}_{-1}(z, \bar z) {\phi}_1(0) {\phi}_1(1) \overline{\phi}_{-1}(\infty) \rangle= |{\cal F}_{-1,1,1,-1|h=0}^{{\rm CA}, c = 3}(z)|^2 = {1\over |  z(1-z)|^2 }\,.
\fe
To see this, note that for a fixed real $z \in (0, 1)$, the difference between the $(2,2)$ superconformal block in two CA channels related by crossing, 
\ie
& {\cal F}_{-1,1,1,-1|h}^{{\rm CA}, c =3}(z) {\cal F}_{-1,1,1,-1|\bar h}^{{\rm CA}, c = 3}(z)
-
(z \to 1-z)
=
{ z^{2\Delta} - (1-z)^{2\Delta} \over (z(1-z))^{\Delta+2} },~~~~~(\Delta=h+\bar h)
\fe
is of a definite sign for all positive $\Delta$ (and vanishes for $\Delta=0$).  The crossing equation relating the two CA channels, which involves a sum of such terms with non-negative coefficients, can be satisfied only if all coefficients for $\Delta > 0$ vanish, hence the claim.  

\begin{figure}[h!]
\centering
\subfloat{
\includegraphics[width=.49\textwidth]{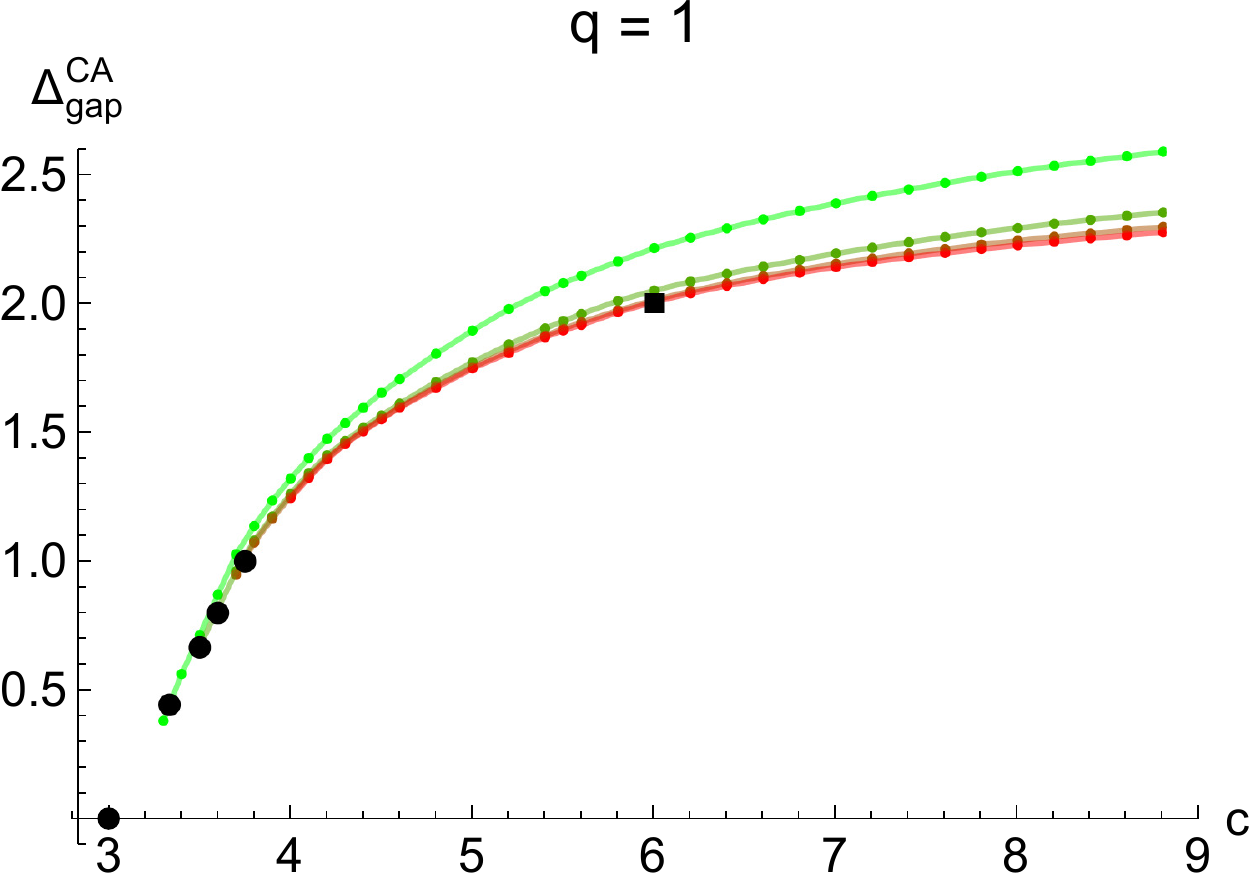}
}
\subfloat{
\includegraphics[width=.49\textwidth]{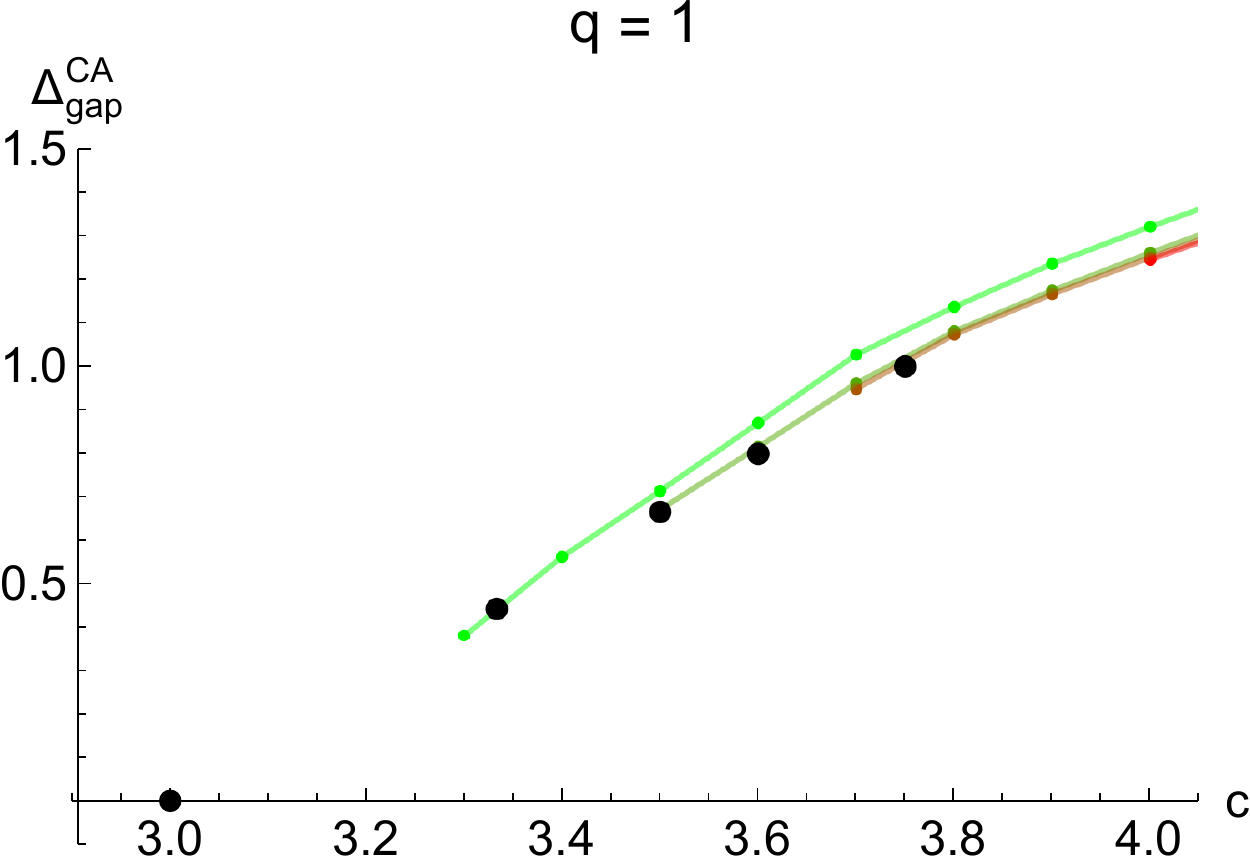}
}
\caption{{\bf Left:} Upper bounds on the gap in the CA channel $\Delta^{CA}_{gap}$, with the minimal assumption \eqref{minimalCCgap} on the gap in the CC channel $\Delta^{CC}_{gap}$, as a function of the central charge $c$, at derivative orders $4, \, 8, \, 12, \, 16$ (from green to red).
{\bf Right:} The same plot zoomed into $3\leq c \leq 4$.
}
\label{Fig:q=1}
\end{figure}

We now proceed to more general central charges. Figure~\ref{Fig:q=1} shows the numerical upper bounds $\Delta^{CA}_{gap}$ on the gap in the CA channel for $3\leq c\leq 9$, taking into account the unitarity constraints on the CC channel gap $\Delta^{CC}_{gap}$ \eqref{minimalCCgap}. More specifically, the $\mathcal{N}=2$ representation theory demands the CC channel gap to be no smaller than $4-{2\over3}c$ for $3\le c< 6$, and does not restrict the CC gap for $c\geq 6$.

Specific examples of four-point functions that saturate the bounds to within numerical precision are marked in black.  Towards the left, we have certain tensor products of ${\cal N} = 2$ minimal models with four-point functions populating the points
$
({10\over3}, {4\over9}), ~ ({7\over2}, {2\over3}), ~ ({18\over5}, {4\over5}), ~ ({15\over4}, 1),
$
as well as the free point $(3, 0)$.  These tensor products of minimal models will be explained in detail in Section~\ref{LG}.  At $c = 6$, we have the point $(6, 2)$, that is realized by a four-point function of fermion bilinears.  In~\cite{Lin:2015wcg}, by extrapolating to infinite derivative order, it was found that $\Delta^{CA}_{gap}$ is likely to be exactly 2, suggesting that the numerical saturation at $(6, 2)$ is exact.

For $3 < c < 3.3$, the numerics do not stabilize even when we truncate the $q(z)$-series up to the maximum order 28 that we consider.  Nonetheless, saturation of the bounds by the tensor products of minimal models as well as the free theory suggests that the bounds could be given by the exact formula
\ie
\Delta^{CA}_{gap} = {4 \over 3}(c-3)
\fe
in the range $3 \leq c \leq {15 \over 4}$.

\subsection{$(2,2)$ Landau-Ginzburg Models with $3<c<4$}
\label{LG}
For small central charges $3<c<4$, we can construct $(2,2)$  Landau-Ginzburg (LG) models defined by quasi-homogeneous superpotentials that possess nontrivial exactly marginal deformations.\footnote{Marginal deformations in $(2,2)$ SCFTs are exactly marginal \cite{Dixon:1987bg}.}   Equivalently, they can be realized by tensor products of $\mathcal{N}=(2,2)$ minimal models.  See \cite{Bertolini:2014ela} for examples of such $(2,2)$ SCFTs.

It is easy to classify such LG models with up to 3 chiral superfields. They are of the following types\footnote{The polynomials that define such superpotentials are known to be of the unimodal quasi-homogeneous type in singularity theory \cite{arnold1985singularities}.}
\ie
&X^3+Y^{3n}+ a XY^{2n}, \quad n\geq 3,\quad c={2(2n-1)\over n} 
\\
& X^4+Y^{2n}+a X^2 Y^n,\quad n=3,5 ,\quad c={3(3n-2)\over 2n} 
\\
& X^4+Y^8+a X^2 Y^4 + b X Y^6 ,\quad c={15\over 4} 
\\
& X^5+Y^5+a X^3 Y^2 + b X^2 Y^3 , \quad c={18\over 5}
\label{LGwm}
\fe
where the superconformal moduli spaces are parametrized by the coefficients $a,b$.

We are interested in the CA and CC gaps in the OPE of the $q=1$ chiral primaries (and their complex conjugates). 
This follows immediately from the fusion rules of $\cN=2$ minimal models \eqref{n2fusion}.  For example for the $\cN=(2,2)$  SCFTs defined by LG superpotential $X^3+ Y^{3n}$ with $n\geq 3$, the lowest non-chiral superconformal primary in the CA channel is given by $\Phi^{k=3n}_{1,0,0}$ (see Appendix \ref{ap:gepner} for notations) in the $\cN=2$ $A_{3n-1}$ minimal model,
\ie
\Delta_{gap}^{CA}=\Delta(\Phi^{k={3n}}_{1,0,0})={4\over 3n}.
\fe
In the CC channel, the gap between the lowest non-chiral superconformal primary whose level-${1\over 2}$ descendant appears and the unitarity bound $h=\bar h =q/2$ for $q=2$ is
\ie
\Delta_{gap}^{CC}=2\left(h(\Phi^{k=3}_{{1\over 2},-{1\over 2},-{1\over 2}})+h(\Phi^{k=3n}_{{n\over 2}+1, {n\over 2}, {n\over 2}})+1-1\right)={4(n+1)\over 3n}.
\fe
Note that the lowest operator appearing in the CC channel here is the product of the level-${1\over 2}$ descendants of $\Phi^{k=3}_{{1\over 2},-{1\over 2},-{1\over 2}}$ and $\Phi^{k=3n}_{{n\over 2}+1, {n\over 2}, {n\over 2}}$ in the respective $\cN=2$ minimal models. The CC and CA gaps for the rest the of $3<c<4$ $\cN=2$ LG models can be found in a similar manner and we summarize them in Table~\ref{tab:tensorgap}. In particular we see that all of these LG models saturate the lower bound $\Delta_{gap}^{CC}\geq 4-{2c\over 3}$ on the CC gap  from $\cN=2$ representation theory. Moreover, the $c={10\over 3},{7\over 2},{18\over 5},{15\over 4}$ models\footnote{For $c={15\over 4}$, the maximal CA gap is achieved when the marginal chiral primary is taken to be $XY^6$ in the LG description (see Table~\ref{tab:tensorgap}).} sit on the numerical CA gap bound along with the $c=3$  $T^2/\bZ_3$ model which has $0$ CA gap (see Figure~\ref{Fig:q=1}).

\begin{table}[htb]
\begin{center}
\begin{tabular}{|c|c|c|c|c|c|}
\hline
LG models & $c$  & marginal chiral primary  & $\Delta_{gap}^{CA}$ & $\Delta_{gap}^{CC}$
\\
\hline 
$X^3 + Y^{3n}$ with $n\geq 3$ & ${2(2n-1)\over n}$ & $XY^{2n}$  & ${4\over 3n}$ & $\frac{4 (n+1)}{3 n}$
\\
\hline 
$X^4 + Y^{2n}$  with $5\geq n \geq 3$ & ${3(3n-2)\over 2n}$ & $X^2Y^{n}$  & ${2\over n}$ &  ${n+2\over n}$
\\
\hline 
$X^4 + Y^{8}$  & ${1 5\over 4}$  & $XY^6 $ & $ 1$ & ${3\over 2}$
\\
\hline 
$X^5 + Y^{5}$  & ${1 8\over 5}$  & $X^2Y^3 $ & ${4\over 5}$ & $8\over 5$
\\
\hline
\end{tabular}
\end{center}
\caption{CA and CC gaps in $\cN=2$ models with $3<c<4$.}
\label{tab:tensorgap}
\end{table}

\section{Dependence on Chiral Ring Data}

In this section we present the numerical bootstrap results on the upper bounds of $\Delta^{CA}_{gap}$ as a function of the chiral ring coefficient $\lambda$, as well as its dependence on $\Delta^{CC}_{gap}$, in $(2,2)$ superconformal theories of various central charges. As motivated in the introduction, we will focus on the case where the external R-charge $q$ is $c/9$. Introducing the chiral ring coefficient explicitly into the crossing equation allows us to probe the dependence of the spectrum on the moduli of exactly marginal deformations. Our bounds will be compared to a number of interesting examples, including the twist field OPE in free orbifolds $T^{2n}/\mathbb{Z}_3$, with $n=1,2,3$, and nonlinear sigma models on Calabi-Yau threefolds.

\subsection{The $c=3$, $q=1/3$ Case}

\begin{figure}[h!]
\centering
\subfloat{
\includegraphics[width=.44\textwidth]{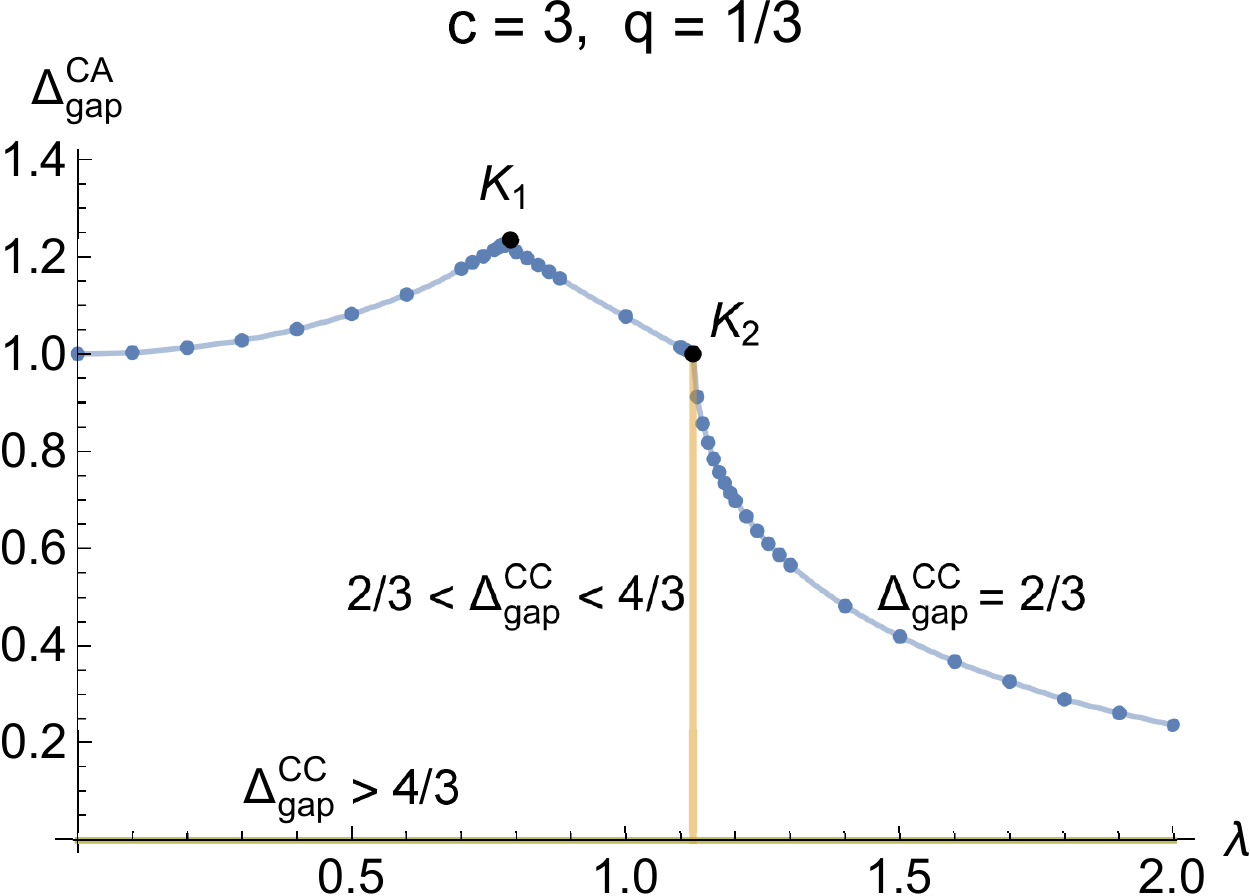}
}
\subfloat{
\includegraphics[width=.54\textwidth]{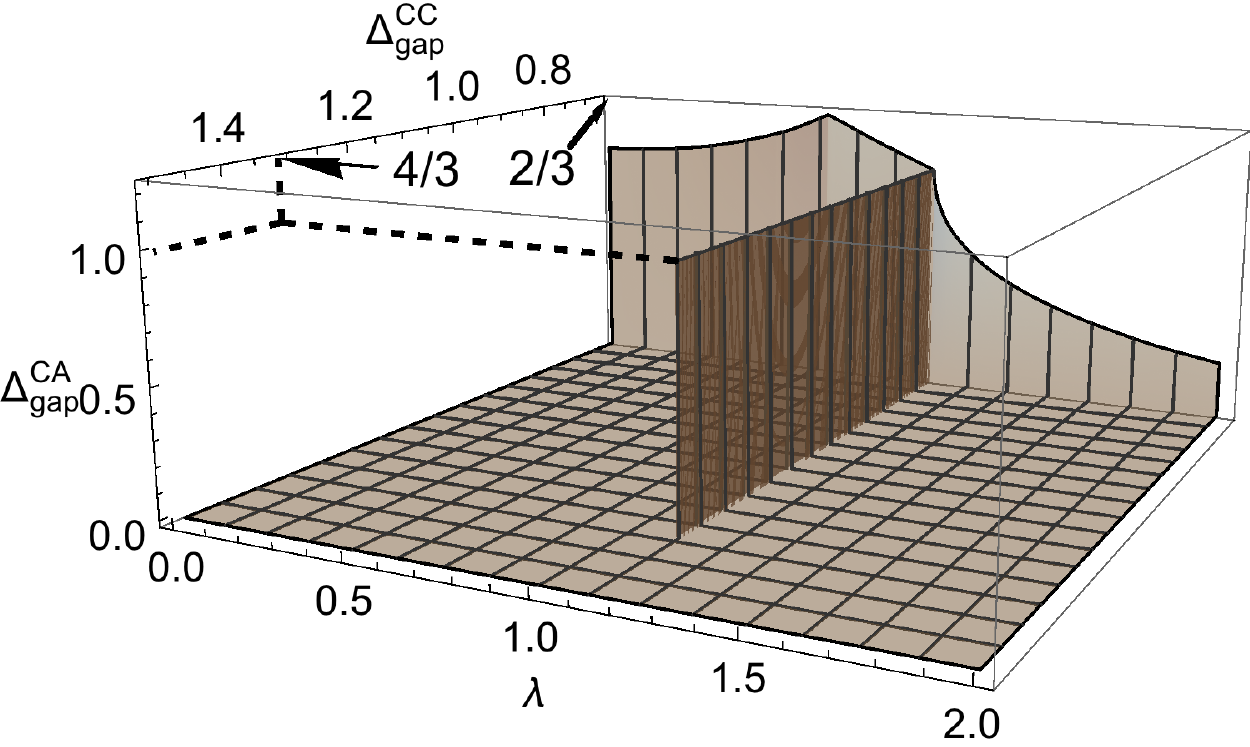}
}
\caption{{\bf Left:} Upper bounds on the gap in the CA channel $\Delta^{CA}_{gap}$  from the numerical bootstrap in the case of $c=3$ and external charge $q=\frac13$.  The bounds depend on the chiral ring coefficient  $\lambda$ and the gap in the CC channel $\Delta^{CC}_{gap}$ we put into the crossing equation. The blue, yellow, and green curves 
are the bootstrap bounds with $\Delta^{CC}_{gap} = {2 \over 3}$, ${2 \over 3} < \Delta^{CC}_{gap} \leq {4 \over 3}$, and $\Delta^{CC}_{gap} > {4 \over 3}$, respectively.  {\bf Right:} The three-dimensional visualization of the same plot.  The peak 
for $\frac23<\Delta^{CC}_{gap} \leq \frac34$ is saturated by the point $(R, b) = (\sqrt{4\over3}, 0)$  on the moduli space of the $T^2/\mathbb{Z}_3$ theory, at which the OPE coefficient for the $(a,a)$ primary $\overline\phi'_{-\frac13}$ vanishes and hence $\Delta^{CC}_{gap}$ increases to $4\over3$. }
\label{Fig:c=3}
\end{figure}

Let us start with the  $c = 3$  case with external R-charge $q = {1\over3}$. In Figure~\ref{Fig:c=3}, we  present the numerical bounds on the gap in the CA  channel $\Delta^{CA}_{gap}$, which depends on the gap in the CC channel  $\Delta^{CC}_{gap}$ and the chiral ring coefficient $\lambda$.  

 The primary example is the $\mathcal{N}=(2,2)$ $T^2 / \bZ_3$ orbifold CFT, which will be reviewed in details in Appendix \ref{Sec:T2Z3}.  The $q=\pm1/3$ BPS primaries $\phi_{\pm\frac13}$ are taken to be twist fields in the Ramond-Ramond sector.  This orbifold CFT has a real two-dimensional moduli space $M$, parametrized by the size  $R$ of the $T^2$ and the $B$-field $b$, with periodicity $b\sim b+1$.\footnote{We will work in the convention $\A'=2$ throughout the paper.}

 \subsubsection{$T^2/\mathbb{Z}_3$ CFT Saturating the Bootstrap Bound}
 
\begin{figure}
\centering
\includegraphics[width= .5\textwidth]{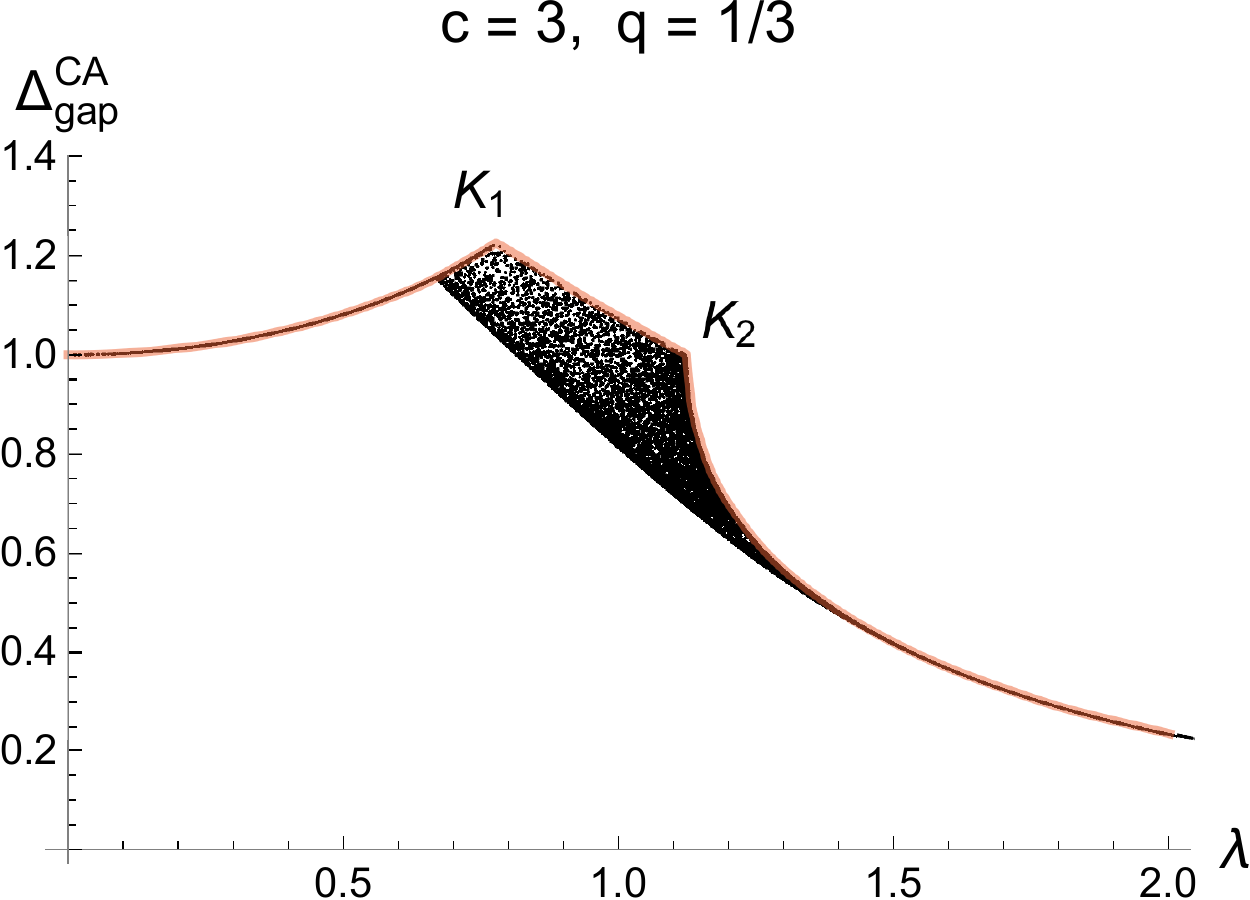}
\caption{The $T^2/\mathbb{Z}_3$ free orbifold theory saturates the $c=3$, $q=\frac13$ $\mathcal{N}=(2,2)$ conformal bootstrap of the four-point function $\langle \phi_{\frac13} \bar \phi_{-\frac13} \bar \phi_{-\frac13} \phi_{\frac13}\rangle$.  Here $\phi_{\frac13}$ is a $q=\bar q= \frac13$ $(c,c)$ primary operator and $\bar \phi_{-\frac13}$ is its $(a,a)$ conjugate.  The vertical axis is the dimension $\Delta^{CA}_{gap}$ of the lowest dimensional non-BPS primary in the chiral-antichiral channel. The horizontal axis is the chiral ring coefficient $\lambda$ in the chiral-chiral channel.   The red curve is the numerical bootstrap upper bound on $\Delta^{CA}_{gap}$, which depends on the value of the chiral ring coefficient $\lambda$ we put into the crossing equation.   The black dots are the analytic results  $(\lambda(R,b),\Delta^{CA}_{gap}(R,b))$ (using \eqref{T2Z3chiralring} and \eqref{CAgapT2Z3}) of the $\mathcal{N}=(2,2)$ $T^2/\mathbb{Z}_3$ orbifold theory, by randomly sampling many points on the moduli space $(R,b)$.  We  see that there are some loci on the moduli space $(R,b)$ of the $T^2/\mathbb{Z}_3$ orbifold theory saturating the numerical bootstrap bound.}\label{fig:T2Z3random}
\end{figure}

  We would like to compare the analytic results of the $T^2/\mathbb{Z}_3$ CFT with the numerical bounds that are presented in Figure~\ref{Fig:c=3}.  This is possible because both the chiral ring coefficient $\lambda(R,b)$ and the CA gap $\Delta^{CA}_{gap}(R,b)$ are known in the $T^2/\mathbb{Z}_3$ CFT, as explicit functions of the two moduli $R,b$ (see  \eqref{T2Z3chiralring} and \eqref{CAgapT2Z3}).   By scanning over the moduli space $M$,  the points $(\lambda,\Delta^{CA}_{gap})$ realized by the $T^2/\mathbb{Z}_3$ CFT are shown as black dots in Figure \ref{fig:T2Z3random}.  The blue curve in Figure \ref{fig:T2Z3random}, on the other hand, is the numerical bootstrap bound on $\Delta^{CA}_{gap}$, assuming the gap in the chiral-chiral channel is $\Delta^{CC}_{gap}=\frac23$.  As discussed in Section \ref{sec:CCgap},  $\frac23$ is the smallest value of $\Delta^{CC}_{gap}$ allowed by the $\mathcal{N}=(2,2)$ representation theory, so we did not impose any non-trivial assumption on the operator spectrum in the chiral-chiral channel of the four-point function $\langle \phi_{\frac13} \bar \phi_{-\frac13} \bar \phi_{-\frac13} \phi_{\frac13}\rangle$.

\begin{figure}
\raisebox{0pt}{\includegraphics[width= .45\textwidth]{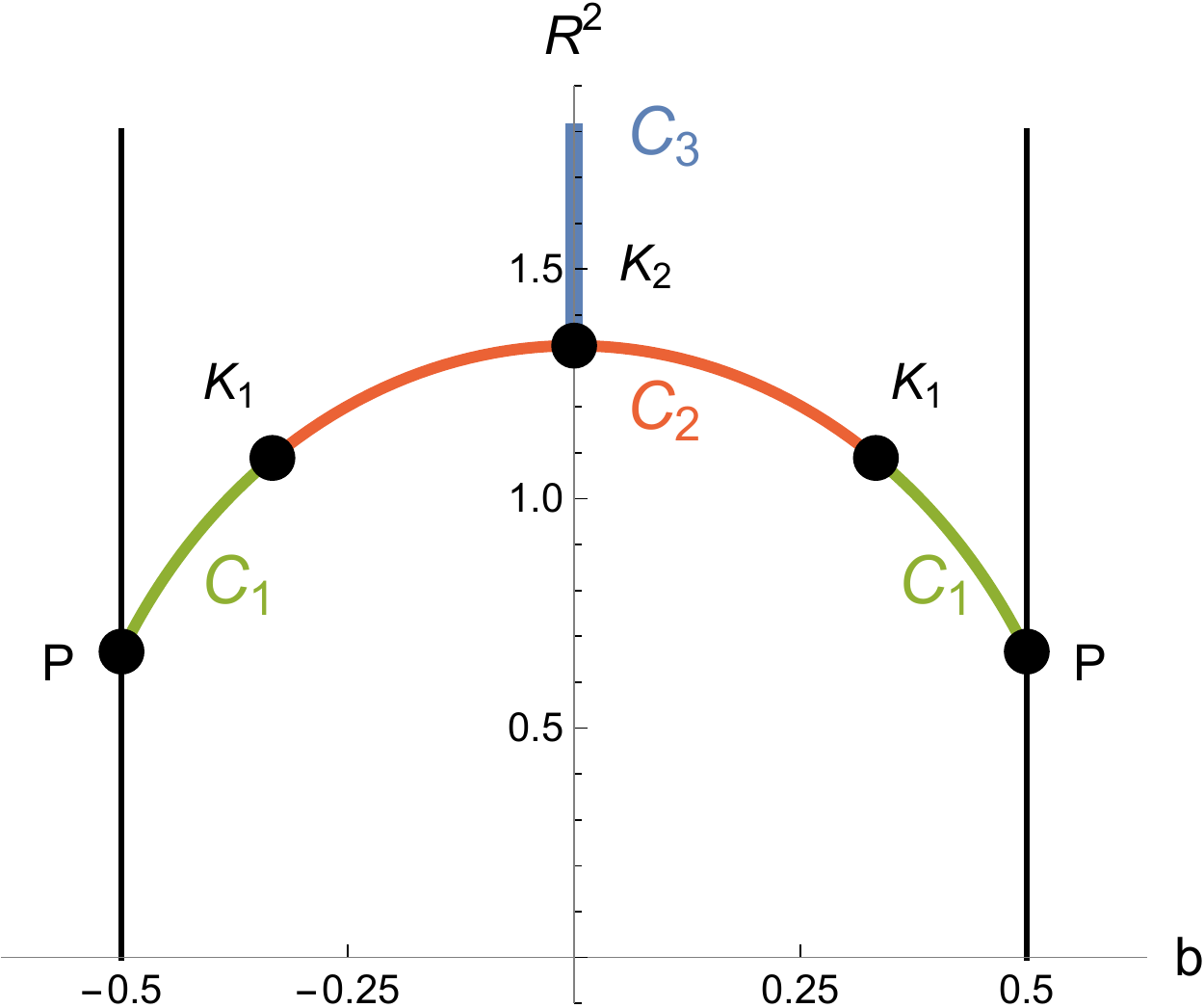}}\quad\quad\quad
\includegraphics[width= .45\textwidth]{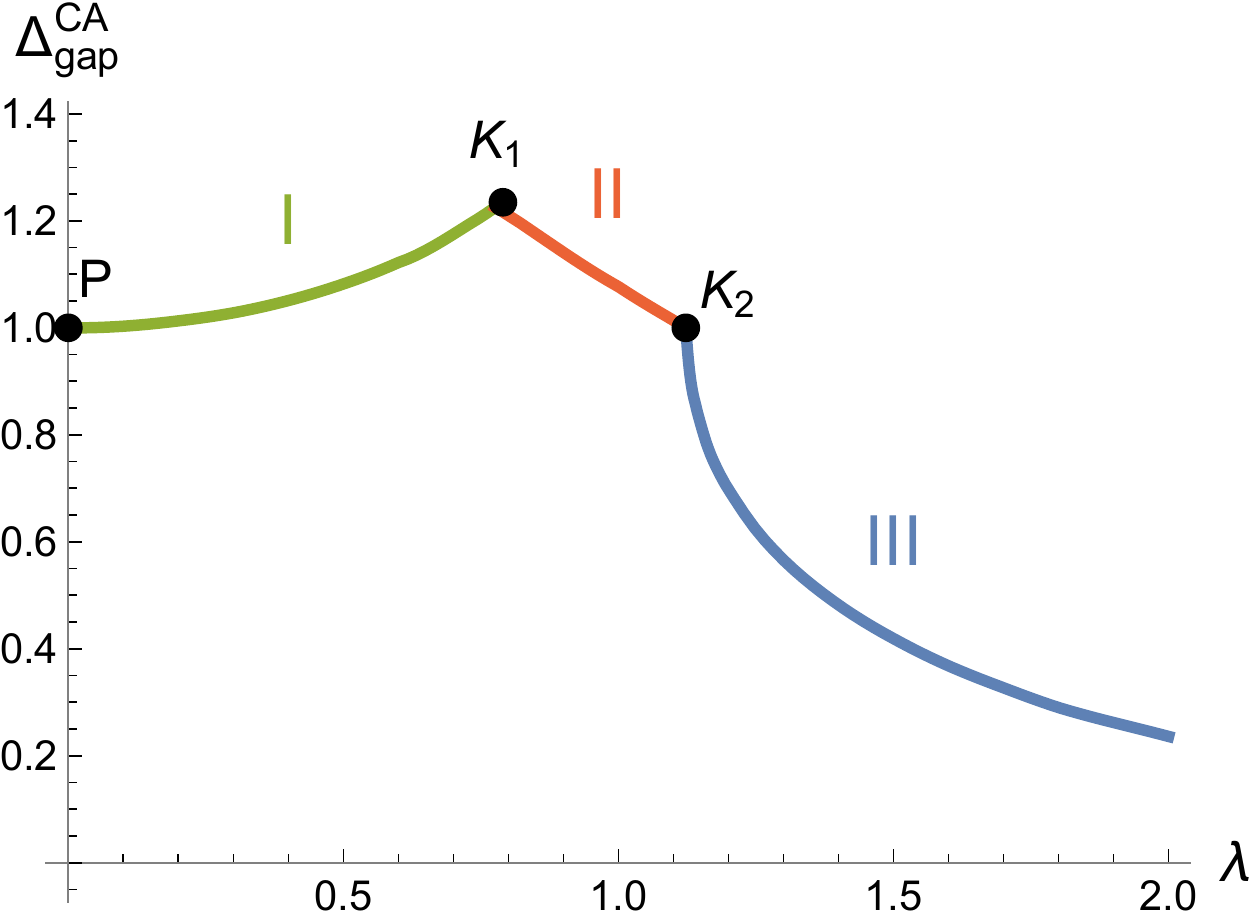}
\caption{Left: The conformal moduli space $M$ of the $\mathcal{N}=(2,2)$ $T^2/\mathbb{Z}_3$ orbifold theory, parametrized by the radius $R$ and the $B$-field $b$ with periodicity $b\sim b+1$.   Right: The numerical bootstrap bound on the lowest dimensional non-BPS operator in the chiral-antichiral channel $\Delta^{CA}_{gap}$ of the four-point function $\langle \phi_{\frac13} \bar \phi_{-\frac13} \bar \phi_{-\frac13} \phi_{\frac13}\rangle$.   The  three real one-dimensional  curves $\mathcal{C}_1,\mathcal{C}_2,\mathcal{C}_3$ on $M$ (left) saturate region I, II, III, respectively, of the numerical bootstrap bound (right). }\label{fig:regions}
\end{figure}

 Rather remarkably, we see that the analytic results of $T^2/\mathbb{Z}_3$ orbifold theory (black dots) saturate  the numerical bootstrap bound (blue curve) for certain loci on the moduli space $M$.  
In the plot of the numerical bootstrap bound, there are two kinks at\footnote{The analytic expressions for the positions of these two kinks are guessed and checked to high numerical precision.} 
\begin{align}
\begin{split}
&K_1: ( \, \lambda \sim 0.776\, ,\, \Delta^{CA}_{gap}=\sqrt{3\over2} \sim1.225 \, )\,,\\
&K_2: (\, \lambda =  2^{\frac16} \sim 1.122\,  ,\,  \Delta^{CA}_{gap}=1 \, )\,.
\end{split}
\end{align}
These two kinks  divide the bootstrap curve into region I, II, and III, ordered from left to right.   We numerically observe that the three regions of the bootstrap curve are saturated by the following loci on the moduli space (see Figure \ref{fig:regions}):\footnote{There are other loci or points on $M$ saturating the bootstrap bound but we will only focus on the loci $\mathcal{C}_1,\mathcal{C}_2,\mathcal{C}_3$ below.}

\noindent $\bullet$ Region I is saturated by the following two disconnected real one-dimensional loci $\mathcal{C}_1$ on the moduli space $M$  of the $T^2/\mathbb{Z}_3$ orbifold theory:\footnote{In fact, since  $\Delta^{CA}_{gap}$ and $\lambda$ are even functions of the $B$-field $b$, each of  the component $\frac13 \le b\le \frac12$ and the component $-\frac12 \le b\le -\frac13$ of $\mathcal{C}_1$ maps to the entire region I in the plot $(\lambda, \Delta^{CA}_{gap})$.}
\begin{align}
\mathcal{C}_1:~ R(b) =  \left[ {16\over 3}  \left( {1\over3} -b^2 \right) \right]^{1\over4}\,,~~~~~\frac13 \le b \le \frac12\,,~~~\text{and}~~~-\frac12 \le b\le -\frac13\,.
\end{align}
In particular at the end point $P$ where $R=\sqrt{2\over 3}$ and $b={1\over 2}$, the orbifold theory factorizes into a tensor product of three $\cN=2$ $A_2$ minimal models. The chiral ring coefficient $\lambda$ vanishes and the CA gap is saturated by the extra conserved currents at this point.

\noindent $\bullet$ Region II is saturated by the real one-dimensional locus $\mathcal{C}_2$ on the moduli space $M$:
\begin{align}
\mathcal{C}_2:~ R(b) =  \left[ {16\over 3}  \left( {1\over3} -b^2 \right) \right]^{1\over4}\,,~~~~~-\frac13 \le b \le \frac13\,.
\end{align}
Even though the curve $\mathcal{C}_1$ and $\mathcal{C}_2$ are smoothly connected on the moduli space $M$, they are separated by a kink $K_1$, which is realized by $(R= {2^{5\over 4} 3^{-{3\over4}}}, b=\pm \frac13)$,  in the plot of $\Delta^{CA}_{gap}$ versus the chiral ring coefficient $\lambda$.  
This is because while  $\lambda(R,b)$ is a continuous function of the moduli $(R,b)$ (as given in \eqref{T2Z3chiralring}),  the gap $\Delta^{CA}_{gap}(R,b)$ is not; the momentum $p_\mu$ and winding number $v^\mu$ that minimize the dimension $h+\bar h$ in \eqref{CAgapT2Z3}  jump  as we vary the moduli from $\mathcal{C}_1$ to $\mathcal{C}_2$.  

\noindent $\bullet$ Region III is saturated by the  real one-dimensional locus $\mathcal{C}_3$ on the moduli space $M$:
\begin{align}
\mathcal{C}_3:~R>0\,,~~~~~b=0\,.
\end{align}
In region III where $b=0$, given a radius $R$, there is a ``dual" radius $R' \equiv {4\over 3R}$ such that $\Delta^{CA}_{gap}(R,0)=\Delta^{CA}_{gap}(R',0)$ and $\lambda(R,0)  =\lambda(R',0)$, and hence mapping to the same point on the plot $(\lambda,\Delta^{CA}_{gap})$.  The kink $K_2$ in the bootstrap bound is realized by the ``self-dual" radius $R= \sqrt{4\over3}$.\footnote{As mentioned below \eqref{zeroC}, there are two other points $R=1/\sqrt{3}$, $b=\pm 1/4$ that also realize the kink $K_2$.}

However, the ``dual" radius $R'$ is \textit{not} the radius obtained by performing $T$-duality twice along the two sides of the torus, which would have been $R^{T\text{-dual}}  =  {4\over \sqrt{3} R}$ (again in the $\alpha'=2$ convention).  Indeed, $T$-duality is not a symmetry of the chiral ring coefficient $\lambda$ of twist fields from a \textit{single} fixed point; rather, $T$-duality would mix twist fields from different fixed points together \cite{Dixon:1986qv}.

\subsubsection{Varying the Chiral-Chiral Gap $\Delta^{CC}_{gap}$}\label{sec:saturation}

In this subsection we will demonstrate how the numerical bootstrap bound for larger values of $\Delta^{CC}_{gap}$ is also saturated by the $T^2/\mathbb{Z}_3$ CFT.

In the case of $c=3$ and external charge $q=\frac13$, we show the bootstrap bound on $\Delta^{CA}_{gap}$ in Figure \ref{Fig:c=3} for various values of the gap $\Delta^{CC}_{gap}$ we assume in the chiral-chiral channel. From the discussion in Section \ref{sec:CCgap},  the minimal gap allowed by the $\mathcal{N}=(2,2)$ representation theory is $\Delta^{CC}_{gap}=\frac23$, which is realized by the descendant of an internal $(a,a)$ primary $\overline\phi_{-\frac13}'$.  In the case of $\Delta^{CC}_{gap}=\frac23$ (the blue curve in Figure \ref{Fig:c=3}), we have discussed how the bootstrap bound is saturated by the $T^2/\mathbb{Z}_3$ theory above in this section.

As we raise the value of $\Delta^{CC}_{gap}$ above $\frac23$ but still below $\frac43$, we observe that the numerical bootstrap bound becomes a peak at the kink $K_2$ (the orange curve in Figure \ref{Fig:c=3}).  Table~\ref{Tab:c=3-peak} suggests that as the derivative order is increased, the peak approaches infinitesimal width and the maximum value of $\Delta^{CC}_{gap}$ approaches $4 \over 3$.  The upper bound on $\Delta^{CA}_{gap}$ is very close to 1 for $\lambda = 2^{1\over6}$ and ${2 \over 3} \leq \Delta^{CC}_{gap} \leq {4 \over 3}$.  The kink $K_2$ is realized by the three points \eqref{zeroC} on the moduli space.  As discussed at the end of Section \ref{sec:CCT2Z3}, these are exactly the points where the OPE coefficient $C(R,b)$ for the $(a,a)$ primary $\overline\phi'_{-\frac13}$ in the chiral-chiral channel vanishes and the $\Delta^{CC}_{gap}$ increases to $\frac43$.

\begin{table}
\centering
\begin{tabular}{|c|c|c|c|}
\hline
$N_\A$ & Upper bound on $\Delta_{gap}^{CC}$ at $\lambda = 2^{1\over6}$ & $\lambda_{min}$ at $\Delta_{gap}^{CC} = {4 \over 3}$ & $\lambda_{max}^*$ at $\Delta_{gap}^{CC} = {4 \over 3}$
\\\hline\hline
4 & 1.37093 & 1.1073 & 1.2993  \\
8 & 1.33396 & 1.1214 & 1.1231 \\
12 & 1.33351 & 1.1225 & 1.1225 \\
16 & 1.33346 & 1.1225 & 1.1225 \\\hline
\end{tabular}
\caption{The width in $\lambda$ and the extent in $\Delta_{gap}^{CC}$ of the allowed region for $\Delta_{gap}^{CC} > {2 \over 3}$ from the numerical bootstrap in the case of $c = 3$ and external charge $q = {1 \over 3}$.  $\lambda_{max}^*$ is defined as the value of $\lambda$ at which $\Delta_{gap}^{CA} = {1 \over 2}$, the reason being that the bounds on $\Delta_{gap}^{CA}$ for $\lambda > 2^{1\over6}$ are nonzero at finite derivative order but approach zero as the derivative order is increased.}
\label{Tab:c=3-peak}
\end{table}

If we further increase the value of $\Delta^{CC}_{gap}$ to be above $\frac43$, the numerical bootstrap bound on $\Delta^{CA}_{gap}$ drops to zero (the green curve), suggesting that there is no zero for the OPE coefficient of the non-BPS operator that is responsible for $\Delta^{CC}_{gap}=\frac43$. 

We therefore reach a satisfying conclusion that the entire three-dimensional  bootstrap bound in Figure \ref{Fig:c=3} is saturated by the $T^2/\mathbb{Z}_3$ orbifold CFT, not just the $\Delta^{CC}_{gap}=\frac23$ slice discussed previously.

\subsection{The $c = 6$, $q=2/3$ Case}

\begin{figure}[h!]
\centering
\includegraphics[width=.6\textwidth]{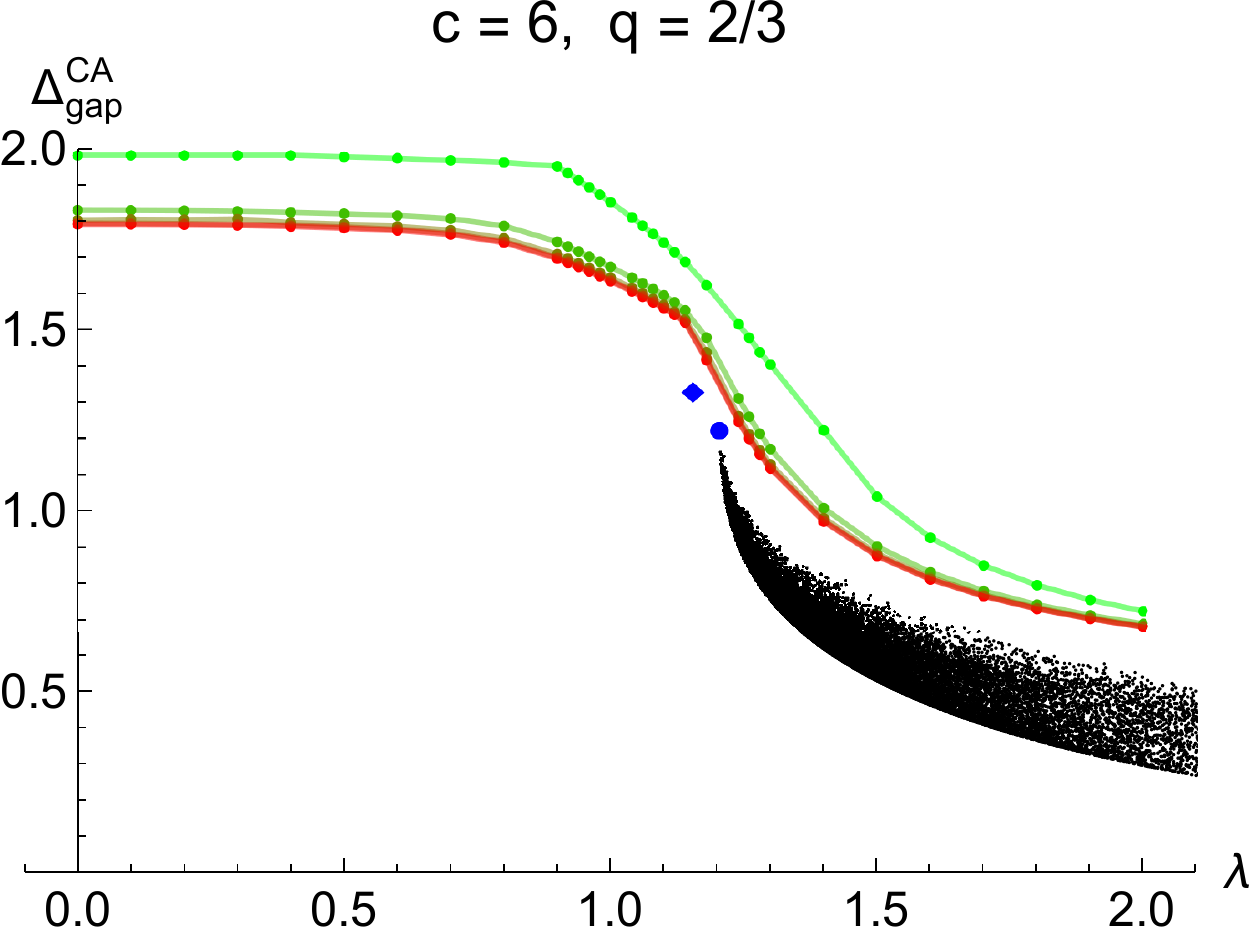}
\caption{Upper bounds on the gap in the CA channel $\Delta^{CA}_{gap}$ in the $c=6$, $q=\frac23$ case. The green to red curves are the numerical bounds obtained from conformal bootstrap, in increasing derivative orders 4, 8, 12, 16, 20. The bounds stabilize to within numerical precision at 20 derivative order.
The black dots are randomly sampled values of $(\lambda, \Delta^{CA}_{gap})$ for the $T^4/\mathbb{Z}_3$ CFT, in the absence of $B$-field. The maximum gap in this case is marked by the blue dot.
The CA gap in the $1^6$ Gepner model is labelled by the blue diamond.
}
\label{fig:T4Z3B=0}
\end{figure}

Next we consider $(2,2)$ SCFTs with $c=6$ and BPS primaries of R-charge $\pm {2\over 3}$. Examples include the twist fields in the $\mathbb{Z}_3$ orbifold of a $T^4$ or K3 CFT that admit $\mathbb{Z}_3$ symmetry. In Figure~\ref{fig:T4Z3B=0} we show the upper bound on the gap in the CA channel $\Delta^{CA}_{gap}$ as a function of the chiral ring coefficient $\lambda$, without any assumption on the gap in the CC channel $\Delta^{CC}_{gap}$.  

We can compare the bounds with the solvable free orbifold $T^4 / \bZ_3$ CFT. The external BPS primaries are taken to be $\mathbb{Z}_3$ twist fields in RR sector.  The metric on the $T^4$ may be written as
\ie
ds^2 = G_{\mu \nu} dx^\mu dx^\nu = M_{ij} (dx^i  +  \omega dx^{i+2}) ( dx^j + \omega^2 dx^{j+2})\,,
\fe
where $\omega= \exp(2\pi i /3)$, $\mu=1,\cdots, 4$ and $i=1,2$.  The periodicities of the coordinates are $x^\mu \sim x^\mu +2\pi$.  $M_{ij}$ is a Hermitian, positive-definite 2$\times$2 matrix, parametrizing the moduli of the $\mathbb{Z}_3$-invariant $T^4$. The $\mathbb{Z}_3$ acts simultaneously on the planes $(x^1,x^3)$ and $(x^2,x^4)$ as rotations by $2\pi /3$.   

In the absence of $B$-field, the chiral ring coefficient $\lambda$ as a function of $M_{ij}$ is given by a direct generalization of \eqref{T2Z3chiralring} 
\begin{align}\label{T4Z3chiralring}
\lambda(M_{ij})   =  \left[\sqrt{3\over 2} {\Gamma(\frac 23)^2 \over \Gamma(\frac13)}\right]^n \sqrt{\det M} \sum_{v^i\in \mathbb{Z}} \exp\left[   - {\sqrt{3} \pi  \over 2 }M_{ij} (v^i + \omega v^{i+2})(v^j + \omega^2 v^{j+2})\right]\,,
\end{align}
with $n=2$ (the complex dimension of the target space torus). 
The weights of the exponential operators in the chiral-antichiral channel are given in \eqref{CAgapT2Z3}.  Values of $(\lambda(M_{ij}) ,\Delta^{CA}_{gap}(M_{ij}))$ for the $T^4/\mathbb{Z}_3$ CFT in the absence of $B$-field are plotted as black dots in Figure~\ref{fig:T4Z3B=0}. They occupy a domain with $\lambda\gtrsim 1.20474$, with the maximal gap at $(\lambda, \Delta^{CA}_{gap})= (1.20474, \sqrt{3\over2})$. When a nonzero flat $B$-field is turned on, all values of $\lambda$ can be realized. At a special value of $B$-field moduli (and the metric moduli), the SCFT is described by a LG model with superpotential $W=\sum_{i=1}^6 X_i^3$. Taking the $q={2\over 3}$ chiral primary to be ${\cal O}=X_1 X_2+X_3 X_4+X_5 X_6$, we see that the CA gap is saturated by the non-BPS primary $X_1 X_2 \overline X_3 \overline X_4 +X_3 X_4 \overline X_5 \overline X_6 +X_5 X_6 \overline X_1 \overline X_2 +c.c$ with $\Delta^{CA}_{gap}={4\over 3}$. With normalizations taken into account, the chiral ring coefficient is determined to be $\lambda={2\over \sqrt{3}}$.

We do not know the precise domain occupied by the twist field OPE in $T^4/\mathbb{Z}_3$ with general nonzero flat $B$-field in the plot of Figure \ref{fig:T4Z3B=0}; random numeric sampling indicates\footnote{The challenge in performing a dense numerical sampling lies in the problem of finding the shortest vector in a high rank lattice of generic pairing matrix (in the $T^{2n}/\mathbb{Z}_3$ case, a $4n$-dimensional lattice of momentum and winding is involved), which is NP-hard \cite{SVP}.} that the bootstrap bound is not saturated by $T^4/\mathbb{Z}_3$ for any value of $\lambda$\footnote{While we do not have a reliable extrapolation of the bounds to infinite derivative order using bounds at derivative order 20 and lower, such attempts with an ansatz that is quadratic in inverse derivative order suggest that 
the infinite derivative order bounds would not be saturated by $T^4/\bZ_3$ at any point in the moduli space we sampled.
}, unlike the previously considered $c=3$, $q=1/3$ case where the bound is saturated by the $T^2/\mathbb{Z}_3$ CFT.

\subsection{The $c=9$, $q=1$ Case}

\begin{figure}[h!]
\centering
\subfloat{
\includegraphics[width=.49\textwidth]{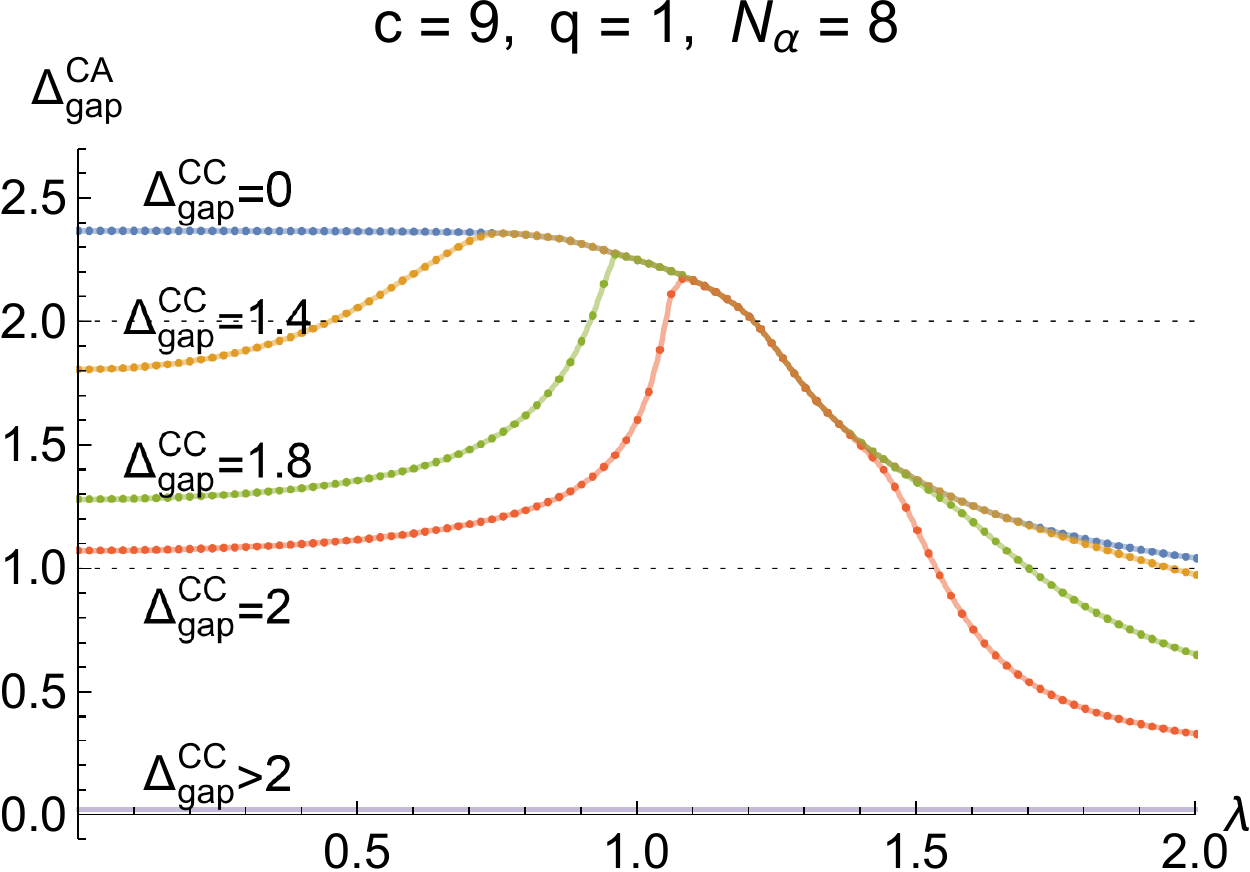}
}
\subfloat{
\includegraphics[width=.49\textwidth]{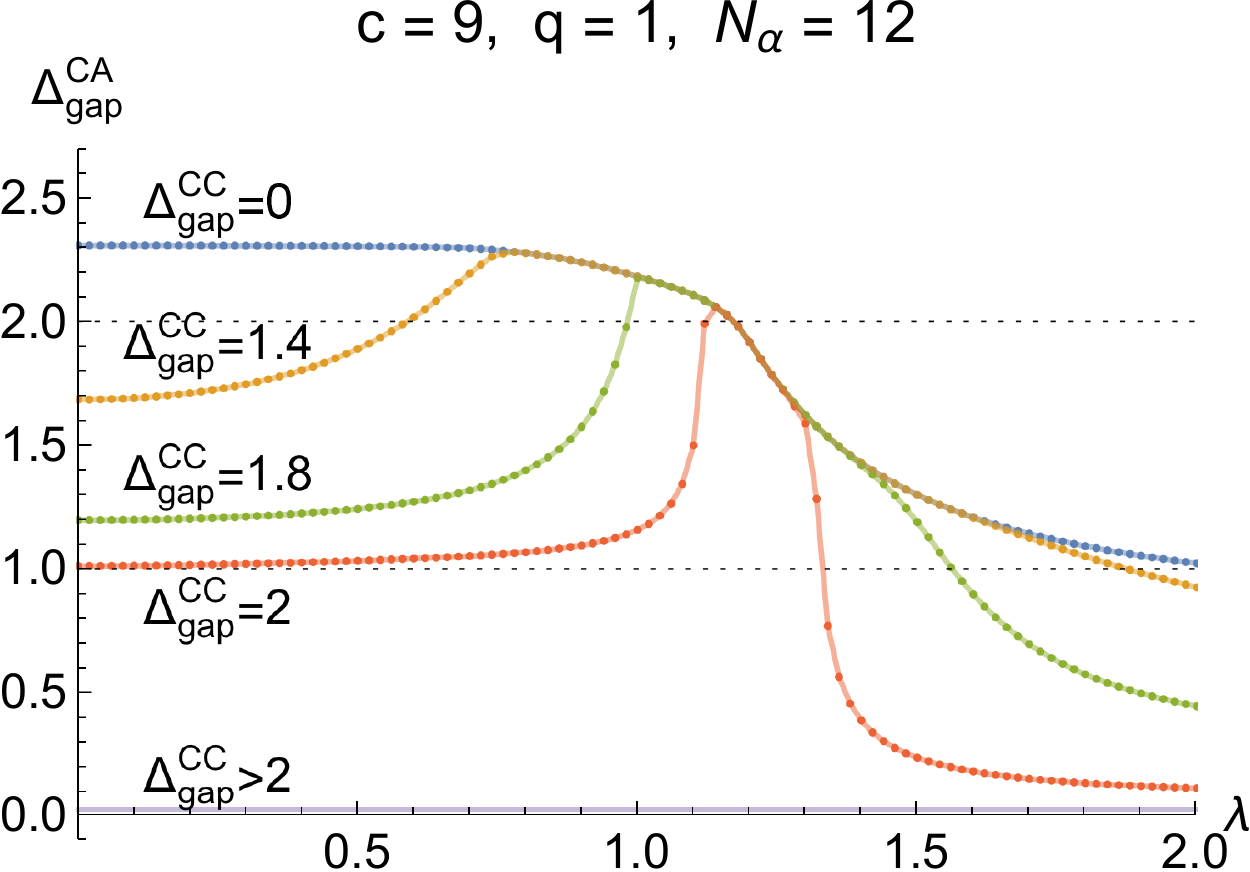}
}
\\\vspace{.2in}
\subfloat{
\includegraphics[width=.49\textwidth]{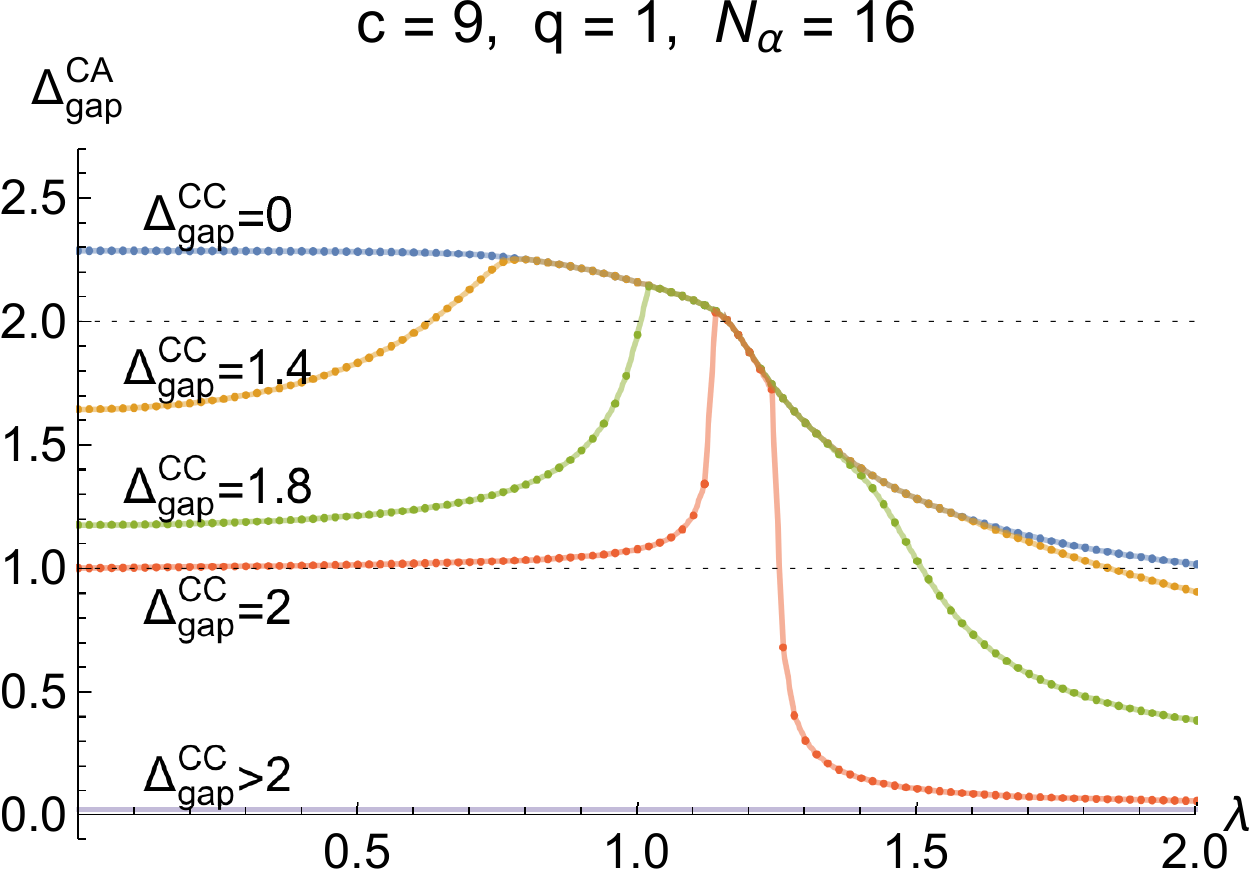}
}
\subfloat{
\includegraphics[width=.49\textwidth]{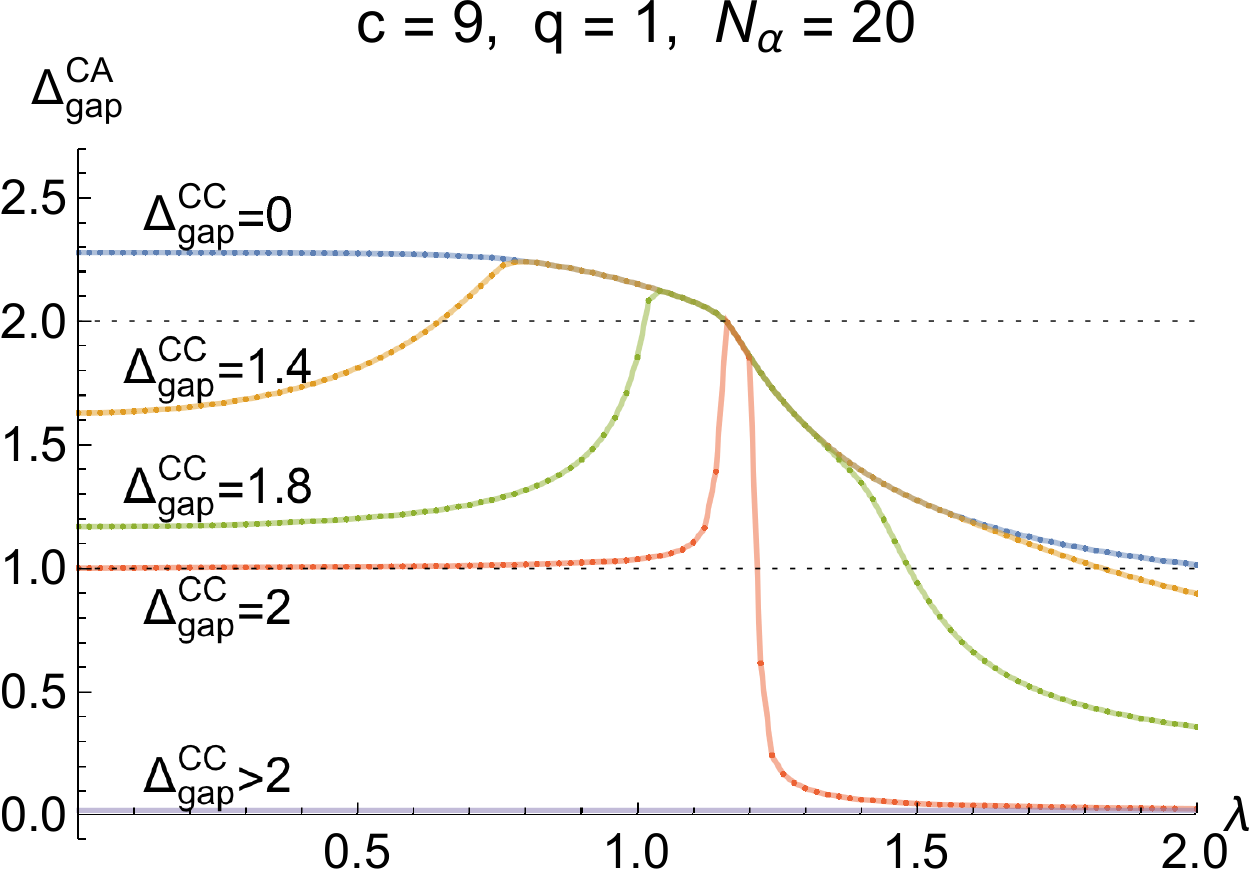}
}
\caption{Upper bounds on the gap in the CA channel $\Delta^{CA}_{gap}$ from the numerical bootstrap, in the case of $c = 9$ and $q = 1$, at derivative orders 8, 12, 16, and 20.  The bounding curves are plotted for different values of the gap in the CC channel $\Delta^{CC}_{gap}$, as functions of the chiral ring coefficient $\lambda$.  The values of $\Delta^{CC}_{gap}$ from top to bottom are $0, \, 1.4, \,1.8, \,2, \, >2$.  We see that the upper bound on $\Delta^{CC}_{gap}$ is 2.  For $\Delta^{CC}_{gap} = 2$, the width of the peak is expected to shrink to zero at infinite derivative order, supported only at $\lambda = {2 \over \sqrt3}$.  The bounding curve (at infinite derivative order) for $\Delta^{CC}_{gap} = 2$ and $\lambda \leq {2 \over \sqrt3}$ is saturated by free fermion correlators (say at the large volume point of the quintic).
}
\label{Fig:c=9}
\end{figure}

\begin{figure}[h!]
\centering
\subfloat{
\includegraphics[width=.49\textwidth]{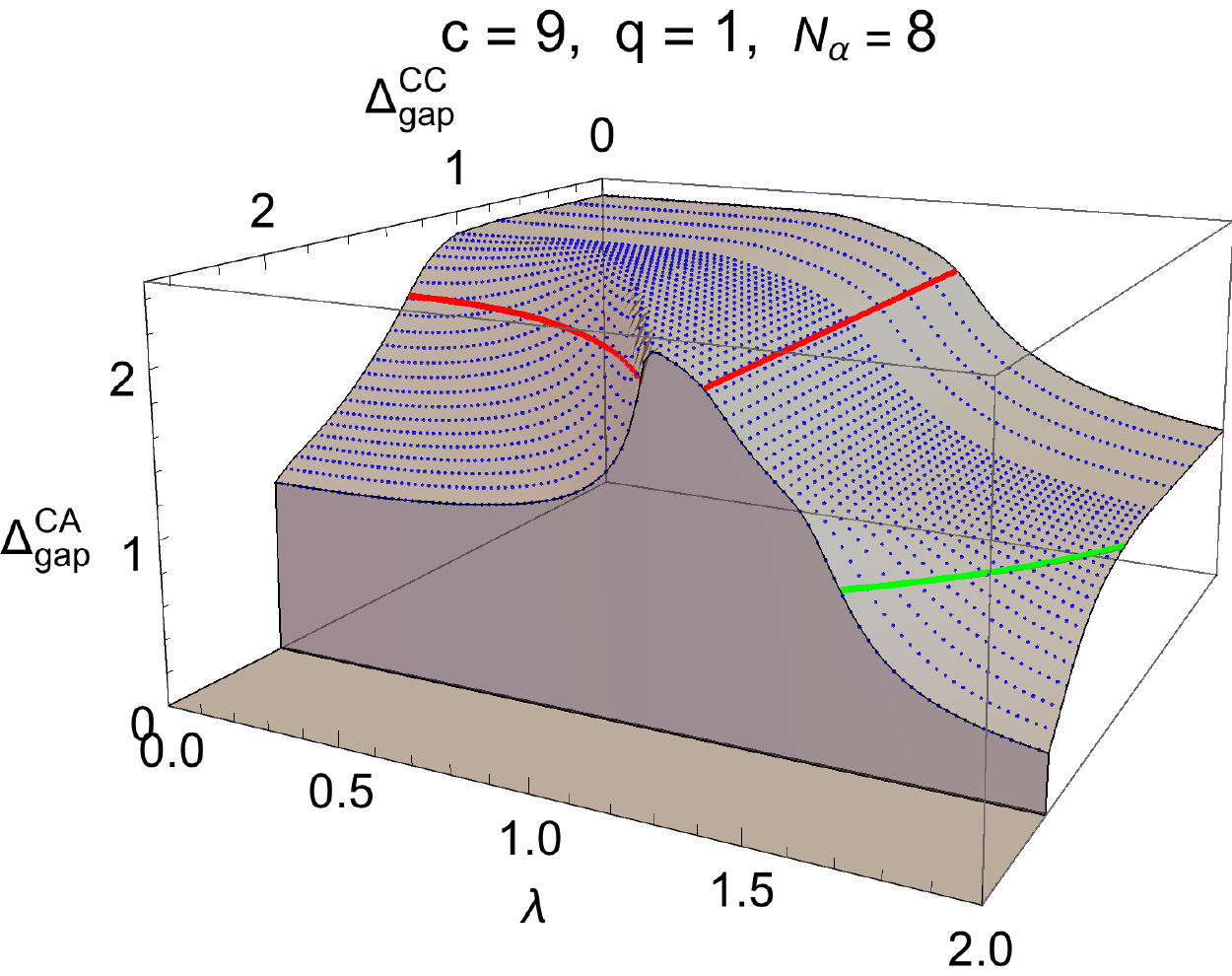}
}
\subfloat{
\includegraphics[width=.49\textwidth]{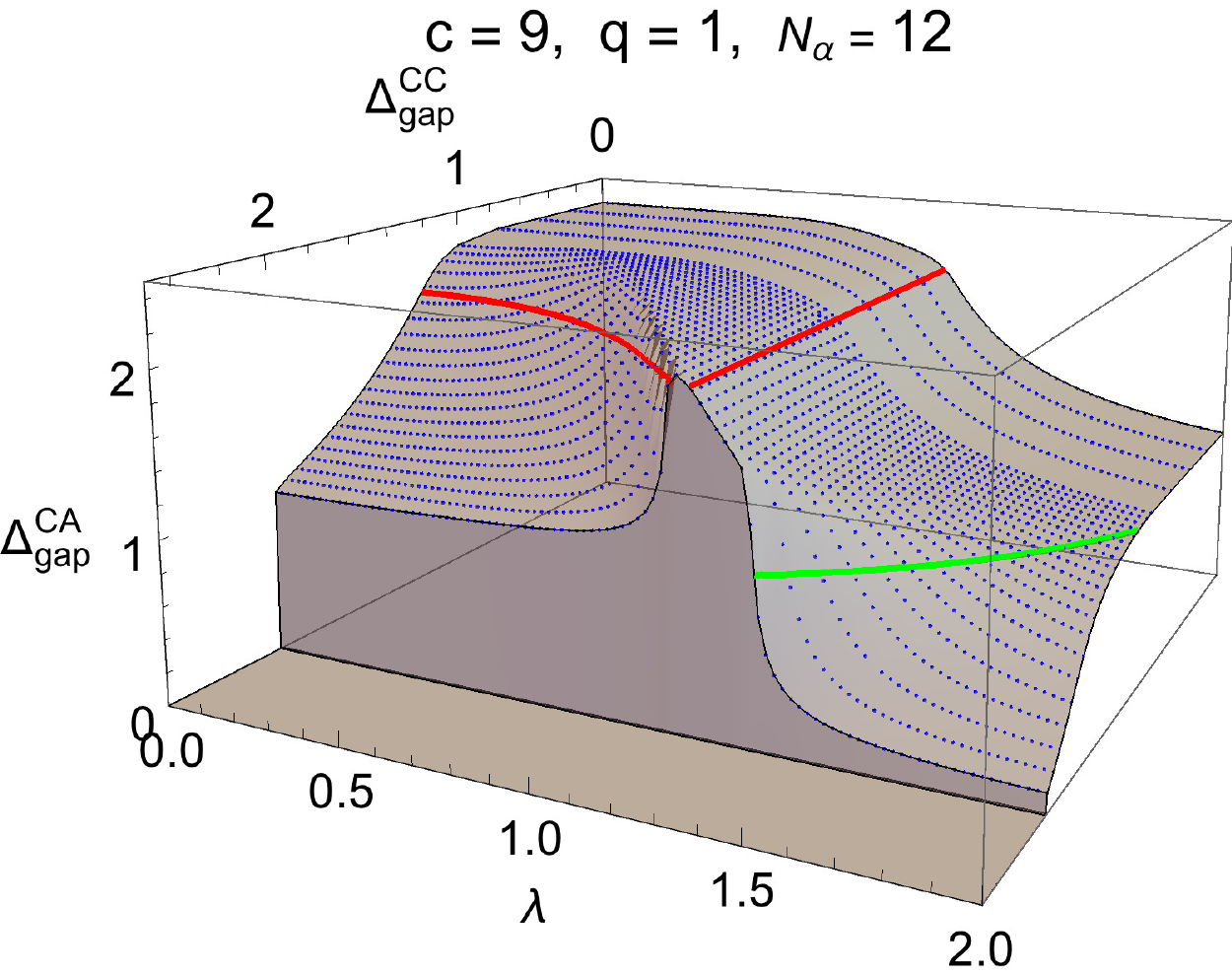}
}
\\\vspace{.2in}
\subfloat{
\includegraphics[width=.49\textwidth]{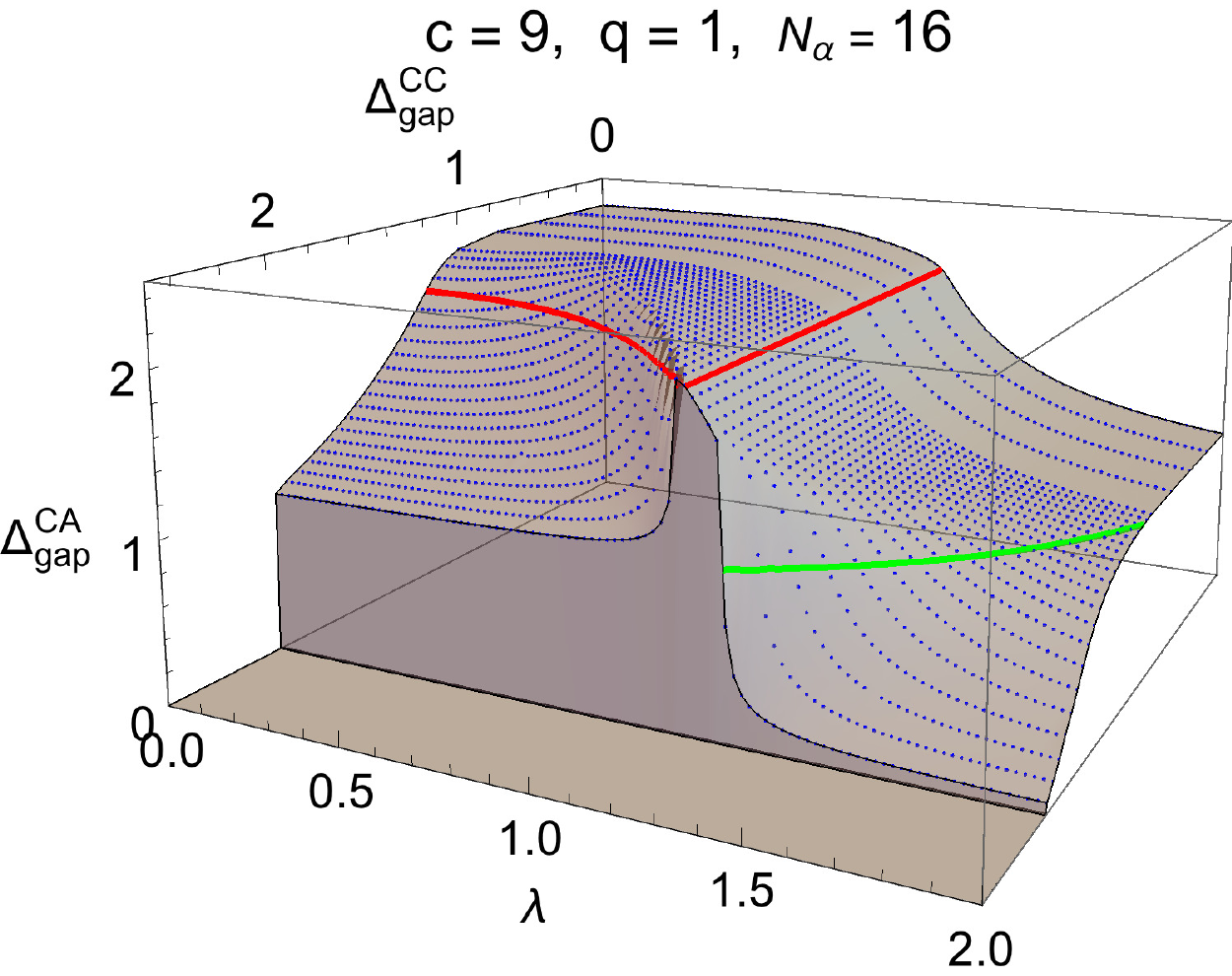}
}
\subfloat{
\includegraphics[width=.49\textwidth]{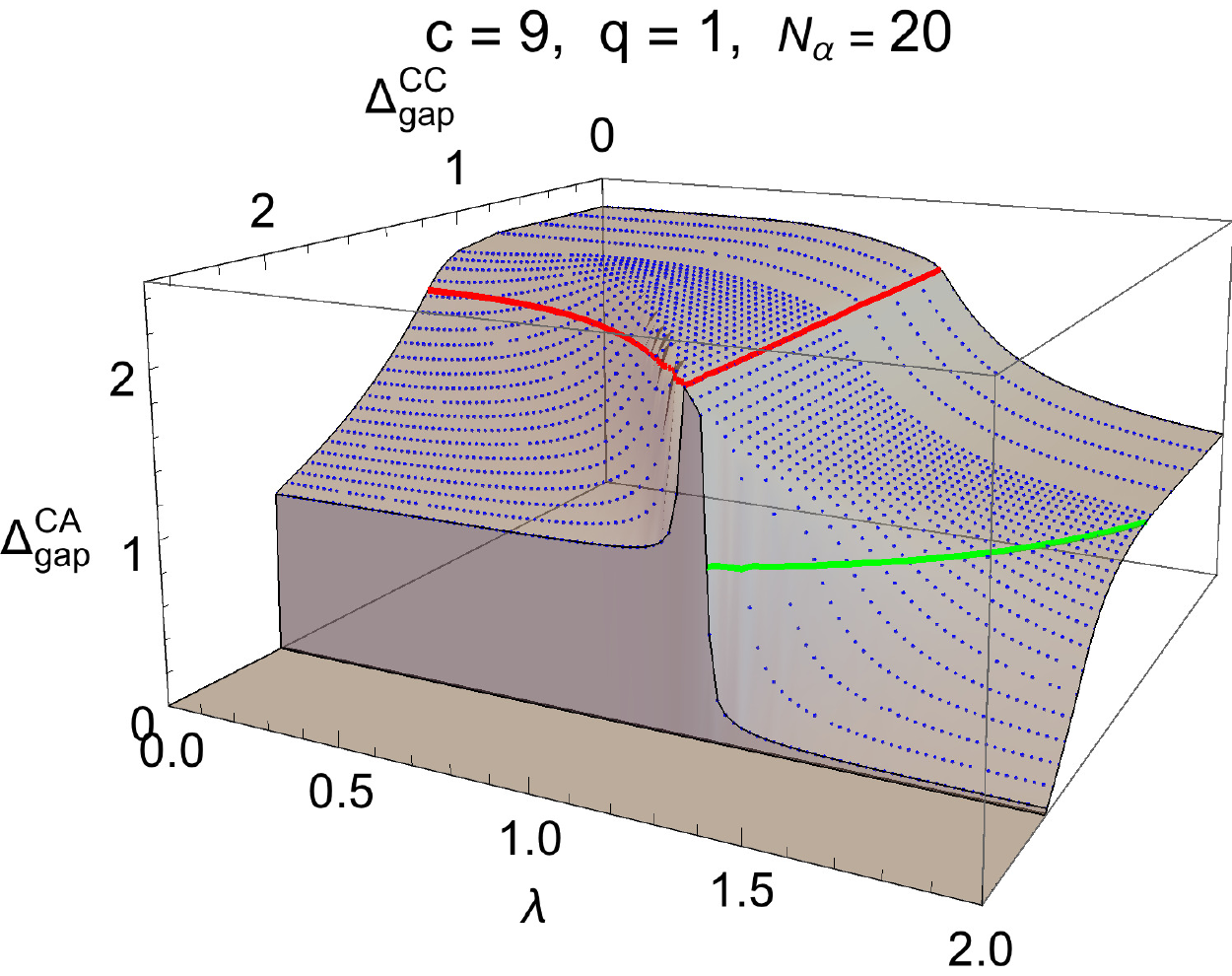}
}
\caption{Upper bounds on the gap in the CA channel $\Delta^{CA}_{gap}$ from the numerical bootstrap, in the case of $c = 9$ and $q = 1$, at derivative orders 8, 12, 16, and 20.  The green and red curves are the contours of $\Delta^{CA}_{gap} = 1$ and 2, respectively, on the bounding surface.  There is an overall upper bound on $\Delta^{CC}_{gap}$ that approaches 2 at large derivative orders, so no data point is displayed for $\Delta^{CC}_{gap} > 2$.
  }
\label{Fig:c=9-3d}
\end{figure}

Finally we turn our attention to $c = 9$ SCFTs, which include supersymmetric nonlinear sigma models on Calabi-Yau threefolds, and consider the OPE of chiral and antichiral primaries with R-charge $\pm1$.  The $G^\mp_{-{1\over2}} \tilde{G}^\mp_{-{1\over2}}$ descendants of these (anti)chiral primaries are associated with exactly marginal deformations of the SCFT. 
In Figures~\ref{Fig:c=9} and~\ref{Fig:c=9-3d} we present the numerical bootstrap bound on the scaling dimension $\Delta^{CA}_{gap}$ of the lightest non-BPS operator in the chiral-antichiral channel, as a function of the chiral ring coefficient $\lambda$ and the gap in the scaling dimension $\Delta^{CC}_{gap}$ of non-BPS operators in the chiral-chiral channel, at derivative orders $N_\A=8, 12, 16, 20$.
For fixed $\Delta_{gap}^{CC}$, The bounding curve $\Delta_{gap}^{CA}(\lambda)$ has a kink at $(\lambda, \Delta^{CA}_{gap}) = ({2\over\sqrt3}, 2)$, which is realized by the OPEs of free fermions (see Appendix \ref{freekink}).

%

\begin{figure}[h!]
\centering
\subfloat{
\includegraphics[width=.49\textwidth]{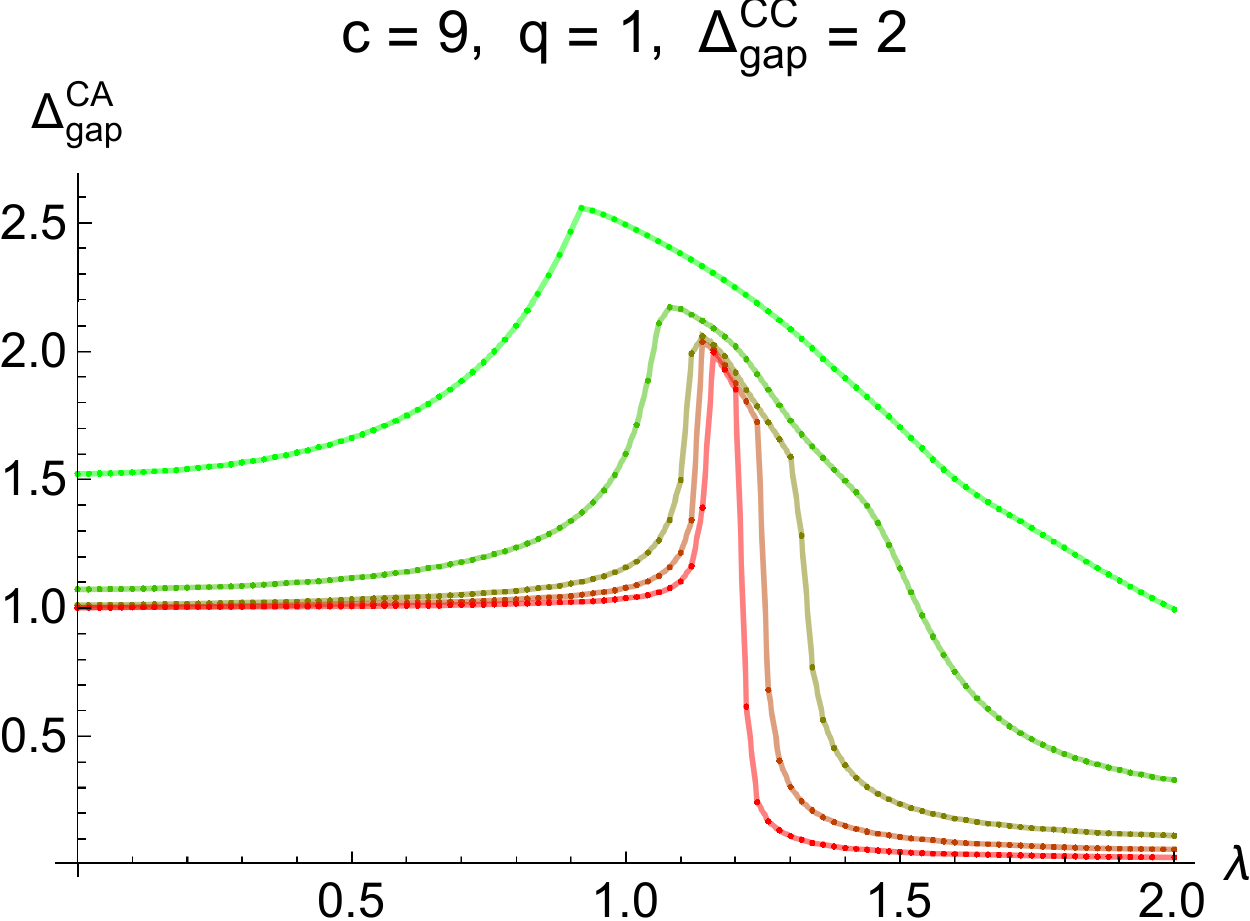}
}
\subfloat{
\includegraphics[width=.49\textwidth]{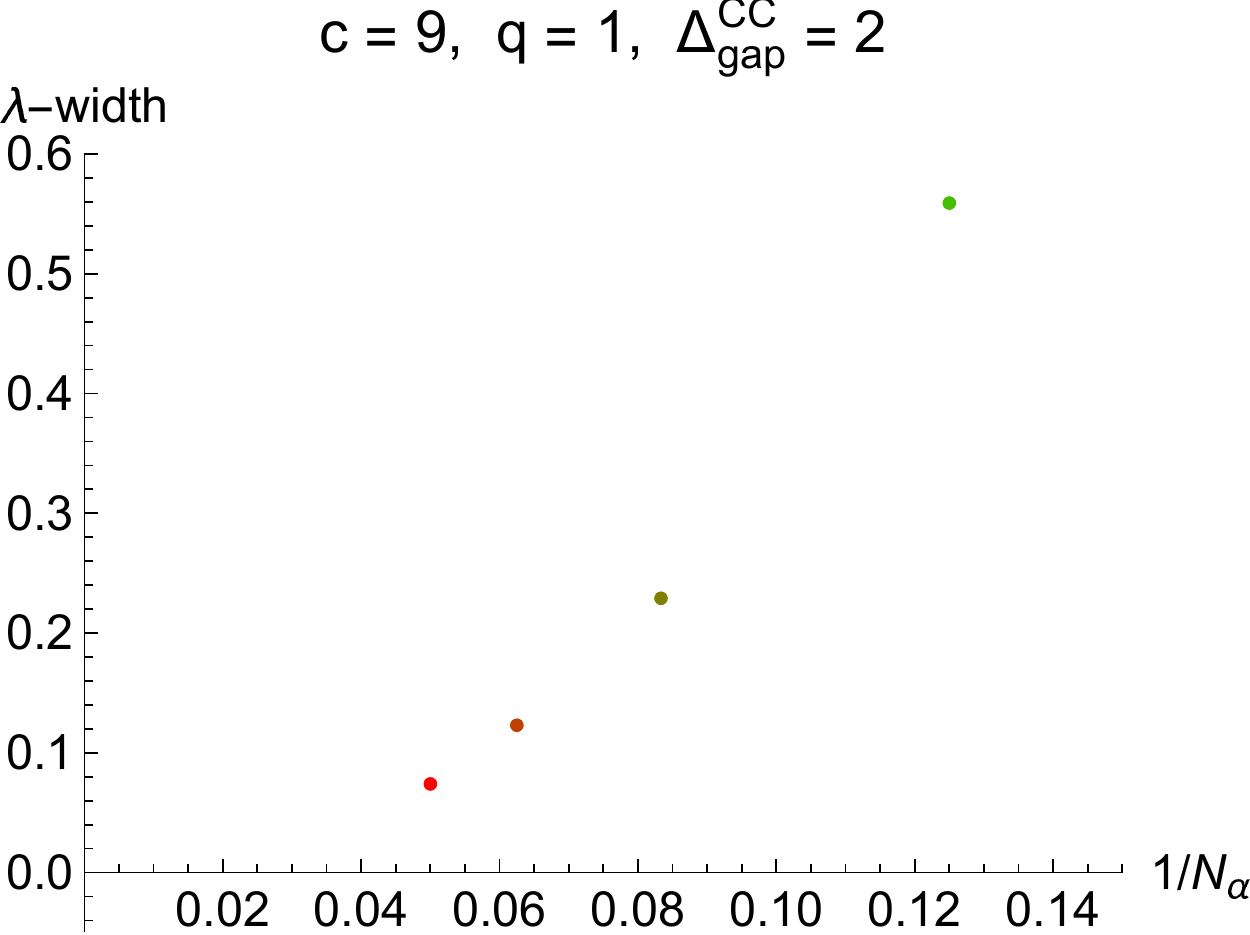}
}
\caption{{\bf Left:} Upper bounds on the gap in the CA channel $\Delta^{CA}_{gap}$ for the gap in the CC channel $\Delta^{CC}_{gap} = 2$, as a function of the chiral ring coefficient $\lambda$,  in the case of $c = 9$ and $q = 1$, at derivative orders 4, 8, 12, 16, 20 (green to red).  {\bf Right:} The width of the peak plotted over inverse derivative order.  The width is defined as $\lambda_r - \lambda_l$, where $\lambda_r$ is the value of $\lambda$ at which $\Delta^{CA}_{gap} = 1$ on the right edge, and $\lambda_l$ is the value of $\lambda$ at which $\Delta^{CA}_{gap} = 1.5$ on the left edge.
}
\label{Fig:c=9-width}
\end{figure}


As the gap $\Delta^{CC}_{gap}$ in the chiral-chiral channel is increased from 0, the bound $\Delta^{CA}_{gap}(\lambda)$ becomes stronger and is no longer monotonic in $\lambda$. There is a maximal value for the chiral-chiral channel gap $\Delta_{gap}^{CC}=2$, above which the crossing equation cannot be satisfied (and the CA gap drops to zero). If we assume $\Delta_{gap}^{CC}=2$, the CA gap bound at increasing derivative orders $N_\A$ (as shown in Figure \ref{Fig:c=9-width}) strongly suggests a convergence to the following bound at $N_\A=\infty$:
\ie\label{c=9q=1CC=2}
\Delta^{CA}_{gap} = \begin{cases}
1, & \lambda < {2\over\sqrt3}
\\
2, & \lambda = {2\over\sqrt3}
\\
0, & \lambda > {2\over\sqrt3}
\end{cases}~~~~~~~~~~~~~~(\Delta^{CC}_{gap} = 2)
\fe
As explained in Appendix \ref{freekink}, this entire curve (in fact, it is part of two lines of kinks in the three-dimensional plot of Figure \ref{Fig:c=9}, along $\Delta^{CC}_{gap}=2$, $\Delta^{CA}_{gap}=1$, and along $\lambda = {2\over\sqrt{3}}$, $\Delta^{CA}_{gap}=2$) is saturated by OPEs in free theories.

\begin{figure}
\centering
\includegraphics[width=.55\textwidth]{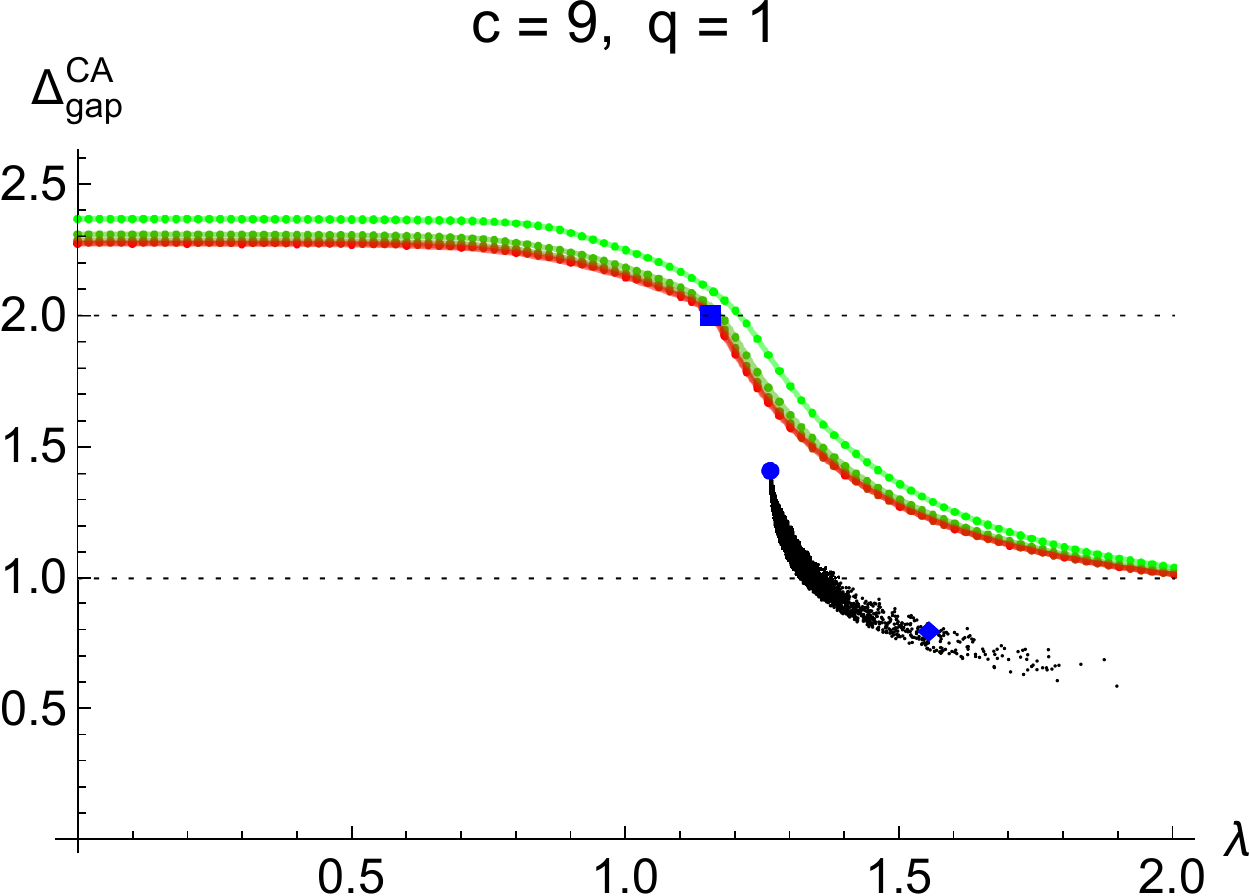}
\caption{Upper bounds on the gap in the CA channel $\Delta^{CA}_{gap}$ in the $c=9$, $q=1$ case. The green to red curves are the numerical bounds obtained from conformal bootstrap, in increasing derivative orders 8, 12, 16, 20, 24. The black dots are randomly sampled values of $(\lambda, \Delta^{CA}_{gap})$ for the $T^6/\mathbb{Z}_3$ CFT, in the absence of $B$-field. The blue circle dot marks the maximal $\Delta^{CA}_{gap}$ in this case. The blue square marks the large volume limit of the quintic that sits at the kink of the bounding curve. The blue diamond (buried in the black dots) marks the $3^5$ Gepner  model, which has $\Delta^{CC}_{gap} = {6 \over 5}$.
}
\label{fig:T6Z3B=0}
\end{figure}

We may compare our numerical bounds with the OPE of K\"ahler moduli operators of the quintic threefold model.  See Appendix~\ref{Sec:Quintic} for a review of the quintic model and the dependence of the chiral ring coefficient on the K\"ahler moduli.  The chiral ring coefficient has a global minimum $\lambda = {2\over\sqrt3} \approx 1.1547$ in the large volume limit, and diverges at the conifold point. As already mentioned, the large volume point has $\Delta^{CA}_{gap} = 2$, which lies precisely at a kink on our bounding curve (see Figure \ref{fig:T6Z3B=0}).  At the conifold point, the CA and CC gaps are $\Delta^{CA}_{gap}={1\over2}$ and $\Delta^{CC}_{gap}= 0$ (see Appendix~\ref{ap:liouville} for details).
The $3^5$ Gepner model (orbifold of five copies of the $c = {9\over5}$ ${\cal N}=2$ minimal model) lies at a point on the K\"ahler moduli space where the chiral ring coefficient takes a local minimum value $\lambda =\frac{\Gamma \left(\frac{3}{5}\right)^{15/2} \Gamma \left(\frac{1}{5}\right)^{5/2}}{\Gamma \left(\frac{2}{5}\right)^{15/2} \Gamma \left(\frac{4}{5}\right)^{5/2}}\sim 1.55532$. In this Gepner model, $\Delta^{CA}_{gap}={4\over5}$ is well within the bootstrap bounds if we assume its value of the gap in the CC channel, $\Delta^{CC}_{gap}={6\over5}$. A discussion of Gepner points that lie inside the moduli space of one parameter Calabi-Yau sigma models
can be found in Appendix~\ref{ap:gepner}.

We can also compare our bounds with the twist field OPE of the Z-manifold, i.e. the $T^6/\mathbb{Z}_3$ CFT \cite{Candelas:1985en}, at the free orbifold point (i.e. without deforming by marginal twist fields, but with generic moduli for the $\mathbb{Z}_3$ invariant $T^6$). The chiral ring coefficient is given by \eqref{T4Z3chiralring} with $n=3$, and the CA gap $\Delta^{CA}_{gap}$ is given in \eqref{CAgapT2Z3}.  The values of $(\lambda ,\Delta^{CA}_{gap})$ for the $T^6/\mathbb{Z}_3$ CFT in the absence of $B$-field are shown as black dots in Figure~\ref{fig:T6Z3B=0}, with the maximal gap realized at $(\lambda,\Delta^{CA}_{gap}) \sim (1.26419, \sqrt{2})$. When a nonzero $B$-field is turned on, all values of $\lambda$ can be realized. We do not know the precise domain in $(\lambda ,\Delta^{CA}_{gap})$ realized by $T^6/\mathbb{Z}_3$ with general nonzero $B$-field, despite having numerically sampled over a large set of points over the (K\"ahler) moduli space. It appears that the twist field OPE of $T^6/\mathbb{Z}_3$ never saturates our bootstrap bound on the CA gap, for any value of $\lambda$.

\begin{figure}[h!]
\centering
\subfloat{
\includegraphics[width=.49\textwidth]{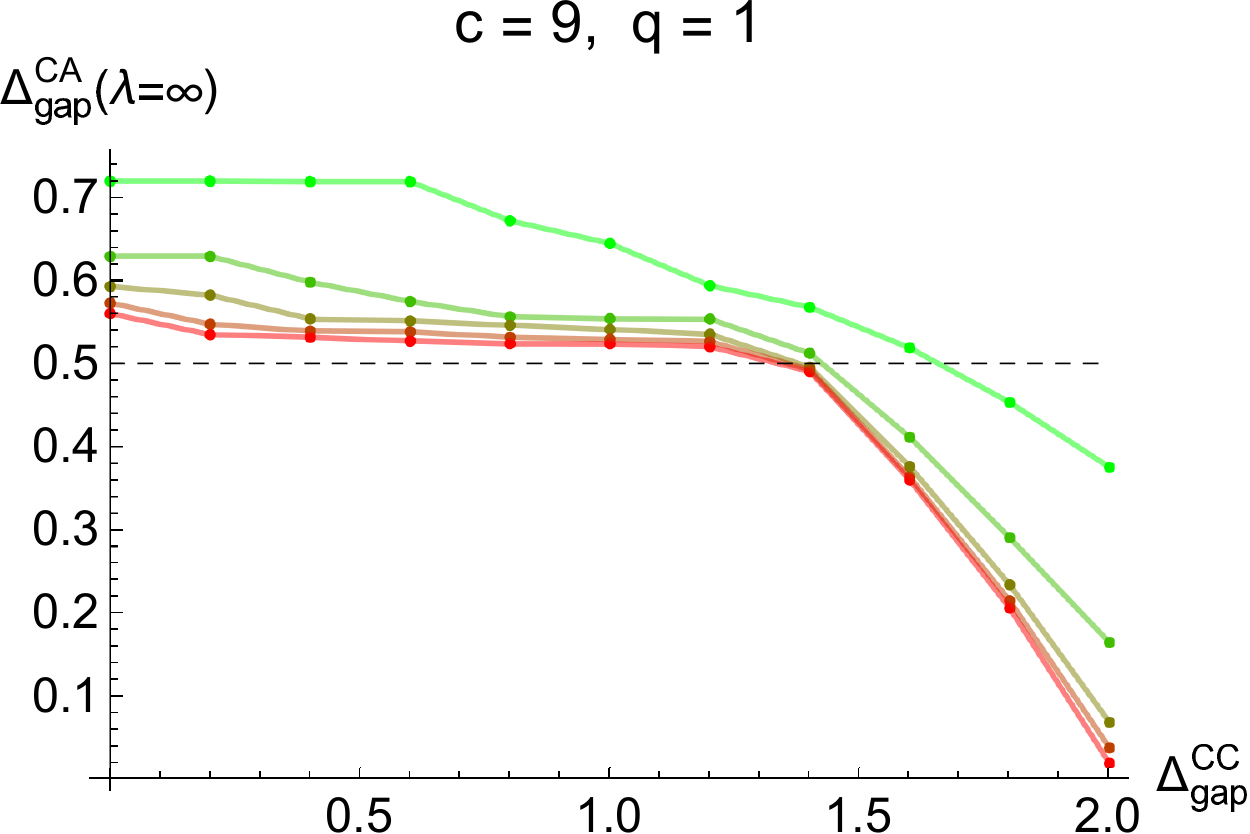}
}
\subfloat{
\includegraphics[width=.49\textwidth]{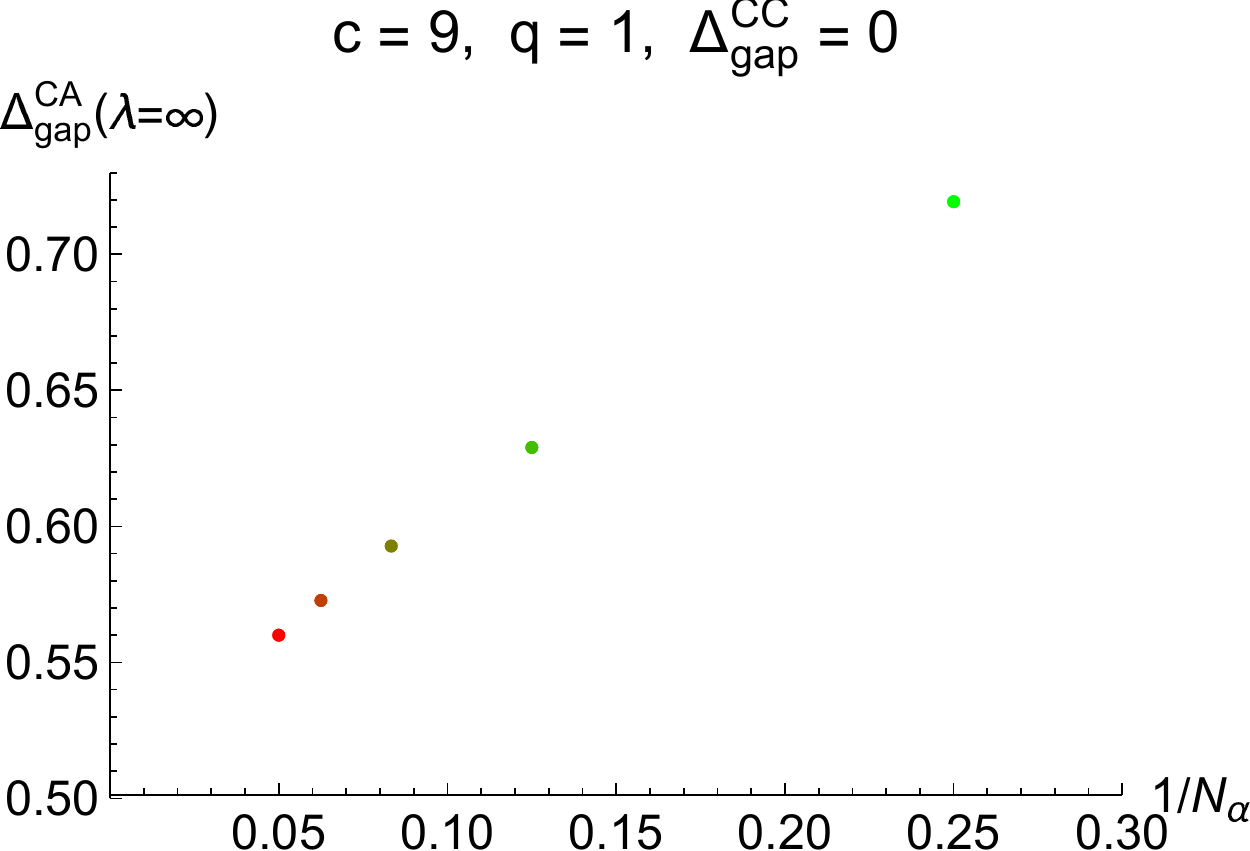}
}
\caption{{\bf Left:} Upper bounds on the gap in the CA channel $\Delta^{CA}_{gap}$ from the numerical bootstrap, as a function of the gap in the CC channel $\Delta^{CC}_{gap}$, in the limit of infinite chiral ring coefficient $\lambda=\infty$, in the case of $c = 9$ and $q = 1$, at derivative orders 4, 8, 12, 16, 20.  
{\bf Right:} The bounds $\Delta^{CA}_{gap}$ at $\lambda = \infty$ and $\Delta^{CC}_{gap} = 0$ plotted over inverse derivative order hints at a convergence towards $\Delta^{CA}_{gap} = {1\over2}$.
}
\label{Fig:c=9-inf}
\end{figure}

Figure~\ref{Fig:c=9-inf} shows the bound on $\Delta^{CA}_{gap}$ in the limit of infinite chiral ring coefficient $\lambda=\infty$, with dependence on $\Delta^{CC}_{gap}$.  For Calabi-Yau models, the infinite $\lambda$ limit corresponds to the conifold point. A continuum of operators with dimension gap ${1\over 2}$ is expected to develop in the vicinity of the conifold point, described by the $c=9$ ${\cal N}=2$ Liouville theory \cite{Ooguri:1995wj} (See Appendix \ref{ap:liouville}). Indeed, our bound on $\Delta^{CA}_{gap}$ at $\lambda=\infty$ is likely saturated by the ${\cal N}=2$ Liouville theory. So far, we have been unable to optimize the bounds of Figure~\ref{Fig:c=9-inf} by a reliable extrapolation to infinite derivative order, due to the limitation of computational power. Unlike at finite $\lambda$, where the bounds stabilize at $d_q = N_\A + 4$, at infinite $\lambda$ it is found empirically that at least $d_q = N_\A + 8$ is required.
Note that there appears to be a transition at $\Delta^{CC}_{gap} \approx 1.4$, above which $\Delta^{CA}_{gap}(\lambda=\infty)$ decreases from ${1\over 2}$, and vanishes as $\Delta^{CC}_{gap}$ exceeds 2.

\section{Summary and Outlook} 

We began with the {\it known knowns}: the chiral ring data, whose moduli dependence is understood, and constrained the {\it known unknowns}: the spectrum of non-BPS operators in Landau-Ginzburg or Calabi-Yau models at generic points on their moduli spaces. We have also probed the {\it unknown unknowns} \cite{Rumsfeld}: the spectra of general $(2,2)$ SCFTs that admit exactly marginal deformations, by constraining the OPE content of  marginal BPS operators. 

We carved out some allowed domains in the space of possible gaps in the CA and CC OPE channels and the chiral ring coefficient. Let us recap some of the main results:

\noindent $\bullet$ An upper bound on the gap in the OPE of a marginal BPS operator and its conjugate was computed for $3\leq c\leq 9$. Interestingly, the bound appears to be saturated by products of ${\cal N}=2$ minimal models for a few special values of the central charge, $c=3, {10\over 3}, {7\over 2}, {18\over 5}, {15\over 4}$.

\noindent $\bullet$ For $c=3$ theories, we considered the gap in the OPE of a pair of BPS primaries of R-charge $\pm {1\over 3}$, as a function of the chiral ring coefficient. The entire bounding curve is saturated by the twist field OPE in the superconformal orbifold $T^2/\mathbb{Z}_3$, along a curve in the K\"ahler moduli space of the latter.

\noindent $\bullet$ For $c=9$ theories, we considered the gaps in the OPE of a pair of marginal BPS primaries (of R-charge $\pm 1$), as a function of the chiral ring coefficient. Without making any assumptions on the CC channel gap, we saw that a kink on the bounding curve, at $\lambda = {2\over \sqrt{3}}$, $\Delta^{CA}_{gap}=2$, is saturated by the OPE of free fields. In the context of K\"ahler deformations of 1-parameter Calabi-Yau models, the kinks corresponds to the large volume limit. The K\"ahler deformations of the quintic model only realizes $\lambda\geq {2\over \sqrt{3}}$. Smaller values of $\lambda$ can be realized on other 1-parameter Calabi-Yau models. In this case, we found that $\Delta^{CA}_{gap}$ may exceed the free field value, namely 2. It remains to be seen whether this larger allowed gap can be realized in the quantum regime of Calabi-Yau models.

\noindent $\bullet$ Various Gepner models and the twist field OPE of the Z-manifold $T^6/\mathbb{Z}_3$ lie well within our bounds. The gap in the continuum that develops at the conifold point, however, appears to saturate our bound in the $\lambda\to \infty$ limit.

\noindent $\bullet$ We observed various kinks on the boundary of the allowed domain in $(\lambda,  \Delta^{CC}_{gap},\Delta^{CA}_{gap})$, some of which are saturated by OPEs of free fields. Many of the features of this plot remain unexplained, and it would be nice to understand whether all of it can be realized by $(2,2)$ SCFTs.

The non-BPS spectrum in Calabi-Yau sigma models has also been constrained from  modular invariance of the torus partition function \cite{Keller:2012mr}.  In that work, an upper bound on the dimension of the lightest non-BPS operator in the \textit{entire spectrum} (rather than in specific OPEs) is obtained numerically as a function of the total Hodge number. The latter plays an analogous role as the chiral ring coefficient $\lambda$ in the crossing equation of four-point functions. In particular, the authors find that there is always a non-BPS primary with dimension less than 2 for all values of the total Hodge number.  On the other hand, our bound (see Figure \ref{Fig:c=9}) constrains the R-charge neutral non-BPS operator in the specific OPE  between a pair of BPS primaries and depends on the conformal moduli through the chiral ring coefficient $\lambda$.  If we do not keep track of the moduli dependence by setting $\lambda=0$ and assume $\Delta^{CC}_{gap}=0$, our bound $\Delta^{CA}_{gap}\sim 2.272$ (at derivative order 24)\footnote{When extrapolated to infinite derivative order, the upper bound on $\Delta^{CA}_{gap}$ is roughly 2.26.} appears to be a weaker bound than that of \cite{Keller:2012mr} as far as the entire spectrum is concerned.

Obvious generalizations of this work include studying the crossing equations for mixed correlators, especially ones that involve simultaneously $(c,c)$ and $(c,a)$ ring operators. For Calabi-Yau models, this is particularly important in that we wish to pin down the point on both the complex and K\"ahler moduli spaces of the theory, and to constrain the spectrum thereof. Further, one would like to extend our analysis to non-BPS 4-point functions, which would require an efficient way to compute the general non-BPS ${\cal N}=2$ superconformal blocks, that is not yet available. Eventually, we wish to combine the crossing equation for the sphere 4-point correlators with the modular crossing equation for the torus partition function and 1-point functions \cite{Hellerman:2009bu, Keller:2012mr, Friedan:2013cba, Qualls:2013eha,Collier:2016cls}. Another potentially fruitful route is to study the crossing equation for disc correlators, subject to boundary conditions that respect spectral flow symmetry (spacetime-BPS D-brane boundary states in the context of string compactification). We are hopeful that much more is to be learned along these lines toward classifying and solving $(2,2)$ superconformal theories.

\section*{Acknowledgments}

We would like to thank Kazuo Hosomichi, Zohar Komargodski, Juan Maldacena, David Poland for discussions, and David Simmons-Duffin for correcting a reference. 
We are grateful to the Tata Institute of Fundamental Research, and the organizers of the workshops {\it Higher Spin Theory and Duality}, MIAPP, Munich, Germany, {\it Conformal Field Theories and Renormalization Group Flows in Dimensions $d>2$}, Galileo Galilei Institute for Theoretical Physics, Florence, Italy, {\it NCTS Summer Workshop on Strings and Quantum Field Theory}, National Tsing Hua University, Hsinchu, Taiwan, and {\it Strings 2016}, YMSC, Tsinghua University, Beijing, China, for their hospitality during the course of this work. This work is supported by a Simons Investigator Award from the Simons Foundation, and in part by DOE grant DE-FG02-91ER40654. YL is supported by the Sherman Fairchild Foundation and the U.S. Department of Energy, Office of Science, Office of High Energy Physics, under
Award Number DE-SC0011632. YL would also like to thank the hospitality the Berkeley Center for Theoretical Physics during the course of this work. SHS is supported by the National Science Foundation grant PHY-1606531. YW is supported by the NSF grant PHY-1620059 and by the Simons Foundation Grant \#488653. The numerical computations in this work are performed using the SDPB package \cite{Simmons-Duffin:2015qma} on the Odyssey cluster supported by the FAS Division of Science, Research Computing Group at Harvard University.

\appendix

\section{$T^2/\mathbb{Z}_3$ Free Orbifold CFT}
\label{Sec:T2Z3}

In this Appendix we will demonstrate that the four-point function of chiral and antichiral primaries in the $\mathcal{N}=(2,2)$ $T^2/\mathbb{Z}_3$ orbifold theory saturates the numerical bootstrap bound for $c=3$ and external charge $q=\frac13$, along a certain loci on the conformal moduli space.

We start by reviewing some basic facts about the torus orbifold CFT. Consider a torus with both sides $2\pi R$ and angle $2\pi /N$.  We denote the target space fields parametrizing the torus by  $X(z,\bar z) $ and $\bar X(z,\bar z)$ with periodicity $X \sim X +2\pi  R  \sim X +2\pi R \omega$.  Here $\omega=  \exp( 2\pi i/N)$.   We will consider the $T^2/\mathbb{Z}_N$ orbifold CFT in which  $X(z,\bar z)$ obeys,\footnote{Recall that to be compatible with the identification on $T^2$, $N$ will be restricted to $2,3,4,6$.}
\begin{align}
\begin{split}
&X(e^{2\pi i }z, e^{-2\pi i } \bar z)  =\omega X(z,  \bar z)   \,,\\
&\bar X(e^{2\pi i }z, e^{-2\pi i } \bar z)  =\bar\omega \bar X(z,  \bar z)\,,
\end{split}
\end{align}
in the twisted sector. 
For each fixed point, there are $N-1$ twist fields  with weights
\begin{align}
 h = \bar h  = {1\over2} { k\over N} \left( 1- \frac k  N  \right)\,,~~~~~~~~k=1,\cdots, N-1\,.
 \end{align}
We will denote  the twist field with $k=1$ and $k=N-1$ by $\sigma_+(z,\bar z)$ and $\sigma_-(z,\bar z)$, respectively.

In the $\mathcal{N}=(2,2)$ superconformal  $T^2/\mathbb{Z}_N$ orbifold CFT, we have in addition two holomorphic fermions $\psi^+(z),\psi^-(z)$ with R-charge $q=\pm1$ and $\bar q=0$, as well as  their antiholomorphic counterparts $\tilde \psi^+(\bar z), \tilde \psi^-(\bar z)$ with R-charge $q=0$ and $\bar q= \pm1$.  Let $H(z)$ and $\tilde H(\bar z)$ be the bosonization of the holomorphic and antiholomorphic fermions,
\begin{align}
J(z) =  \psi^+ \psi^-  (z)=  i \partial H(z)\,,~~~~~ \tilde J(\bar z) =  \tilde \psi^+\tilde {\psi}^-(\bar z) =  i \partial \tilde H(\bar z)\,,
\end{align} 
where $J(z)$ and $\tilde J(\bar z)$ are the holomorphic and antiholomorphic $U(1)_R$ currents, respectively.  
The $\mathbb{Z}_N$ spin field $s_\pm  (z)= \exp (\pm ikH(z)/N)$ has weight
\begin{align}
h= {1\over2} \left( k\over N\right)^2\,.
\end{align}

From now on we will concentrate on the $\mathcal{N}=(2,2)$ $T^2/\mathbb{Z}_3$ orbifold CFT. We will consider the OPE and the four-point function of the $q=\bar q=\frac13$ $(c,c)$ primary $\phi_{\frac13}(z,\bar z)$ and its $(a,a)$ conjugate primary $\bar \phi_{-\frac13}(z,\bar z)$,
\begin{align}
\begin{split}
&\phi_{\frac13} (z,\bar z) =  e^{ \frac i3 H(z)  +  \frac i3 \tilde H(\bar z) } \sigma_+(z,\bar z)\,,\\
&\bar \phi_{-\frac13}(z,\bar z) =  e^{ - \frac i3 H(z)  -  \frac i3 \tilde H(\bar z) } \sigma_-(z,\bar z)\,.
\end{split}
\end{align}
Note that the weights of $\phi_{\frac13}(z,\bar z)$ and $\bar \phi_{-\frac13}(z,\bar z) $ are both $h = \bar h = \frac 12 (\frac 13)^2 + \frac 12 \frac 13 \frac 23= \frac16$.

The K\"ahler moduli space $M$ of the $\mathcal{N}=(2,2)$ $T^2/\mathbb{Z}_3$ orbifold CFT is parametrized by two real moduli, the radius $R$ and the $B$-field $b$. Note that there is no  complex structure moduli because the shape of the torus is fixed.  We will normalize the $B$-field to have periodicity 1, i.e., $b\sim b+1$.

For arbitrary moduli, the four-point function of $\phi_{\frac13}$ and $\overline\phi_{-\frac13}$ has been computed in \cite{Dixon:1986qv,Chun:1989se}
\begin{align}\label{4ptT2Z3}
& \langle \phi_{\frac13} (z,\bar z) \overline\phi_{-\frac13} (0,0)\bar \phi_{-\frac13}(1,1)\phi_{\frac13} (\infty,\infty)\rangle
 &=  {| z(1-z)  |^{-2/3 } \over |F(z)|^2} \sum_{p\in \Lambda^* \,, v\in \Lambda_c} w^{{1\over2} (p+v/2+B\cdot v/2)^2}\bar w^{{1\over 2} (p-v/2+B\cdot v/2)^2}\,,
\end{align}
where 
\begin{align}
w(z) =  \exp\left[\,-  { 2 \pi \over \sqrt{3}}  {F(1-z)\over F(z) } \right]\,,~~~~
\bar w(z) =  \exp\left[\,-  { 2 \pi \over \sqrt{3}}  {\bar F(1- \bar z)\over \bar F(\bar z) } \right]\,,
\end{align}
with
\begin{align}
F(z) = \,_2F_1({1\over 3}  ,\frac23 ;1;z)\,.
\end{align}
Here $\Lambda$ is the lattice for the original target space torus.\footnote{The lattice $\Lambda$ is normalized such that there is no factor of $2\pi$. For example,  $\Lambda$ for a square torus with sides $2\pi R$ is normalized to be $\Lambda = \{ ( nR, m R) \,|\, n,m\in \mathbb{Z}\}$.} $\Lambda^*$ is the dual lattice of $\Lambda$. $\Lambda_c $ is a sublattice of $\Lambda$ defined as $\Lambda_c = (1-\theta) \Lambda \equiv \{ (1-\theta)u, u\in \Lambda\}$, where $\theta$ is the rotation by $2\pi /3$.  There is a selection rule in the OPE between $\phi_{\frac13}$ and $\overline\phi_{-\frac13}$ that restricts the winding number $v$ to live in $\Lambda_c$ but not the full $\Lambda$.

In the following we will study the chiral-chiral channel $ \phi_{\frac13} (z,\bar z) \phi_{\frac13} (\infty,\infty) $ and the  chiral-antichiral channel $ \phi_{\frac13} (z,\bar z)\bar \phi_{-\frac13} (0,0) $
of the four-point function \eqref{4ptT2Z3}.  We will see that this free orbifold theory saturates the numerical bootstrap bound on the gap in the chiral-antichiral channel along certain loci on the moduli space $M$.

\subsection{Chiral-Chiral Channel}\label{sec:CCT2Z3}

Let us first  consider the chiral-chiral OPE channel between $\phi_{\frac13}$ and $\phi_{\frac13}$.  There are two types of $\frac12$-BPS primaries, two types of $\frac14$-BPS primaries, and one type of non-BPS primary allowed by the $\mathcal{N}=(2,2)$ selection rule (see Section \ref{Sec:Block}) to appear in the chiral-chiral channel:\footnote{As noted in Section \ref{sec:CCgap}, all the non-BPS primaries in the chiral-chiral channel are in fact degenerate in the $c=3$ theory if the external R-charge $q\neq1/2$. The non-BPS degenerate primaries in the case of $c=3$ and external $q=1/3$ are labeled by a half-integer $r\neq \pm 1/2$ with weight given by $h=  r/3$.}
\begin{enumerate}
\item
The lowest dimensional operator in this channel is a $(c,c)$ primary with $q=\bar q=\frac23$ and $h=\bar h=\frac13$,
\begin{align}
\phi_{\frac23}(z,\bar z) \equiv e^{ i\frac 23 H(z)  +i  \frac 23 \tilde H(\bar z) }  \sigma_{-}(z,\bar z)\,.
\end{align}
The chiral ring coefficient, i.e. the OPE coefficient for $\phi_{\frac23}$ in the chiral-chiral channel,  has been computed in\footnote{Throughout this paper we adopt the $\alpha'=2$ convention.
} \cite{Dixon:1986qv,Chun:1989se}
\begin{align}\label{T2Z3chiralring}
\lambda(R,b)   =\left|   \sqrt{3\over2} {\Gamma\left( \frac23 \right)^2\over \Gamma\left( \frac13\right)} R \sum_{v^1,v^2\in \mathbb{Z}} \exp\left[   - {\sqrt{3} \pi R^2 \over 2 } \left(1-i {4 \over \sqrt{3}R^2} b\right) |(v^1 + \omega v^2)|^2\right]\right|\,.
\end{align}

\item  The $G^+_{-\frac12}\tilde G^+_{-\frac12}$ descendant of an $(a,a)$ primary\footnote{We add a prime to distinguish this internal $(a,a)$ primary from the external $(a,a)$ primary $\phi_{-\frac13}$ which has the same charges and weights.}  $\overline\phi'_{-\frac13}(z,\bar z)$ with $q=\bar q=-\frac13$ and $h=\bar h = \frac16$.

 \item The  $G^+_{-\frac12}\tilde G^+_{-\frac12}$ descendant of a $\frac14$-BPS primary that is \textit{antichiral} on the left and non-BPS on the right, with $q=\bar q=-\frac13$ and $h=\frac16$, $\bar h > \frac16$ as well as their conjugates.

Notice that the $(a,a)$ primary and $\frac14$-BPS primary are forbidden by the $\mathcal{N}=(2,2)$ selection rule in the case when the external $(c,c)$ primary has $q=\bar q>\frac12$ (e.g.   the Calabi-Yau CFT).  They are the BPS limits of the non-BPS operators discussed below.

\item The level $\tilde G^+_{-\frac12}$ descendant of a different type of $\frac14$-BPS primary that is \textit{chiral} on the left and non-BPS on the right, with $q=\frac23, \bar q=-\frac13$ and $h=\frac13$, $\bar h > \frac16$, as well as their conjugates.  

  \item The level $G^+_{-\frac12}\tilde G^+_{-\frac12}$ descendant of a non-BPS operator  on the left and right with charge $q=\bar q= -\frac13$ and $h,\bar h> \frac 16$.  
  \end{enumerate}
  
  There is another constraint on the weights of the $\frac14$-BPS primaries and the non-BPS primaries.  In the OPE  between two identical scalars $\phi_{\frac13}(z,\bar z)$, only even spin operators can appear.  This further constrains the (antichiral, non-BPS) $\frac14$-BPS primary to have $\bar h \ge\frac{13}{6}$, and the (chiral, non-BPS) $\frac14$-BPS primary to have $\bar h \ge\frac{11}{6}$.  Similar constraints apply to their conjugates.  In particular, this constraint on the spin forbids the $\tilde G^+_{-\frac12}$ descendant of a (chiral, antichiral) primary with $q=\frac23$ and $\bar q=-\frac13$, as well as its conjugate, to appear in the chiral-chiral channel.   We summarize the quantum numbers of the allowed internal multiplets in the chiral-chiral channel in Table \ref{table:CCT2Z3}.

\begin{table}[h]\label{table:CCT2Z3}
\centering
\begin{tabular}{|c|c|c|c|}
  \hline \text{Primary} & $\substack{\text{Quantum Numbers}\\\text{of the Primary}}$&~$\substack{\text{Level of the Operators} \\\text{that Appear}}$~ & $\Delta^{CC}_{gap} $\\
  \hline $(c,c)$ &  $q=\bar q = \frac23,~h=\bar h = \frac13$ &  $(0,0)$  & 0  \\
  \hline $(a,a)$ &$q=\bar q =- \frac13,~h=\bar h = \frac16$   & $(\frac12 ,\frac12)$   & $\frac23 $  \\
  \hline $(a,n)$   &$q=\bar q =- \frac13,~h= \frac16,~\bar h\ge \frac{13}{6}$ & $(\frac12 ,\frac12) $   & $\bar h+ \frac12\ge \frac83$   \\
  \hline $(c,n)$  &$q=\frac23, ~\bar q= -\frac13,~h= \frac13,~\bar h\ge\frac{11}{6}  $   & $(0 ,\frac12) $  &$\bar h+ \frac16 \ge 2 $  \\
    \hline   $(n,n )$ &$q=\bar q= -\frac13 ,~ h,\bar h>\frac16$&$ (\frac12,\frac12) $& $h+\bar h +\frac13 \ge \frac23 $  \\
  \hline 
\end{tabular}
\caption{The allowed $\mathcal{N}=(2,2)$ multiplets that can appear in the chiral-chiral channel of the four-point function $\langle \phi_{\frac13} \overline\phi_{-\frac13} \bar \phi_{-\frac13} \phi_{\frac13}\rangle$ in the $c=3$ case.  Here $c,a,n$ stands for chiral, antichiral, and non-BPS, respectively.  Note that the level $(\frac12 ,\frac12)$ descendant of an $(a,a)$ primary minimizes the $\Delta^{CC}_{gap}$, which is defined as the gap in the scaling dimensions of the operator that appears in the OPE and of the $(c,c)$ primary.  We omitted the conjugates (i.e. $(n,a)$ and $(n,c)$) of the $\frac14$-BPS primaries in the above table.}
\label{table:CCT2Z3}
\end{table}

\subsubsection*{The Gap in the Chiral-Chiral Channel}

Recall that  $\Delta^{CC}_{gap}$  is defined as the gap between the dimension of the lightest operator  that does not belong to a $(c,c)$ multiplet, and that of a charge $2q$ $(c,c)$ primary $\phi_{2q}$.  In the current case, $\Delta^{CC}_{gap}$ is  the scaling dimension of this  lightest operator  subtracted by  $\frac23$, the scaling dimension of the lowest dimensional operator $\phi_{\frac23}$.  Note that this  lightest operator is always a superconformal descendant while its  primary does not show up in the chiral-chiral channel due to R-charge conservation (see Table~\ref{table:CCT2Z3}).

 We summarize the $\Delta^{CC}_{gap}$ for various internal channels in Table \ref{table:CCT2Z3}. 
In particular, the level $(\frac12 , \frac12)$ descendant  
\begin{align}\label{aadescendant}
G_{-\frac12}^+  \tilde G_{-\frac12}^+ \overline\phi'_{-\frac13}(z,\bar z)\,
\end{align}
of an $(a,a)$ primary $\overline\phi'_{-\frac13}(z,\bar z)$ with $q=\bar q=-\frac13$ minimizes the gap  in the chiral-chiral channel with
\begin{align}
\Delta^{CC}_{gap}= \frac 2 3\,.
\end{align} 
Note that the $(a,a)$ primary $\overline\phi'_{-\frac13}$  itself does not appear in the chiral-chiral OPE.  
  We will assume this minimal gap in the crossing equation when we do the numerical bootstrap.

Over a generic point on the moduli space $M$, the OPE coefficient of this $(a,a)$ primary $\overline\phi'_{-\frac13}(z,\bar z)$ in the chiral-chiral channel is non-vanishing and hence $\Delta^{CC}_{gap}=\frac23$.  However, over special points this OPE coefficient might vanish and $\Delta^{CC}_{gap}$  would be bigger than $\frac23$.  The OPE coefficient $C(R,b)$ of this $(a,a)$ primary $\overline\phi'_{-\frac13}$ can be extracted from the subleading terms in the large $z$ expansion of the exact expression of the four-point function \eqref{4ptT2Z3} 
We can read off the low-lying multiplets in the chiral-chiral channel from the above expansion,
\begin{align}\label{expansion}
&(c,c),~~(a,a),~~\, (n,n)~h=\bar h= \frac12,~~(n,n)~h=\bar h= \frac56,\notag\\
&(c,n)~\bar h ={11\over6},~~
(a,n) ~ \bar h = {13\over 6},~~~\,~~~(n,n)~h=\bar h= \frac76, \cdots.
\end{align}
Here $h$ and $\bar h$ denote the weights of the primaries, not  the actual operators that appear in the OPE.  $c,a,n$ stand for chiral, antichiral, and non-BPS primaries, respectively. For the $\frac14$-BPS primaries, their conjugates are also implicitly included. We see that all possible types of multiplets in Table \ref{table:CCT2Z3} allowed by the $\mathcal{N}=(2,2)$ representation theory are present in the four-point function in the $T^2/\mathbb{Z}_3$ CFT.   Furthermore, as mentioned in Section \ref{sec:CCgap}, all the non-BPS multiplets in the $c=3$ and $q=1/3$ case are degenerate $g_r(h,-1/3)=0$, with the weight of the primary given by $h=r/3$.  Indeed, all the weights of the non-BPS primaries in \eqref{expansion} are of the above form.

As another consistency check, note that the lowest $\frac14$-BPS operators that appear  in the chiral-chiral channel of the four-point function  have quantum numbers  $(h,\bar h)=(1/3,11/6),~  (q,\bar q)=(2/3,-1/3)$ and $(h,\bar h)=(1/6,13/6),~  (q,\bar q)=(-1/3, -1/3 )$ respectively. The latter is also related by a diagonal spectral flow to a $\frac14$-BPS operator of $(c,n)$ type with quantum numbers $(h,\bar h)=(1/3,7/3),~  (q,\bar q)=(2/3, 2/3 )$. We will see that both of them are captured by the elliptic genus in Appendix \ref{App:EG}.

While the gap in the chiral-chiral channel is $2/3$ generically, $\Delta^{CC}_{gap}$ can jump to higher value at some special points over the moduli space where the OPE coefficient  $C(R,b)$ for the $(a,a)$ primary $\overline\phi_{-\frac13}'$ vanishes. From the next to leader term in the expansion of the four-point function \eqref{4ptT2Z3}, we obtain the analytic expression for  $C(R,b)$
\ie
C(R,b)={2^3  3^{\frac54} \pi^{11\over 4}\over \Gamma\left({1\over 6}\right)^{9/2}}R \sum_{v^1,v^2\in \mathbb{Z}} \exp\left[   - {\sqrt{3} \pi R^2 \over 2 }  \left(1-i {4\over \sqrt{3}R^2} b\right)   |(v^1 + \omega v^2)|^2\right]
(1-\sqrt{3}\pi R^2  |(v^1 + \omega v^2)|^2)
\fe
which is proportional to $\partial_R \lambda(R,b)$.  The OPE coefficient $C(R,b)$ has zeroes at 
\begin{align}\label{zeroC}
\begin{split}
&R={2\over \sqrt{3} }\,,~~~~~b=0\,,\\
&R={1\over \sqrt{3}}\,,~~~~~b=\pm{1\over 4}\,.
\end{split}
\end{align}
Furthermore, all these points have the same chiral ring coefficient $\lambda(R,b) = 2^{1\over 6}$ and $\Delta^{CA}_{gap}=1.$\footnote{The analytic expression for the gap in the chiral-antichiral channel is given in \eqref{CAgapT2Z3}.}   
At these points, the gap in the chiral-chiral channel is saturated by the $G^+_{-\frac12}\tilde G^+_{-\frac12}$ descendant of a  non-BPS primary.  The weight of this descendant is $h=\bar h = 1$, which gives a gap $\Delta^{CC}_{gap} = \frac43$.    In Section \ref{sec:saturation}, we saw   how the above jump in the chiral-chiral gap can be seen from the numerical bootstrap bound.

\subsection{Chiral-Antichiral Channel}

In the chiral-antichiral OPE channel between $\phi_{\frac13}$ and $\overline\phi_{-\frac13}$, the internal primaries are the exponential operator $\mathcal{O}_{p,v}(z,\bar z)$ in the untwisted sector,
\begin{align}\label{exp}
\mathcal{O}_{p,v}(z,\bar z)=\mathcal{N}\sum_{\alpha \in \mathbb{Z}_3 }  \exp\left[\, i( p+v/2) \cdot \alpha X_L (z)+ i (p-v/2)\cdot \alpha X_R(\bar z) \,\right] \,,
\end{align}
where the sum in $\alpha$ over the $\mathbb{Z}_3$ images is to project to the $\mathbb{Z}_3$ invariant combinations. The constant $\mathcal{N}$ is chosen such that the two-point function of $\mathcal{O}_{p,v}$ is one.  The exponential operator is labeled by the momentum $p\in \Lambda^*$ and the winding $v$, with the weight given by $h={1\over 2} (p+v/2)^2$ and $\bar h = \frac12(p-v/2)^2$ in the absence of $B$-field.  A priori, the winding $v$  can be any lattice point in $\Lambda$.  However, a selection rule \cite{Dixon:1986qv} in the chiral-antichiral channel allows only those $v\in \Lambda_c$ to appear in the OPE between $\phi_{\frac13}$ and $\overline\phi_{-\frac13}$.

We can parametrize the weights of these exponential operators more explicitly.  
Let us write the metric $ds^2 = G_{\mu\nu}dx^\mu dx^\nu$ $(\mu,\nu=1,2)$ of the target space torus  as
\begin{align}
\begin{split}
&ds^2 =  (dx^1 + \omega dx^2 ) (dx^1 +\omega^2 dx^2 )\,,\\
&G_{\mu\nu} = \left(\begin{array}{cc}~1 ~& -\frac12 \\-\frac12 & ~1~\end{array}\right)\,,~~~~~~~
G^{\mu\nu} = \left(\begin{array}{cc}~\frac43~ & \frac23~ \\ ~\frac23~ & \frac43~\end{array}\right)\,.
\end{split}
\end{align}
Here $x^\mu \sim x^\mu +2\pi R$. 
The $B$-field background is
\begin{align}
B_{\mu\nu} =b{2\over R^2}  \left(\begin{array}{cc}0 & 1 \\-1 & 0\end{array}\right)\,,
\end{align}
with $b$ normalized to have periodicity $b\sim b+1$.  The weight of the exponential operator $\mathcal{O}_{p,v}(z,\bar z)$ with momentum $p\in \Lambda$ and winding $v\in \Lambda$ is
\begin{align}
\begin{split}\label{CAgapT2Z3}
&h   = \frac12 G^{\mu\nu} \left[  { p_\mu \over R}+ \frac12 (  G_{\mu\rho} +B_{\mu\rho}) v^\rho R\right]\left[ { p_\nu \over R}+  \frac12  (  G_{\nu\sigma} +B_{\nu\sigma}) v^\sigma R\right]\,,\\
&\bar h  = \frac12 G^{\mu\nu} \left[ { p_\mu \over R}-  \frac12  (  G_{\mu\rho} -B_{\mu\rho}) v^\rho R\right]\left[ { p_\nu \over R}-  \frac12 (  G_{\nu\sigma} -B_{\nu\sigma}) v^\sigma R\right]\,,
\end{split}
\end{align} 
 with $p_\mu, v^\mu \in \mathbb{Z}$.  
The selection rule in the chiral-antichiral channel that $v\in \Lambda_c \subset \Lambda$ is translated into the requirement that
\begin{align}\label{selectionT2Z3}
v^1+v^2 \in 3\mathbb{Z}\,.
\end{align}
Given a point on the moduli space $M$  of the $T^2/\mathbb{Z}_3$ orbifold CFT, the minimum of \eqref{CAgapT2Z3} with the restriction \eqref{selectionT2Z3} is our final formula  for the lowest dimension  $\Delta^{CA}_{gap}$  of non-BPS primaries in the chiral-antichiral channel of the four-point function $\langle \phi_{\frac13} \bar \phi_{-\frac13} \bar \phi_{-\frac13} \phi_{\frac13}\rangle$. Importantly, $\Delta^{CA}_{gap}$ is \textit{not} a continuous function of the moduli $(R,b)$. This is because the  the momentum $p_\mu$ and winding number $v^\mu$ that minimize the dimension $h+\bar h$ in \eqref{CAgapT2Z3}  might jump  as we vary the moduli.

\section{Elliptic Genus of the $T^2/\bZ_3$ Orbifold CFT}\label{App:EG}

 At a special point on the moduli space of the $T^2/\bZ_3$ CFT, the theory factorizes into the tensor product of three $\cN=(2,2)$ $A_2$ minimal models.  The elliptic genus of the $T^2/\bZ_3$ orbifold CFT can be computed at this point to be \cite{Witten:1993jg}
\ie
Z_{T^2/\bZ_3}= (Z_{A_2})^3=\left( {\theta_1(Q,y^{2/3})\over \theta_1(Q,y^{1/3})}\right)^3\,.
\label{T2Z3EG}
\fe 
Here we define $Q=e^{2\pi i \tau}$. 

The NS sector elliptic genus is related by (diagonal) spectral flow,
\ie
Z^{NS}_{T^2/\bZ_3}(Q,y)=Z_{T^2/\bZ_3}(Q, y Q^{1/2})y^{1/2}Q^{1/4}\,.
\label{T2Z3EGNS}
\fe
To see if there are $\frac14$-BPS operators (BPS on the (anti)holomorphic side only) at a generic point on the moduli space, we will expand \eqref{T2Z3EGNS} in terms of  (twisted) $\cN=2$ characters. Recall from \eqref{nonBPS2} that all the $U(1)_R$ charged $c=3$ $\cN=2$ NS representations are {\it degenerate} satisfying $g_r(h,q)=0$. They are labeled by a half-integer $r$ and the $U(1)_R$ charge $q$. For $q> 0$ and $r>0$, the characters are given by \cite{Dobrev:1986hq,Kiritsis:1986rv,Matsuo:1986cj}
\ie
&ch_{r,q=1}(Q,y)=  {(1-Q)Q^{r}y \over (1+Q^{r}y)(1+Q^{r+1}y )}F_{NS}(Q,y) \,,
\\
&ch_{r,0<q<1}(Q,y)= {Q^{qr} y^q \over (1+Q^{r}y)} F_{NS}(Q,y)\,,
\fe
and the identity character
\ie
&ch_{0}(Q,y)=  {(1-Q) \over (1+Q^{1/2}y)(1+Q^{1/2}y^{-1})} F_{NS}(Q,y) \,,
\fe
where 
\ie
F_{NS}(Q,y)=\prod_{k=1}^\infty {(1+Q^{k-1/2}y)(1+Q^{k-1/2}y^{-1})\over (1-Q^k)^2}\,,
\fe
and similarly for characters with $q<0$ and $r<0$ (the representations are charge conjugate to those with $q>0$).

The twisted NS characters are defined by 
\ie
\widetilde{ch}_{r,q}(Q,y)\equiv & (-1)^{-r} ch_{r,q}(Q,y^{-1})\,.
\fe
We thus have the decomposition
\ie
&Z^{NS}_{T^2/\bZ_3}(Q,y)=\widetilde{ch}_{0}(Q,y)+3\widetilde{ch}_{{1\over 2},{1\over 3}}(Q,y)+3\widetilde{ch}_{{1\over 2},{2\over 3}}(Q,y)+ \widetilde{ch}_{{1\over 2},1}(Q,y)
\\
& 
-3  \widetilde{ch}_{-{5\over 2},-{1\over 3}}(Q,y)
+3  \widetilde{ch}_{{7\over 2},{1\over 3}}(Q,y)
+ \widetilde{ch}_{-{3\over 2},-1}(Q,y)
-3  \widetilde{ch}_{-{5\over 2},-{2\over 3}}(Q,y) 
-3  \widetilde{ch}_{-{11\over 2},-{1\over 3}}(Q,y)
\\
&
+3 \widetilde{ch}_{{13\over 2},{1\over 3}}(Q,y) 
+3 \widetilde{ch}_{{7\over 2},{2\over 3}}(Q,y) 
+\dots \,,
\label{T2Z3EGdc}
\fe
where the first line comes from the BPS operators ($(c,c)$ ring elements), while the second and third lines are associated to ${1\over 4}$-BPS operators (in the non-BPS degenerate representations on one side). As we have seen in the previous subsection, some of the latter operators appear in the  chiral-chiral channel  of the four point function $\langle \phi_{\frac13} \overline\phi_{-\frac13} \overline\phi_{-\frac13} \phi_{\frac13}\rangle$.

\section{Free Fermion OPEs at the Kinks}
\label{freekink}

In this Appendix, we show that the bound \eqref{c=9q=1CC=2} for OPEs of marginal BPS operators in $c=9$ SCFTs at $\Delta^{CC}_{gap}=2$ is exactly saturated by free fermion correlators (say at the infinite volume point of the quintic).   In the $\mathbb{C}^3$, $T^6$, or their orbifold CFTs, we have three holomorphic fermions $\psi^\mu(z)$ with $q=+1$ and three holomorphic fermions $\psi^{\bar \mu}(z)$ with $q=-1$. Here $\mu,\bar \mu=1,2,3$. Similarly, we have three antiholomorphic fermions $\tilde\psi^\mu(\bar z)$ with $\bar q= -1$  and three antiholomorphic fermions $\tilde\psi^{\bar \mu}(\bar z)$ with $\bar q=+1$.  They satisfy the OPE 
\ie
\psi^\mu(z)  \psi^\nu(0) \sim0\,,~~~~~\psi^{\bar\mu}(z)  \psi^{\bar\nu}(0) \sim0\,,~~~~~~\psi^\mu(z)  \psi^{\bar\nu}(0) \sim{\delta^{\mu\bar \nu} \over z}\,,
\fe
and similarly for $\tilde\psi^\mu(\bar z)$ and $\tilde\psi^{\bar \mu}(\bar z)$.  
Let us consider the $(c,c)$ primary with $q=\bar q=1$,
\ie\label{charge1}
\phi_1 (z,\bar z)  \equiv {1\over \sqrt{\text{Tr} (MM^\dagger)}}\,M_{\mu\bar\nu}\,  \psi^\mu(z)\tilde\psi^{\bar\nu}(\bar z)\,,
\fe
 where $M_{\mu\bar\nu}$ is some general complex 3$\times$3 matrix.  The operator $\phi_1(z,\bar z)$ normalized to have unit two-point function, $\langle \phi_1(z,\bar z) \overline\phi_{-1}(0,0) \rangle= 1/|z|^4$.  
 
We first note that $\Delta^{CC}_{gap}$ is  2, which is our assumption in the bootstrap bound in Figure~\ref{Fig:c=9-width}, as is realized by operators of the form
\ie
M_{\mu\bar\nu} M_{\rho\bar \sigma} \partial^2\psi^\mu(z)\psi^\rho(z)\tilde \psi^{\bar \nu}(\bar z) \tilde \psi^{\bar \sigma}(\bar z)\,.
\fe
   Note that the  dimension 3 operators $M_{\mu\bar\nu} M_{\rho\bar \sigma} \partial\psi^\mu(z) \psi^\rho(z)\tilde \psi^{\bar \nu}(\bar z) \tilde \psi^{\bar \sigma}(\bar z)$ and its complex conjugate are descendant of $\phi_2$ (defined below).

In the chiral-chiral channel $\phi_1\phi_1$, the lightest operator is the $q=\bar q=2$ $(c,c)$ primary,
\ie\label{charge2}
\phi_2 (z,\bar z)  \equiv {1\over \sqrt{2}\sqrt{(\Tr M M^\dagger)^2 - \Tr MM^\dagger M M^\dagger}}
M_{\mu\bar\nu} M_{\rho\bar \sigma} \psi^\mu(z) \psi^\rho(z)\tilde \psi^{\bar \nu}(\bar z) \tilde \psi^{\bar \sigma}(\bar z)\,,
\fe
normalized such that it has unit two-point function. Combining \eqref{charge1} and \eqref{charge2}, we  have computed the chiral ring coefficient $\lambda$ for $\phi_2$ in the $\phi_1\phi_1$ OPE,
\begin{align}\label{c=9lambda}
\lambda = \sqrt{2}  \sqrt{1 - {  \Tr MM^\dagger M M^\dagger\over \Tr(MM^\dagger)^2}  }\,.
\end{align}
By choosing different matrices $M_{\mu\bar \nu}$, we will see that the four-point function of $\phi_1$, $\overline\phi_{-1}$ saturates the bound \eqref{c=9q=1CC=2} with $\Delta^{CC}_{gap}=2$.  

To start with, note that $\lambda\le2/\sqrt{3}$ by the Cauchy-Schwarz inequality, where the equality holds if and only if $MM^\dagger =Id$.  Hence the four-point function of $\phi_1,\overline\phi_{-1}$ realizes the region $\lambda>2/\sqrt{3}$ in the bootstrap bounding curve \eqref{c=9q=1CC=2}.

Next, in the chiral-antichiral channel, the lightest non-identity operator is
\ie
\mathcal{O}(z,\bar z)\equiv(MM^\dagger)_{\mu\bar\nu} \psi^\mu(z) \tilde\psi^{\bar\nu}( z)\,,
\fe
with scaling dimension 1. The operator $\mathcal{O}$ is a superconformal primary unless the matrix $M$ is chosen such that $MM^\dagger=Id$.  This explains the region $\lambda<2/\sqrt{3}$ in the bounding curve \eqref{c=9q=1CC=2}.  Finally, when $MM^\dagger=Id$, we have $\lambda=2/\sqrt{3}$ and the operator $\mathcal{O}$ is the R-current. It follows that the lightest non-BPS primary is replaced by the normal-ordered operator  $:\phi_1\overline\phi_{-1}:$, which has dimension 2. This explains the peak at $\lambda=2/\sqrt{3}$ in the bounding curve \eqref{c=9q=1CC=2}.  

In summary, we see that by choosing different linear combinations of free fermion bilinears, the numerical bootstrap bound is realized for all values of the chiral ring coefficient $\lambda$ in the case of $\Delta^{CC}_{gap}=2$.\footnote{We can alternatively consider the Gepner models of the type $1^9$ (e.g. a tensor product of 9 $\cN=2$ $A_2$ minimal models) which is conveniently described by the LG model with superpotential $W=\sum_{i=1}^9 X_i^3$. A particular $\bZ_3$ orbifold of the tensor product theory describes the mirror of the Z-manifold \cite{Candelas:1993nd}. By taking $\cal O$ to be a linear combination of the chiral primaries $X_1 X_2 X_3, X_4 X_5 X_6, X_7 X_8 X_9$, and using the $\cN=2$ fusion rules \eqref{n2fusion}, we see $\Delta_{gap}^{CC}=3\times {2\over 3}=2$ and the numerical bootstrap bound is again saturated.  
}

\section{The  Quintic Threefold}
\label{Sec:Quintic}

In this section we review some basic facts about the $\mathcal{N}=(2,2)$ nonlinear sigma model on the quintic Calabi-Yau threefold. In particular, we will review the exact formula of the chiral ring coefficient  and discuss various special points on the K\"ahler moduli space.

Let 
 $x_0,x_1,x_2,x_3,x_4$
 be the homogeneous coordinates of $\mathbb{P}^4$.  A quintic threefold $\mathcal{M}$ is   a hypersurface defined by  the vanishing locus of a  quintic polynomial of $x_i$'s in $\mathbb{P}^4$.  The coefficients in the quintic polynomials, modulo linear redefinitions of the coordinates $x_i$, parametrize the complex structure moduli space.  Hence the dimension of the complex structure moduli space for the quintic $\mathcal{M}$ is $h^{2,1}(\cM)=  {9!\over 5!4! }-  25=101$.  On the other hand, there is  one parameter  associated to the choice of the  K\"ahler class, which can be thought of as the size of $\mathbb{P}^4$,  i.e. $h^{1,1}(\cM)=1$.


To construct its mirror, we consider a one-parameter family of the quintic $\mathcal{M}$ that is given by
\ie\label{mirror}
p=x_0^5+ x_1^5+x_2^5+x_3^5+x_4^5  - 5 \psi \,x_0  x_1 x_2 x_3 x_4= 0,
\fe
in $\mathbb{P}^4$.  The mirror quintic $\mathcal{W}$ is obtained by performing the following $\mathbb{Z}_5^3$ orbifold action,
\ie
& (x_0, x_1,x_2,x_3,x_4) \mapsto ( \A^{4}x_0 ,x_1  , \A x_2 ,  x_3 ,  x_4)\,,\\
& (x_0, x_1,x_2,x_3,x_4) \mapsto ( \A^{4}x_0 ,x_1  , x_2 , \A x_3 , x_4)\,,\\
& (x_0, x_1,x_2,x_3,x_4) \mapsto ( \A^{4}x_0 ,x_1  , x_2 ,  x_3 , \A x_4)\,.
\fe
Here $\alpha= e^{ 2\pi i /5}$.  The mirror quintic has one complex structure modulus $\psi$ and 101 K\"ahler moduli, $h^{2,1}(\cW) = 1$ and $h^{1,1}(\cW)=101$.  Note that the true coordinate on the complex structure moduli space of the mirror quintic $\mathcal{W}$ is $\psi^5$, since  the replacement $\psi\mapsto \A \psi$ in \eqref{mirror} can be undone by coordinate redefinitions of $x_i$.  Hence the complex structure moduli space of the mirror quintic can be taken to be the fundamental region $0\le \text{arg}\, \psi <2\pi/5$ on the $\psi$-plane.

\subsection{Chiral Ring Coefficient of the K\"ahler Moduli Space}

The chiral ring coefficient and the metric on the K\"ahler moduli space of the quintic was obtained in the seminal work of \cite{Candelas:1990rm} using mirror symmetry.  In this subsection we review the results.

The K\"ahler potential  $K$ on the complex structure moduli space of the mirror quintic $\mathcal{W}$, or equivalently, on the K\"ahler moduli space of the quintic $\mathcal{M}$, is given by \cite{Candelas:1990rm},
\ie
e^{-K} = {(2\pi)^6 i\over 5^7} \left(\begin{array}{cccc}\varpi_2^* & \varpi_1^* & \varpi_0^* & \varpi_4^*\end{array}\right)
\left(\begin{array}{cccc}0 & 1 & 3 & 1 \\-1 & 0 & 3 & 3 \\-3 & -3 & 0 & 1 \\-1 & -3 & -1 & 0\end{array}\right)
\left(\begin{array}{c}\varpi_2 \\\varpi_1 \\\varpi_0 \\\varpi_4\end{array}\right)
\fe
where  the functions $\varpi_j(\psi)$ are defined as
\ie\label{varpi}
&\varpi_0(\psi ) = \sum_{n=0}^\infty {(5n)!\over (n!)^5 (5\psi )^{5n}}\,,\\
&\varpi_j (\psi) = \sum_{r=0}^3 \log^r (5\psi) \sum_{n=0}^\infty b_{jrn} { (5n)!\over(n!)^5 (5\psi)^{5n} },~~~j=1,2,3\,,
\fe
in the region $|\psi|>1$ of the fundamental domain on the the $\psi$-plane, and
\ie\label{varpi2}
&\varpi_j(\psi ) = -{1\over 5} \sum_{m=1}^\infty {\A^{2m} \Gamma(m/5) (5\A^j\psi)^m \over \Gamma(m) \Gamma(1-m/5)^4},~~~|\psi|<1,~~~j=0,\cdots,4\,.
\fe
in the region $|\psi|<1$.  The coefficients $b_{jrn}$ are defined in Appendix B of \cite{Candelas:1990rm}.  The metric on the complex structure moduli space is given by $g_{\psi \bar\psi  } =   {\partial^2 K / \partial \psi \partial\bar\psi}$. 
Going back to the original quintic threefold $\mathcal{M}$, the K\"ahler modulus $t =t_1+it_2$ of $\mathcal{M}$  is related to the complex structure modulus $\psi$ of the mirror quintic $\mathcal{W}$ by the mirror map,
\begin{align}\label{mirrormap}
t=  {2\varpi_1  - 2\varpi_0 +\varpi_2 - \varpi_4 \over 5\varpi_0}\,.
\end{align}

In the large volume limit of the quintic, the exactly marginal $(c,c)$ primary operator with $q=\bar q=1$ corresponding to the K\"ahler modulus $t$ can be written as
\begin{align}
\phi_1(z,\bar z)  \sim J_{\mu \bar\nu} \,\psi^\mu (z)\widetilde {\psi}^{\bar \nu} (\bar z)\,,
\end{align}
where $J_{\mu\bar \nu}$ is the harmonic representative of the K\"ahler class.  The chiral ring coefficient $\lambda$ for this $(c,c)$ primary is given by the following combination
\ie
\lambda (t) =     g_{\psi \bar\psi }^{-\frac32}   \,e^K |\kappa_{\psi\psi\psi}|\,,
\fe
which is invariant under the coordinate and K\"ahler transformations.   The latter is given by rescaling the holomorphic three-form $\Omega$ of the mirror quintic by a holomorphic function of $\psi$.  Here $\psi$ is thought of as a function of the K\"ahler modulus $t$ through the inverse of the mirror map \eqref{mirrormap}. The ``Yukawa coupling" is defined as  $\kappa_{\psi\psi\psi}  =  \int_{\mathcal{W}} \Omega \wedge {\partial^3 \over \partial^3\psi } \Omega$, which equals to
\ie
\kappa_{\psi\psi\psi} =  \left( {2\pi i \over 5}\right)^3  {5\psi^2 \over 1-\psi^5}\,.
\fe

\begin{figure}[h!]
\centering
\includegraphics[width=.6\textwidth]{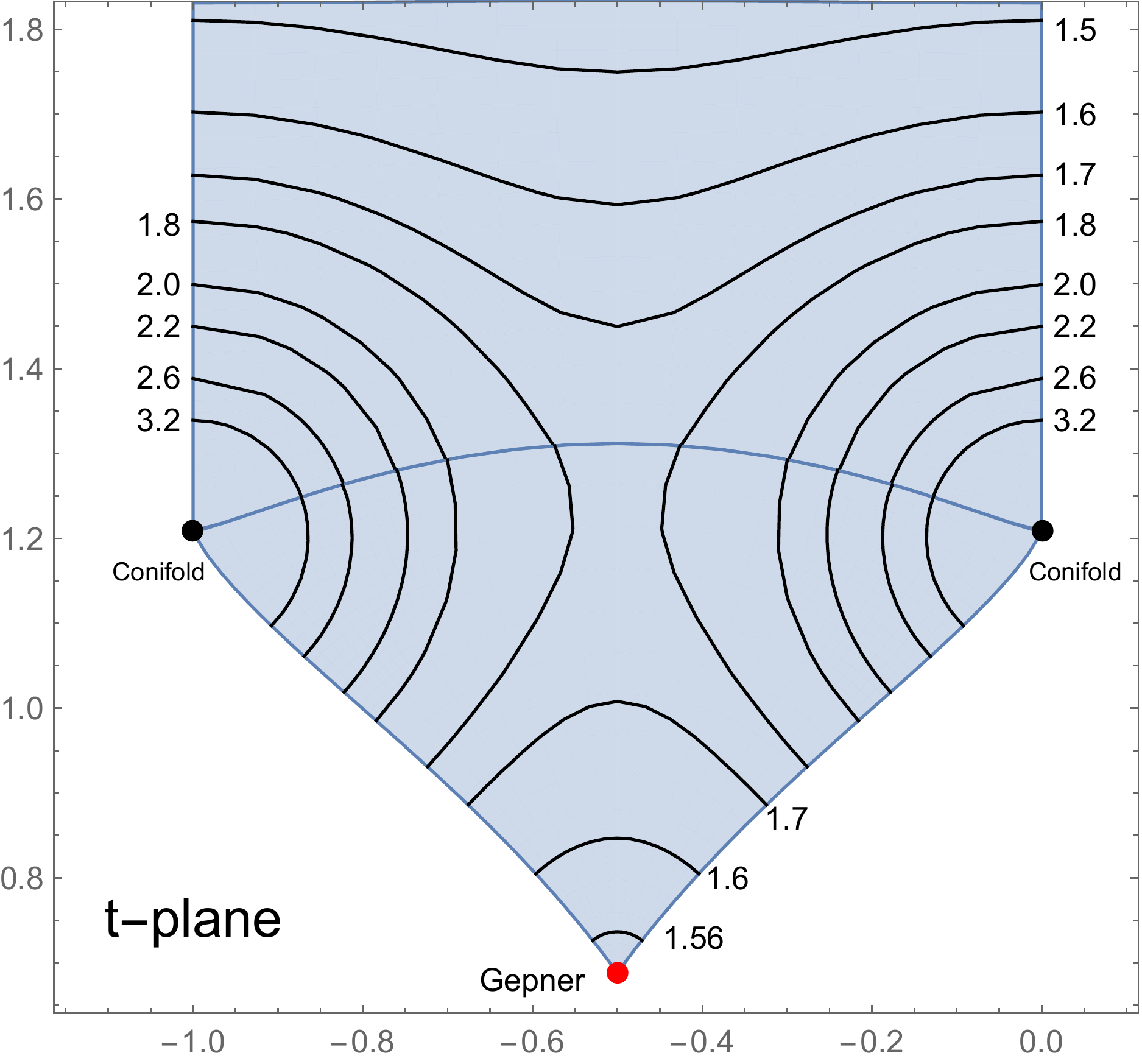}
\caption{The contour plot for the chiral ring coefficient $\lambda$ of the K\"ahler modulus for the quintic $\mathcal{M}$ in the $t$-coordinates.  The blue shaded region is the K\"ahler moduli space of the quintic. The black curves are the constant  $\lambda$ loci, with the values of $\lambda$ indicated at the ends of the curves. The  blue curve in the middle corresponds to $|\psi|=1$, with two ends being the conifold point $\psi=1$.  The Gepner point $\psi=0$ is shown in red.}
\label{Fig:scorpion}
\end{figure}

In Figure \ref{Fig:scorpion} we present the contour plot of the chiral ring coefficient $\lambda(t)$ of the K\"ahler modulus of $\mathcal{M}$ in the $t$-coordinates.  There are a few special points on the K\"ahler moduli space that we will pay special attention to:

\begin{itemize}
\item The large volume point $t= i \infty$ of the quintic $\mathcal{M}$, or equivalently the large complex structure point $\psi=\infty$ of the mirror quintic $\mathcal{W}$. In this limit, the mirror map simplifies to
\ie
t\sim {5i \over 2\pi }\log 5\psi,~~~~~\psi\to \infty~~\text{or}~~t\to i \infty.
\fe
The K\"ahler potential on the moduli space expanded  around the large volume point is 
\ie
e^{-K} =  {20\over 3} t_2^3 + {50\over \pi^3} \zeta(3) +\cdots\,,
\fe
where the $\dots$ stand for the worldsheet instanton corrections that are powers of $e^{-2\pi t_2}$.   The value of the chiral ring coefficient at the large volume point is the global minimum on the whole moduli space:
\begin{align}
\lambda(t=i\infty) = {2\over\sqrt{3}} \sim 1.15\,.
\end{align}
The gaps in the CC and CA channel at the large volume limit have been discussed in Appendix \ref{freekink}.

\item The conifold point $t\sim 1.21i$ or $\psi=1$.\footnote{The other point $t\sim -1+1.21i $ or $\psi=\exp(2\pi i /5)$ in Figure \ref{Fig:scorpion} is identified as the same point as $t\sim 1.21i$ or $\psi=1$ on the moduli space.}  At this point the quintic CFT becomes singular and the chiral ring coefficient diverges $\lambda \to \infty$.  The gaps at the conifold CFT can be determined through the $\mathcal{N}=2$ Liouville theory description to be $\Delta^{CA}_{gap}=1/2$ and $\Delta_{gap}^{CC}=0$.  We will have more to say about this in the subsequent subsection.

\item The point $t=-\frac12 +\frac45 i \sin^3({2\pi\over5})\sim -0.5 + 0.69i$ or $\psi=0$ is where the $3^5$ Gepner point is located  at on the K\"ahler moduli space. The $3^5$ Gepner model, realized at a specific point on the K\"ahler and complex structure moduli space, is exactly solvable and is given by an orbifold of five copies of the $c=9/5$ $\mathcal{N}=(2,2)$ minimal model \cite{Gepner:1987vz}.  The value of the chiral ring coefficient at this point is a local minimum:
\begin{align}
\lambda(t=-\frac12 +\frac45 i \sin^3({2\pi\over5}))=\frac{\Gamma \left(\frac{3}{5}\right)^{15/2} \Gamma \left(\frac{1}{5}\right)^{5/2}}{\Gamma \left(\frac{2}{5}\right)^{15/2} \Gamma \left(\frac{4}{5}\right)^{5/2}}\sim 1.56\,.
\end{align}
The gaps in the CC and the CA channel are $\Delta^{CC}_{gap}=6/5$ and $\Delta^{CA}_{gap}=4/5$.

\end{itemize}

\subsection{The Conifold Point and the $\cN=2$ Liouville Theory}
\label{ap:liouville}
Approaching the conifold point of the quintic moduli space, the $(2,2)$ Calabi-Yau sigma model becomes singular and develops a continuum in the operator spectrum. 
The continuum states in the singular limit are believed to be captured by the $\cN=2$ Liouville theory of central charge $c=9$ \cite{Ooguri:1995wj}. 
 More generally, the $\cN=2$ Liouville theory (T-dual to the $\cN=2$ $SL(2)_k/U(1)$ cigar CFT) is labelled by $k\in \bZ_+$ with central charge given by $c={3(k+2)\over k}$. The relevant $\cN=2$ superconformal primaries in the NS sector are denoted as $V_{j,m,\bar m}$ with quantum numbers
\ie
h=-{j(j+1)\over k}+{m^2\over k},\quad q={2 m\over k}\,,
\fe
and similarly for the anti-holomorphic part. In the spectrum of the $\cN=2$ Liouville theory, there are  continuous representations with $j=-{1\over 2}+i \bR$ which are  non-degenerate, discrete BPS representations with $j=|m|-1,~|m|=1,{3\over 2},\dots, {k\over 2}$, and  discrete non-BPS degenerate representations with $-{1\over 2}<j<{k-1\over 2}$ and $|m|-j=2,3,\dots$. 

The marginal chiral primary we consider in the four point function is $V_{{k\over 2}-1,{k\over 2},{k\over 2}}$. In the CA channel, the gap is saturated by the bottom of the continuum representations with $q=0$ at $j=-{1\over 2}$
\ie
\Delta_{gap}^{CA}=2 h_{gap}^{CA}={1\over 2k}={c-3\over 12}.
\fe
Similarly in the CC channel, the non-BPS primaries (whose $G^+_{-1/2}$ descendant appears) develop a  gap above the unitarity bound $h=1/2$ for $q=1$, 
\ie
\Delta_{gap}^{CC}=2 h_{gap}^{CC}=2\left({1\over 4k}+{k\over 4}-{1\over 2}\right)={(k-1)^2\over 2k}={(c-9)^2\over 12(c-3)}.
\fe
We have written the gaps in terms of $c$ because although the gaps were derived for integral $k$, they are expected to hold for general $k$.\footnote{For discussions about $\cN=2$ Liouville theories with rational $k$, see \cite{Eguchi:2003ik,Eguchi:2004ik} for examples.}

\subsection{Gepner Points of One-Parameter Calabi-Yau Models}
\label{ap:gepner}
A simple class of one-parameter (i.e. with only one complex K\"ahler modulus) Calabi-Yau manifolds  generalizing the quintic are given by hypersurfaces in weighted projective space:
\ie
& \mathbb{WP}^4_{1,1,1,1,1} \quad x_1^5+x_2^5+x_3^5+x_4^5+x_5^5=0  
\\
&\mathbb{WP}^4_{1,1,1,1,2}\quad x_1^6+x_2^6+x_3^6+x_4^6+x_5^3=0 
\\
&\mathbb{WP}^4_{1,1,1,1,4}\quad x_1^8+x_2^8+x_3^8+x_4^8+x_5^2=0
\\
&\mathbb{WP}^4_{1,1,1,2,5}\quad x_1^{10}+x_2^{10}+x_3^{10}+x_4^5+x_5^2=0 .
\fe
The relevant Gepner points are described by the orbifold of tensor products of $\cN=2$ minimal models $SU(2)_k/U(1)$ of the type $\prod_i (k_i-2): 3^5,4^4 1,6^4, 8^3 3$. Let us denote the chiral ring generators by $X_i$. 
Then the marginal chiral primaries are given by $\prod_{i=1}^5 X_i,\prod_{i=1}^5 X_i,\prod_{i=1}^4 X_i^2$ and $\prod_{i=1}^4 X_i^2$ respectively.

The chiral ring coefficients $\lambda$ are determined by the three point function coefficients in $\cN=2$ minimal models, which are given by
\ie
\la  \Phi_{j,j,j}\Phi_{j,j,j}\Phi_{2j,-2j,-2j}\ra 
=
 \sqrt{\frac{\Gamma \left(\frac{1}{k}\right) \Gamma \left(\frac{4 j+1}{k}\right)}{\Gamma \left(\frac{k-1}{k}\right) \Gamma \left(\frac{-4 j+k-1}{k}\right)}} 
 {\Gamma \left(\frac{-2 j+k-1}{k} \right)\over \Gamma \left(\frac{2 j+1}{k}\right)}
\fe
for $4j\leq k-2$, after identifying $X_i^j=\Phi_{j,m=j,\overline m=j}$ \cite{Distler:1988ms}. Therefore we have
\ie
3^5:\quad \lambda=&\left(\Gamma \left( {3\over 5}\right)\over \Gamma \left({2\over 5} \right) \right)^{15\over 2}\left(\Gamma \left( {1\over 5}\right)\over \Gamma \left({4\over 5} \right) \right)^{5\over 2}
\sim 1.56
\\
4^4 1:\quad \lambda=&0
\\
6^4:\quad \lambda=&\frac{\Gamma \left(\frac{1}{8}\right)^2 \Gamma \left(\frac{5}{8}\right)^6}{\Gamma \left(\frac{3}{8}\right)^6 \Gamma \left(\frac{7}{8}\right)^2}\sim 2.35
\\
8^3 3:\quad \lambda=&0.
\fe

We  will denote the superconformal primaries in the $\cN=2$ $SU(2)_k/U(1)$ theory (with $A$-type modular invariants) by  $\Phi_{j,m}$, with quantum numbers
\ie
h={j(j+1)\over k}-{m^2\over k},\quad q={2m\over k}.
\fe
To identify the relevant CA and CC gaps, let us first recall the fusion rules for $\cN=2$ minimal models (we will focus on the holomorphic side for simplicity) \cite{Mussardo:1988av}. \ie
\Phi_{j_1,m_1} \Phi_{j_2,m_2} =  \sum_{j=|j_1-j_2|}^{min(j_1+j_2, (k-2)-j_1-j_2)} [\Phi_{j,m_1+m_2}]
\label{n2fusion}
\fe
where
\ie
{}[\Phi_{j,m_1+m_2}]\equiv \begin{cases}
 \Phi_{j,m_1+m_2}&~~~~\text{if}~~~~|m_1+m_2|\leq j \,,
\\
\Phi_{{k-2\over 2}-j,m_1+m_2 - {k\over 2}}&~~~~\text{if}~~~~m_1+m_2>j \,,
\\
 \Phi_{{k-2\over 2}-j,m_1+m_2+ {k\over 2}}&~~~~\text{if}~~~~m_1+m_2<-j \,.
\end{cases}
\fe
 It is clear from the above fusion rules that the lightest operator in the OPE of a chiral primary $\Phi_{j,j,j}$ with its conjugate is  $\Phi_{1,0,0}$, with the following gap,
 \ie
 \Delta^{CA}_{gap}=\Delta(\Phi_{1,0,0})={4\over k}.
 \fe
Similarly, in the OPE of a pair of $\Phi_{j,j,j}$ , the lightest non-BPS superconformal primary (whose descendant appears) is $\Phi_{{k\over 2}-2j,2j-{k\over 2},2j-{k\over 2}}$, leading to a gap
\ie
 \Delta^{CC}_{gap}=2\left({1\over 2}+h(\Phi_{{k\over 2}-2j,2j-{k\over 2},2j-{k\over 2}})-{2j\over k }\right)
 =2\left({1\over 2}+{k-4j\over 2k}-{2j\over k }\right)
 ={2(k-4j)\over k}.
\fe
We summarize the results in Table~\ref{Tab:Gep}.

\begin{table}[htb]
\begin{center}
\begin{tabular}{|c|c|c|c|}
\hline
Gepner model &$\lambda$ & $\Delta_{gap}^{CA}$ & $\Delta_{gap}^{CC}$
\\
\hline
$3^5$ & $\left(\Gamma \left( {3\over 5}\right)\over \Gamma \left({2\over 5} \right) \right)^{15\over 2}\left(\Gamma \left( {1\over 5}\right)\over \Gamma \left({4\over 5} \right) \right)^{5\over 2}$ & ${4\over 5}$ & ${6\over 5}$
\\
\hline
$4^4 1$ & $0$ & ${2\over 3}$ & ${2\over 3}$
\\
\hline
$6^4$ & $\frac{\Gamma \left(\frac{1}{8}\right)^2 \Gamma \left(\frac{5}{8}\right)^6}{\Gamma \left(\frac{3}{8}\right)^6 \Gamma \left(\frac{7}{8}\right)^2}$ & ${1\over 2}$ & $1$
\\
\hline
$8^3 3$ & 0 & ${2\over 5}$  & ${2\over 5}$
\\
\hline
\end{tabular}
 \end{center}
 \caption{CA and CC gaps in $c=9$ Gepner models}
 \label{Tab:Gep}
 \end{table}

\bibliographystyle{JHEP}
\bibliography{N2draft}

 \end{document}